\documentclass[12pt]{article}
\usepackage{url}
\usepackage{enumitem}
\usepackage{amssymb} 
\usepackage{amsmath} 
\usepackage{latexsym} 
\usepackage{amsfonts}
\usepackage{natbib}
\usepackage{geometry}                		
\usepackage[parfill]{parskip}    		% Activate to begin paragraphs with an empty line rather than an indent									
\usepackage{caption}
\usepackage{color}
\usepackage{float}
\usepackage{setspace}
\usepackage{lscape}
\usepackage[colorlinks=false,linkcolor=black,citebordercolor={1 0 0}]{hyperref}
\usepackage[pdftex]{graphicx}
\usepackage{enumitem}
\usepackage{lscape}
\usepackage[center]{subfigure}
\usepackage{captcont}
\usepackage[font=small,labelfont=bf]{caption}
\usepackage{tabularx}
\usepackage{rotating}
\usepackage{comment}
\usepackage[flushleft]{threeparttable}

\usepackage{booktabs}

\newcommand{\sign}{{\rm sign}}
\newcommand{\Var}{{\rm Var}}
\newcommand{\Cov}{{\rm Cov}}
\newcommand{\Corr}{{\rm Corr}}

\newtheorem{example}{Example}[section]
\newtheorem{thm}{Theorem}[section]

\newtheorem{lemma}{Lemma}[section]

\newtheorem{rem}{Remark}[section]
\renewcommand{\theequation}{\arabic{equation}}
\bibpunct{(}{)}{;}{a}{,}{,}

\begin{document}

\smallskip

\smallskip

\renewcommand{\baselinestretch}{1.2}
\centerline{\textbf{New Approaches to Robust Inference on Market (Non-)Efficiency,}} \centerline{\textbf{Volatility Clustering and Nonlinear Dependence}}
\medskip \medskip {\centerline{Rustam Ibragimov}}\smallskip

{\centerline{\textit{Imperial College Business School}}}\smallskip %{and St. Petersburg State University}}}\smallskip

\bigskip {\centerline{Rasmus S\o ndergaard Pedersen}}\smallskip

{\centerline{\textit{Department of Economics, University of Copenhagen}}}

\bigskip {\centerline{Anton Skrobotov}}\smallskip

{\centerline{\textit{Russian  Presidential Academy of National Economy and Public Administration (RANEPA)}}} {\centerline{\textit{and St. Petersburg State University}}}

%\begin{center}
%CONFERENCE DRAFT PAPER -- PLEASE DO NOT CITE OR CIRCULATE WITHOUT %THE AUTHOR'S PERMISSION
%\end{center}

\medskip \medskip
\par \bigskip
{\small{{\centerline{\bf Abstract}}
Many financial and economic variables, including financial returns, exhibit nonlinear dependence, heterogeneity and heavy-tailedness. These properties may make problematic the analysis of (non-)efficiency and volatility clustering in economic and financial markets using traditional approaches that appeal to asymptotic normality of sample autocorrelation functions of returns and their squares.

This paper presents new approaches to deal with the above problems. We provide the results that motivate the use of measures of market (non-)efficiency and volatility clustering based on (small) powers of absolute returns and their signed versions. 

We further provide new approaches to robust inference on the measures in the case of general time series, including GARCH-type processes. The approaches are based on robust $t-$statistics tests and new results on their applicability are presented. In the approaches, parameter estimates (e.g., estimates of measures of nonlinear dependence) are computed for groups of data, and the inference is based on $t-$statistics in the resulting group estimates. This results in valid robust inference under heterogeneity and dependence assumptions satisfied in real-world financial markets. Numerical results and empirical applications confirm the advantages and wide applicability of the proposed approaches.
 
 %Because this is the first time the econometrically justified robust methods have been used in the analysis of market (non-)efficiency and volatility clustering, the exposition is presented in some detail with illustrations of the problems with definition and inference on the standard ACF-based dependence measures in financial and economic markets,  motivation for the use of their analogues based on powers of the returns and robust inference approaches for the latter measures.

%{\color{blue}\textbf{RI: The abstract seems too long.}}\\
 \medskip

\medskip}}

%\textit{JEL Classification:} \medskip C12, C14, G12, G14

%\renewcommand{\baselinestretch}{2}
\section{Introduction}\label{intro}

%\subsection{Stylized facts of financial markets} 

Many studies argue that time series of financial returns, $R_t$, and other key economic and financial variables and indicators like foreign exchange rates exhibit several common statistical properties, often referred to as stylized facts; see e.g. \citet[Ch. 2]{CLM}, \cite{cont2001empirical}, \citet[Ch. 1-2]{Taylor}, \citet[Ch. 1-3]{Tsay}, \citet[Ch. 1]{ChrisRM}, \citet[Ch. 3]{MFE}, and references therein. The following three properties are the most important stylized facts that much of the empirical literature agrees upon, together with the standard mean-zero property, $E(R_t)=0$, implying the absence of systematic gains or losses:
\begin{enumerate}[label=(\roman*)]
  \item \label{Autocorr} Absence of linear dependence and linear autocorrelations: 
  $\Corr(R_t, R_{t-h})\approx 0,$ even for small lags $h=1, 2, ...$
 % \begin{eqnarray} \label{corr} \nonumnber
 % \Corr(R_t, R_{t-h})\approx 0,
  %\end{eqnarray} even for small lags $h=1, 2, ...,$
  
  \item \label{AutocorrSq} The presence of nonlinear dependence and volatility clustering, captured by significant positive autocorrelation in simple nonlinear functions of the returns and different measures of volatility, such as squared returns: $\Corr(R_t^2, R_{t-h}^2)>>0,$   even for large lags $h>0.$ %This property implies, in particular, that financial returns are not i.i.d. %and thus the \emph{strong market efficiency hypothesis} does not hold.
 % \begin{eqnarray} \label{corrSq} \nonumber
 % \Corr(R_t^2, R_{t-h}^2)>>0,
%  \end{eqnarray} 
    %even for large lags $h>0.$ This property implies, in particular, that financial returns are not i.i.d. and thus the \emph{strong market efficiency hypothesis} does not hold.
  \item \label{Heavy} Heavy tails: The (unconditional) returns distributions exhibit heavy power-law tails, $\lim_{x\to + \infty}x^{\zeta}P(|R_t|>x)= C$, with a constant $C>0$ and the tail index $\zeta>0.$ %(e.g., \cite{cont2001empirical}).
  %\begin{eqnarray} \label{Heavy1} 
 % P(|R_t|>x)\sim C/x^{\zeta}, 
%  \end{eqnarray}
 
  %\footnote{Following the standard definition and notation, for two positive functions $f,g$, $f(x) \sim g(x)$, if $f(x)/g(x) \to 1$ as $x\to \infty$.}\footnote{The above properties and other stylized facts have been established for different frequencies, including, e.g., weekly, monthly and high-frequency returns (see \cite{cont2001empirical}); in this paper, we focus on daily returns for simplicity of presentation.}
\end{enumerate}
%\textcolor{red}{RSP: Perhaps write instead: $\lim_{x\to + \infty}x^{\zeta}P(|R_t|>x)= C$, so that we do not have to define $\sim$?} 
%%\textbf{Put efficiency refs in footnote or the above okay?}
%%\textcolor{green}{\textbf{We can add if referee ask - now it's ok. A.S.}}

%The analysis and modelling of properties \ref{Autocorr}-\ref{AutocorrSq} have been central to the development of modern financial theory and financial econometrics, including the development of market efficiency hypotheses by E. Fama and (G)ARCH time series by R. Engle and C. Granger. 
%{\color{blue}\textbf{The standard approach to modeling the above stylized facts, including nonlinear dependence, volatility clustering and heavy-tailedness property \ref{Heavy}, in financial returns is based on the use of GARCH-type processes (op. cit.).}} 

%\textcolor{red}{I have edited the following paragraph. /RSP}

Note that property \ref{Autocorr} is often cited as support for the \emph{market efficiency hypothesis} (see, among others, \citeauthor{cont2001empirical}, \citeyear{cont2001empirical}, and references therein). %, \textcolor{red}{that is, for the martingale difference property of financial returns. %Note that mds implies uncorrelatedness, but not the other way around. 
%Property \ref{AutocorrSq} implies that financial returns are not i.i.d.
A standard way of testing properties \ref{Autocorr}-\ref{AutocorrSq} for a given sample $(R_t)_{t=1,\ldots,T}$ is to compute the sample counterparts of the correlations and rely on limiting Gaussian distributions of these (suitably scaled and standardized) statistics together with some ``robust" (heteroskedasticity and autocorrelation consistent, HAC-type) estimates of their asymptotic variances; see, e.g., \cite{baltussen2019} for a recent application in the analysis of linear autocorrelations in returns on major world stock indices.  Such approaches may not be reliable under  heavy-tailedness and nonlinear dependence (see, e.g., \citeauthor{GO}, \citeyear{GO}, \citeauthor{DM1998AS}, \citeyear{DM1998AS}, \citeauthor{MS2000AS},  \citeyear{MS2000AS}; Sec$.$ 5.3 in \citeauthor{cont2001empirical}, \citeyear{cont2001empirical}; Sec$.$ 3.3.3 in \citeauthor{IIW}, \citeyear{IIW}, and references therein). Properties \ref{Autocorr}-\ref{AutocorrSq} are commonly modelled using the much celebrated class of GARCH-type processes. Depending on the tail index $\zeta$, characterizing the degree of heavy-tailedness in property \ref{Heavy} of a given (GARCH-type) process, the correlations may not be defined and/or the sample correlations may have limiting distributions given by functions of multivariate non-Gaussian stable distributions, or may be inconsistent. For instance, $\Corr(R_t, R_{t-h})$ ($\Corr(R_t^2, R_{t-h}^2)$) is defined only if $\zeta > 2$ ($\zeta > 4$), and asymptotic normality of the sample counterpart requires that $\zeta > 4$ ($\zeta > 8$). The latter condition is hardly justified empirically, as it is typically found that $\zeta \in (2, 4)$ for financial returns in developed markets, whereas emerging markets' returns may even have $\zeta<2$ and, hence, infinite variances (e.g., \citeauthor{LoretanAndPhillips94}, \citeyear{LoretanAndPhillips94}, \citeauthor{cont2001empirical}, \citeyear{cont2001empirical}, Sec. 1.2 and 3.2 in \citeauthor{IIW}, \citeyear{IIW},  and references therein). Note that the applicability of HAC inference approaches typically relies on moments of even higher order to be finite. Moreover,  HAC-based inference methods often have poor finite sample properties, even in rather standard inference problems, especially with data with pronounced dependence and heterogeneity; see, among others, \cite{Andrews}, \cite{AM}, \cite{DL}, %\cite{KVB}, \cite{KV1},  the discussion in \cite{Ph}, \cite{Muller0, Muller}, 
and \cite{IM2010} [IM (2010), henceforth]. %, and \cite{LS}.

In this paper, we present new robust approaches for dealing with the issue of heavy tails in testing for (non-)efficiency, volatility clustering and nonlinear dependence in financial return series. We exploit the property that if $R_t$ has power-law tails with the index $\zeta >0$, as in \ref{Heavy}, then for any $p>0$, $|R_t|^p$ has the tail index $\zeta/p$. This suggests that, even under pronounced heavy tails, correlations are well-defined and Gaussian limiting distributions for sample correlations can be obtained under suitable power transformations of the original return time series. Specifically, as a natural analogue to nonlinear dependence property \ref{AutocorrSq}, we consider the property
\begin{enumerate}[label=(\roman*')]
\setcounter{enumi}{1} \item  \label{AbsPower} $\Corr(|R_t|^p, |R_{t-h}|^p)>>0,$  even for large lags $h>0$ for some $p>0$. %(\textbf{sufficiently} small)  
\end{enumerate}

%\textcolor{red}{In order to shorten down the introduction, I removed the paragraph starting with "In (ii'),...". /RSP}
Under heavy-tailedness property \ref{Heavy}, the correlations in \ref{AbsPower} are well-defined whenever $2p<\zeta,$ and, as is shown in this paper, under general conditions, asymptotic normality of the corresponding sample correlations holds if $4p<\zeta$.

We further propose the correlations $\Corr(R_t, |R_{t-h}|^s \sign(R_{t-h})),$ $s>0,$ of `signed' powers of absolute returns as measures of market (non-)efficiency. Similar to the power transformations in \ref{AbsPower}, these measures lead to formulation of natural analogues of property \ref{Autocorr} under heavy-tailed and conditionally heteroskedastic time series:
%The analogue of property \ref{Autocorr} has the form
% \begin{enumerate}[label=(\roman*')]\setcounter{enumi}{0} \item \label{Autocorr1}  $ corr(R_t, |R_{t-h}|^s sign(R_{t-h}))\approx 0,$ even for small lags $h=1, 2, ...,$ \end{enumerate} and
\begin{enumerate}[label=(\roman*')]\setcounter{enumi}{0} 
\item \label{Autocorr2}  $\Corr(R_t, |R_{t-h}|^s \sign(R_{t-h}))\approx 0,$ even for small lags $h=1, 2, ...$ for some $s>0$.
\end{enumerate}
Similar to property \ref{AbsPower}, the correlations in \ref{Autocorr2} are well-defined for $\zeta > 2 \max(1, s)$, and, under suitable conditions, asymptotic normality of the corresponding sample correlations holds if $\zeta > 2(1+s)$.

To the best of our knowledge, property \ref{AbsPower} for powers of absolute returns was originally considered by \cite{DGE} in relation to detecting long-memory in returns. The purpose of power transformations in the present manuscript is different in the sense that power transformations of returns serve as a \emph{necessary} step for making the corresponding correlations well-defined %and their estimators \textbf{asymptotically normal} (or converge to scale mixtures of normals such as stable distributions in the case of signed measures), 
and for carrying out reliable inference in the presence of heavy-tailedness and conditional heteroskedasticity in returns.

The main contribution of this paper is the development of robust approaches to inference on measures of market (non-)efficiency, nonlinear dependence, and volatility clustering, such as the correlations in \ref{Autocorr2} and \ref{AbsPower}. Firstly, we establish asymptotic normality of sample auto(cross)correlations of arbitrary transformations of a time series under general mixing conditions for the data-generating process (DGP). Further, in order to avoid (HAC-based) estimation of limiting variances of the statistics, we propose robust $t-$statistic inference approaches in the spirit of IM (\citeyear{IM2010}, \citeyear{IM2016}), and prove their asymptotic validity for general classes of DGPs, including GARCH-type time series. A similar approach was recently considered in \cite{Pedersen} in relation to inference about an autoregressive coefficient in linear autoregressive models in the presence of heavy-tailed \emph{symmetric} GARCH-type errors. In contrast, the robust inference approaches proposed in this paper do not impose any symmetry restriction on the DGPs. This is a desirable feature, as it quite is common for financial time series to have skewed marginal distributions (gain/loss asymmetry) as well as leverage effects (e.g., \citeauthor{cont2001empirical}, \citeyear{cont2001empirical}).

We provide a numerical analysis that demonstrates appealing finite sample properties of the robust $t-$statistic inference approaches. Lastly, we revisit the aforementioned study by \cite{baltussen2019} and illustrate the applicability of the approaches in relation to inference on properties \ref{Autocorr2} and \ref{AbsPower} in major stock market indices. Importantly, we document that all the associated return series are likely to have tail indices $\zeta <4$, i.e. infinite fourth moments. Moreover, when applying our robust approaches, taking into account return heavy-tailedness, we find weak evidence of negative serial dependence in returns, in contrast to the conclusions made by \cite{baltussen2019}.

The $t-$statistic approaches to robust inference in IM (\citeyear{IM2010}, \citeyear{IM2016}) and those proposed in this paper complement and are related to inference approaches based on self-normalization (see the review in \citeauthor{PLS}, \citeyear{PLS}, \citeauthor{Shao}, \citeyear{Shao}; Remark \ref{tHAC11} and references therein) and fixed-smoothing (fixed-$b$) heteroskedasticity and autocorrelation robust (HAR) methods that do not rely on consistency of limiting variance estimators and use nonstandard ``fixed-b" asymptotics or Student-$t$ or $F$ distributional approximations (see \citeauthor{KVB}, \citeyear{KVB}, \citeauthor{KV}, \citeyear{KV}, \citeyear{KV1}, \citeauthor{Jansson}, \citeyear{Jansson}, %\citeauthor{Ph1}, \citeyear{Ph1},  
\citeauthor{Muller0}, \citeyear{Muller0}, 
\citeyear{Muller}, \citeauthor{SPJ}, \citeyear{SPJ}, \citeauthor{Sun1}, \citeyear{Sun1}, \citeyear{Sun3, Sun2};  \citeauthor{LS}, \citeyear{LS}, \citeauthor{LS1}, \citeyear{LS1}, and Remark \ref{tHAC1} in this paper).

The paper is organized as follows. Section \ref{GARCH} introduces and discusses measures of serial and nonlinear dependence based on autocorrelations of powers of absolute returns, and provides asymptotic theory for estimators of these measures. We further propose robust $t-$statistic approaches for reliable inference on the measures, and show that the approaches are asymptotically valid under general conditions. Section \ref{sec:simulations} investigates the finite-sample properties of the inference methods. Section \ref{sec:empirical} provides an empirical illustration of the inference methods. Section \ref{sec:conclusion} concludes and discusses suggestions for future research. Proofs and additional simulation results can be found in the Online Appendix.

\section{Inference on measures of market (non-)efficiency, nonlinear dependence and volatility clustering} \label{GARCH}
\subsection{Autocovariances and -correlations for transformed returns} \label{Sample} The results in this paper hold for a wide class of stationary time series processes that satisfy mixing and moment conditions stated below. (Throughout, ``stationarity" refers to the notion of strict stationarity.) This includes heavy-tailed GARCH($p, q$) time series, generalized GARCH processes (e.g., \citeauthor{Pedersen}, \citeyear{Pedersen}), and heavy-tailed stochastic volatility processes (e.g., \citeauthor{davis:mikosch:2001}, \citeyear{davis:mikosch:2001}), among others.  To present the main ideas, we focus on the GARCH(1,1) process as an ongoing example.
%{\color{blue}\textbf{Very minor: maybe remove 'strictly'. changed 'illustrate' to 'present'}}
%{\color{blue} \textbf{What processes should we consider/present the results for? Maybe GARCH(p, q) or general GARCH as in Rasmus' paper? Follow-up projects on tests of market efficiency (strong and weak), new intertemporal dependence measures based on characteristic functions... }}
%{\color{red}\textbf{RSP: For the sake of clarity, we should focus on GARCH(1,1) as our main example. But all our results hold for a huge class of processes, as we essentially need a CLT for mixing processes, and the coupling argument that holds for strongly mixing processes. Even the stable limit theory holds for a wide range of processes [see e.g. the high-level conditions in Hsing and Davis (1995) and Davis and Mikosch (1998).]}}

Let $\mathbb{Z}=\{..., -2, -1, 0, 1, 2, ...\}.$ A GARCH($1,1$) process, $(R_t)_{t\in \mathbb{Z}}$, is given by
\begin{eqnarray} \label{ReturnsEq} 
R_t=\sigma_t Z_t, \; t\in \mathbb{Z}, \end{eqnarray}
where $(Z_t)_{t\in \mathbb{Z}}$ is a sequence of i.i.d. random variables (r.v.'s) with mean zero and unit variance, $E(Z_t)=0$ and $\Var(Z_t)=1$, and $(\sigma_t^2)_{t\in \mathbb{Z}}$ is a conditional volatility process,
\begin{eqnarray}
\label{VolEq} \sigma_t^2=\omega + \alpha R_{t-1}^2 + \beta \sigma_{t-1}^2, \quad \omega >0,\ \ \alpha,\beta \ge 0.  \end{eqnarray}
%with $\omega >0$ and $\alpha,\beta \ge 0$. %In the following we give a brief overview of the most important properties of the GARCH(1,1) process.

%Throughout the paper, $Z$ denotes a r.v. that has the same distribution as that of the r.v.'s $(Z_t).$

%\textbf{I think this sentence is need to be precise with notation}

As is well-known, the process in (\ref{ReturnsEq})-(\ref{VolEq}) has a  stationary and ergodic version if and only if $E[\log(\alpha Z_t^2+\beta)]<0$ (e.g., \citeauthor{Nelson}, \citeyear{Nelson}). 
%Also, in this case, stationarity of $(\sigma_t^2)$ implies stationarity of the GARCH(1, 1) process $(R_t).$
%\footnote{In the case of a GARCH($p$, $q$) model (\ref{ReturnsEq}) with the volatility process $\sigma_t^2=\omega + \sum_{i=1}^p \alpha_i R_{t-i}^2 + \sum_{j=1}^q \beta_j \sigma_{t-j}^2,$ where $\omega, \alpha_i, \beta_j\ge 0,$ sufficient conditions for existence of a stationary solution $(R_t),$ $\sigma_t^2$ are given by $\omega>0,$ $\sum_{i=1}^p \alpha_i + \sum_{j=1}^q \beta_j\le 1,$ certain restrictions on the distribution of $Z,$ and some further technical conditions (see \cite{BP}, the discussion in \cite{DM1} and references therein). The condition $\sum_{i=1}^p \alpha_i + \sum_{j=1}^q \beta_j<1$ is also necessary and sufficient for the finiteness of $Var(R_t)<\infty,$ and thus for the second order stationarity of GARCH($p$, $q$) process $(R_t).$ }
%\footnote{In the case of ARCH(1) process with $\beta=0$ and the standard normal $Z,$ the condition $E[\log(\alpha_1 Z^2)]<0$ holds if $0<\alpha<2e^{\gamma_0}\approx 3.568,$ where $\gamma_0$ is Euler's constant.} 
In addition, under mild conditions on the distribution of $Z_t$, e.g., if it has a Lebesgue density, the GARCH process is $\beta$-mixing with geometric rate; e.g., \citet[Thm. 3]{francq:zakoian:2006}. This implies that the process is also $\alpha$-mixing with geometric rate (see, e.g., \citeauthor{rio2017}, \citeyear{rio2017}, for additional details on mixing processes). 
%This in turn ensures that the process is also strongly, or $\alpha$-, mixing with geometric decay, which (under suitable conditions) enables us to apply a central limit theorem (CLT) to general transformations of $R_t$. We refer to Appendix \ref{sec:mixing} for additional details about $\alpha$- and $\beta$-mixing processes.  
%{\color{blue}\textbf{Are the conditions $\omega>0$ and $E[\log(\alpha_1Z^2+\beta_1)]<0$ necessary and sufficient for stationarity? Example 1 in \cite{DM2}}} {\color{red}\textbf{RSP: Yes.}}
 %define $Z:$ a r.v. with the same distribution as that of
Under, essentially, the conditions listed above, the stationary solution  to (\ref{ReturnsEq})-(\ref{VolEq}) satisfies Kesten's theorem (e.g., \citeauthor{MS2000AS}, \citeyear{MS2000AS}). Specifically, the unconditional distribution of $R_t$ has power-law tails as in \ref{Heavy} with the tail index $\zeta>0$ given by the unique positive solution to the equation
\begin{eqnarray} \label{KestenEq} 
E[(\alpha Z_t^2 + \beta)^{\zeta/2}]=1.
\end{eqnarray}
For instance, $\zeta \in (2,4)$ if $1-(\kappa_Z-1)\alpha^2<(\alpha + \beta)^2 < 1$, %and $\alpha^2 \kappa_Z + \beta^2 +2\alpha\beta > 1$, 
where $\kappa_Z \equiv E[Z_t^4].$ %is the kurtosis of $Z_t$.

Given a stationary process, $(R_t)_{t\in\mathbb{Z}}$, we consider the following population autocovariance and autocorrelation functions of order $h$ for measuring nonlinear dependence and volatility clustering in the process. We emphasize that the measures are non-zero if the process is conditionally heteroskedastic. For $p>0$ and $E[|R_t|^{2p}]<\infty$, let
\begin{align}
    \gamma_{|R|^p}(h) & = \Cov(|R_t|^p, |R_{t-h}|^p),\quad h=0,1,\ldots, \label{eq:autocov:power}\\
    \rho_{|R|^p}(h) & = \Corr(|R_t|^p, |R_{t-h}|^p) = \frac{\gamma_{|R|^p}(h)}{\gamma_{|R|^p}(0)}, \quad  h=1,2\ldots \label{eq:autocorr:power}
\end{align}

To quantify the degree of efficiency, i.e. if $R_t$ is predictable with respect to its lagged values, we define, for $s>0$ and $E[|R_t|^{1+s}]<\infty$,
\begin{align}
    \gamma'_{R, |R|^s \sign(R)}(h) & = \Cov(R_t, |R_{t-h}|^s \sign(R_{t-h})),\quad h=0,1,\ldots, \label{eq:autocrosscov:power}
\end{align}
and for $\max\{ E[|R_t|^{2s}], E[|R_t|^{2}]  \} < \infty$, denote
\begin{align}
    \rho'_{R, |R|^s \sign(R)}(h) & = \Corr(R_t, |R_{t-h}|^s \sign(R_{t-h})) = \frac{\gamma'_{R, |R|^s \sign(R)}(h)}{\sqrt{\gamma_{R}(0) \gamma_{|R|^s \sign(R)}(0)}}  , \quad  h=1,2,\ldots, \label{eq:autocrosscorr:power}
\end{align}

where $\gamma_{R}(0)=\Var(R_t)$ and $\gamma_{|R|^s \sign(R)}(0)=\Var(|R|^s \sign(R))$.
%(see the general notation in the Online Appendix).
\begin{example} \label{Example1}
As indicated in the introduction, in the presence of heavy tails, e.g., when $(R_t)_{t\in\mathbb{Z}}$ follows a GARCH(1,1) process with the tail index $\zeta > 0$, the quantities in \eqref{eq:autocov:power} and \eqref{eq:autocorr:power} are defined if $\zeta > 2p$. Likewise, the covariances in \eqref{eq:autocrosscov:power} are defined if $\zeta > 1+s$, and the correlations in  \eqref{eq:autocrosscorr:power} are defined if $\zeta > 2\max\{1,s\}$.
\end{example}
%{\color{blue}\textbf{Should we say 'unconditional' tail index or 'unconditional power law distribution or okay as is - included 'unconditional' before Kesten's result before. Changed from GARCH to general heavy-tailed proceses.}}
\begin{rem}
For $s=1$, \eqref{eq:autocrosscov:power} and \eqref{eq:autocrosscorr:power} are identical to the usual linear autocovariances and autocorrelations, respectively.  For $s\ne 1$, \eqref{eq:autocrosscov:power} and \eqref{eq:autocrosscorr:power} are also able to detect market (non-)efficiency. In particular, if $(R_t)_{t\in\mathbb{Z}}$ is a martingale difference sequence, e.g. if it is a GARCH process, the quantities in  \eqref{eq:autocrosscov:power} and \eqref{eq:autocrosscorr:power} are equal to zero, exactly like the usual linear autocovariances and autocorrelations.
\end{rem}

%{\color{blue}\textbf{Should we add 'provided they are defined needed at the end of the sentence - don't want to consider IGARCH...}}
%In the above notation, analogues \ref{Autocorr2}, \ref{AbsPower} of stylized facts \ref{Autocorr}, \ref{AutocorrSq}, \ref{AbsVol} on absence of linear autocorrelations and presence of nonlinear dependence and volatility clustering in financial returns become

%\begin{enumerate}[label=(\roman*')]\setcounter{enumi}{0}  \item \label{Autocorr2New}  $\gamma'_{R, |R|^s sign(R)}(h), \rho'_{R, |R|^s sign(R)}(h)\approx 0,$ even for small lags $h=1, 2, ...$ \end{enumerate}

%\begin{enumerate}[label=(\roman*'')]\setcounter{enumi}{1} \item  \label{AbsPowerNew} $\gamma_{|R|^p}(h), \rho_{|R|^p}(h)>>0,$  even for large lags $h>0.$ \end{enumerate}

In the next section, we consider estimation of the dependence measures and provide large sample theory for their estimators.

\subsection{Limit theory for sample \textbf{dependence} measures}\label{AsNorm}
Let $(R_t)_{t=1,\ldots,T}$ be a sample of observations. Denote by $\hat{\mu}_{R}$, $\hat{\mu}_{|R|^p}$ and $ \hat{\mu}_{|R|^s \sign(R)}$, respectively, the sample means of $R_t$, $|R_t|^p$, and $|R_t|^s \sign(R_t)$, for $p,s>0$, i.e.
\begin{align} \label{eq:sample_means}
  & \hat{\mu}_{R}  = \frac{1}{T} \sum_{t=1}^T R_t, \quad \hat{\mu}_{|R|^p}  =\frac{1}{T} \sum_{t=1}^T  |R_t|^p, \quad \hat{\mu}_{|R|^s \sign(R)}  =\frac{1}{T} \sum_{t=1}^T  |R_t|^s \sign(R_t).
\end{align}
The sample versions of \eqref{eq:autocov:power} and \eqref{eq:autocorr:power} are given, respectively, by
\begin{align}
    \hat{\gamma}_{|R|^p}(h) &= \frac{1} {T} \sum_{t=h+1}^{T} (|R_t|^p-\hat{\mu}_{|R|^p}) (|R_{t-h}|^p-\hat{\mu}_{|R|^p}), \label{eq:sample:autocov:power} \\
    \hat{\rho}_{|R|^p}(h) &= \frac{\hat{\gamma}_{|R|^p}(h)}{\hat{\gamma}_{|R|^p}(0)} . \label{eq:sample:autocorr:power}
\end{align}
Likewise, the sample versions of \eqref{eq:autocrosscov:power} and \eqref{eq:autocrosscorr:power} are
\begin{align}
    \hat{\gamma'}_{R, |R|^s \sign(R)}(h) &= \frac{1} {T} \sum_{t=h+1}^{T} (R_t-\hat{\mu}_{R}) (|R_{t-h}|^s \sign(R_{t-h})-\hat{\mu}_{|R|^s \sign(R)}), \label{eq:sample:autocrosscov:power} \\
    \hat{\rho'}_{R, |R|^s \sign(R)}(h) &= \frac{\hat{\gamma'}_{R, |R|^s \sign(R)}(h)}{\sqrt{\hat{\gamma}_{R}(0)\hat{\gamma}_{|R|^s \sign(R)}(0)}}, \label{eq:sample:autocrosscorr:power}
\end{align}
where $\hat{\gamma}_{R}(0)$ and $\hat{\gamma}_{|R|^s \sign(R)}(0)$ denote the sample variances of $R_t$ and $|R_t|^s \sign(R_t)$ defined in the usual way similar to $\hat{\gamma}_{|R|^p}(0)$ in (\ref{eq:sample:autocov:power}).

%In the case $s=1,$ the estimators $\hat{\rho}_{R, |R|^s sign(R)}(h)$ become the usual sample linear autocovariances and autocorrelations $\hat{\gamma}_{R}(h)=\frac{1} {T} \sum_{t=h+1}^{T} (R_t-\hat{\mu}_{R}) (R_{t-h}-\hat{\mu}_{R}),$ $\hat{\rho}_{R}=\hat{\gamma}_{R}(h)/\hat{\gamma}_{R}(0)$ of $R_t.$
%As in \cite{DM1998AS,  DMBACF, DM2000ACF, MS2000AS}, we note that usually, centering around the population/sample mean is included in the definition of the population/sample ACVF and ACF's; however, in the heavy-tailed
%case, centering with the sample mean is not relevant for asymptotic results. Arguments similar to those in the paper yield the same limits in the case of the centered versions of the ACVF and ACF's considered.

% Also cite and consider \cite{DMBernoulli} in a review and applications of t-statistics

The following Lemmas \ref{NormalAsymp} and \ref{NormalAsymp1} provide a basis for asymptotic inference on the properties \ref{AbsPower} and \ref{Autocorr2}, respectively. The lemmas follow from the general results in the Online Appendix for sample autocovariances and autocorrelations of \emph{arbitrary} functions of $\alpha$-mixing processes. 
%{\color{red}\textbf{Maybe should include (non-)efficiency at the end of the sentence.}}

\begin{lemma} \label{NormalAsymp} Let $(R_t)_{t\in \mathbb{Z}}$ be a stationary $\alpha$-mixing process. For $p>0,$ assume that there exists a value $\delta >0$ such that $E[|R_t|^{4p+\delta}]<\infty,$ and such that the mixing coefficients $\alpha(n)$ satisfy $\sum_{n=1}^\infty \alpha(n)^{\delta/(2+\delta)}<\infty$. Then, with $\hat{\gamma}_{|R|^p}(h)$ and $\hat{\rho}_{|R|^p}(h)$ defined in \eqref{eq:sample:autocov:power} and \eqref{eq:sample:autocorr:power}, respectively, for a fixed integer $m$, one has 
\begin{eqnarray} \label{Autocov1} \sqrt{T}(\hat{\gamma}_{|R|^p}(h)-{\gamma}_{|R|^p}(h))_{h=0, 1,\ldots, m}\rightarrow_d (G_{h, p})_{h=0, \ldots, m},
\end{eqnarray}
\begin{eqnarray} \label{Autocor1} \sqrt{T}(\hat{\rho}_{|R|^p}(h)-{\rho}_{|R|^p}(h))_{h=1,\ldots, m}\rightarrow_d (H_{h, p})_{h=1,\ldots, m},
\end{eqnarray}
where the limits are multivariate Gaussian with mean zero. %Asymptotic normality in (\ref{Autocor1})  further holds for all $0<p<\zeta/4,$ if ${\rho}_{|R|^p}(h))_{h=0, 1,\ldots, m}=0.$  % The result for autocorrelations follows from continuous mapping
\end{lemma}

\begin{lemma} \label{NormalAsymp1} Let $(R_t)_{t\in \mathbb{Z}}$ be a stationary $\alpha$-mixing process. For $s>0$, assume that there exists a value $\delta >0$ such that $E[|R_t|^{2(1+s)+\delta}]<\infty,$ and such that the mixing coefficients $\alpha(n)$ satisfy $\sum_{n=1}^\infty \alpha(n)^{\delta/(2+\delta)}<\infty$. Then, with $\hat{\gamma'}_{R, |R|^s \sign(R)}(h)$ defined in \eqref{eq:sample:autocrosscov:power}, one has 
\begin{eqnarray} \label{signs} 
\sqrt{T}(\hat{\gamma'}_{R, |R|^s sign(R)}(h)-{\gamma'}_{R, |R|^s \sign(R)}(h))_{h=0, 1,\ldots, m}\rightarrow_d (G'_{h, s})_{h=0, 1,\ldots, m},
\end{eqnarray}
where $(G'_{h, s})_{h=0, 1,\ldots, m}$ is multivariate Gaussian with mean zero.

If ${\gamma'}_{R, |R|^s \sign(R)}(h))_{h=1,\ldots, m} = (0,\ldots,0)$, with $\hat{\rho'}_{R, |R|^s \sign(R)}(h)$ defined in \eqref{eq:sample:autocrosscorr:power}, one has
\begin{eqnarray}
\sqrt{T}(\hat{\rho'}_{R, |R|^s \sign(R)}(h))_{h=1,\ldots, m}\rightarrow_d ((\gamma_{R}(0)\gamma_{|R|^s \sign(R)}(0))^{-1/2}G'_{h, s})_{h=1,\ldots, m}. \label{eq:conv:autocrosscorr:nullity}
\end{eqnarray} 
If $\max\{E[|R_t|^{4+\delta}], E|R_t|^{4s+\delta}]\}<\infty$, then
\begin{eqnarray}
\sqrt{T}(\hat{\rho'}_{R, |R|^s \sign(R)}(h)-\rho'_{R, |R|^s \sign(R)}(h) )_{h=1,\ldots, m}\rightarrow_d (H'_{h, s})_{h=1,\ldots, m}, \label{eq:conv:autcrosscorr:general}
\end{eqnarray}
where $(H'_{h, s})_{h=1,\ldots, m}$ is multivariate Gaussian with mean zero.
\end{lemma}

%{\color{red}{maybe drop 'for a fixed integer $m$ - somewhat cumbersome formulation}}
In Lemma \ref{NormalAsymp1}, the moment conditions for asymptotic normality of sample covariances in (\ref{signs}) are weaker than those of sample correlations in (\ref{eq:conv:autcrosscorr:general}). The reason is that asymptotic normality of sample correlations relies on joint asymptotic normality of sample covariances and variances of $R_t$ and $|R_t|^s \sign(R_t)$, if the true correlations are non-zero. If the true correlations are zero, as in \eqref{eq:conv:autocrosscorr:nullity}, the convergence of the sample correlations only relies on asymptotic normality of the sample covariances and consistency of the sample variances, which in turn requires the same moment conditions as for convergence of sample covariances.
%{\color{red}{Dropped 'generally' in 'generally weaker', used 'auto(cross)covariances and auto(cross)correlations. Maybe this is not needed for shortness.}}

%\textcolor{red}{I combined the old Example 2.2 and Remark 2.3. }

\begin{example} \label{ex:garch:appl:limit}
 As discussed in Section \ref{Sample}, under suitable conditions, GARCH(1,1) processes are $\alpha$-mixing with geometric decay, and hence satisfy the conditions on the mixing coefficients in Lemmas \ref{NormalAsymp} and \ref{NormalAsymp1}. In particular, Lemma \ref{NormalAsymp} holds if the tail index $\zeta > 4p$. Whenever $\zeta > 2(1+s),$ the normal asymptotics in (\ref{signs}) and (\ref{eq:conv:autocrosscorr:nullity}) hold for the GARCH(1, 1) processes. If the moment conditions in Lemmas \ref{NormalAsymp} and \ref{NormalAsymp1} are not satisfied, the rates of convergence of the sample covariances and correlations are slower than $\sqrt{T}$ and the limits are given by functions of r.v.'s with non-Gaussian (in general, asymmetric) stable distributions. Importantly, the rates of convergence and the limiting distributions depend on the (unknown) tail index $\zeta$ and the powers $p$ and $s$. E.g., from the results in \cite{DM1998AS} and \cite{MS2000AS} (see also \citeauthor{DM2}, \citeyear{DM2}) it follows that, in the case $p=1$ and $\zeta \in (2, 4),$ $T^{1-2/\zeta}(\hat{\gamma}_{|R|}(h)-{\gamma}_{|R|}(h))$ has an infinite variance asymmetric stable limiting distribution with the index of stability given by $\zeta/2.$ A similar result applies to $T^{1-2/\zeta}(\hat{\gamma}_{|R|^2}(h)-{\gamma}_{|R|^2}(h))$ when $\zeta \in (4, 8),$ where the index of stability of the limiting asymmetric stable distribution is $\zeta/4$. Moreover, for the cases $\{p = 1 \ \text{and} \ \zeta \in (0,2)\}$ and $\{p = 2 \ \text{and} \ \zeta \in (0, 4)\}$, $\hat{\gamma}_{|R|^p}(h)$ is inconsistent and has a non-Gaussian asymmetric stable limit. As the rate of convergence and the limiting distributions depend on the unknown value of the tail index, $\zeta$, the results on convergence of full-sample covariance and correlation estimators are not directly applicable in terms of hypothesis testing and other inference problems. However, as discussed in Remark \ref{rem:scalemix} and Example \ref{ex:garch:independence} below, the stable limit theory may be used (under suitable conditions) in inference using the robust $t$-statistic approaches considered in the next section. %of little use %to functions of stable r.v.'s 
%are practically useless for hypotheses testing and other inference problems.} 
%\textbf{See also Remark \ref{rem:coupling} and the discussion and Example \ref{ex:garch:independence} in the next section on applicability of robust $t-$statistic approaches proposed in this paper in the context.} } %We refer to \cite{IIW} and the references therein for additional details about stable distributions. %The same reasoning applies to the sample correlations, $\hat{\rho}_{|R|^p}(h)$, \textbf{where the limits are given by functions of non-Gaussian stable vectors.} Similarly, if the returns follow a stochastic volatility process with heavy-tailed noise, it holds (under suitable conditions) that the sample covariances and correlations have stable limiting distributions but with a rate of convergence faster than $\sqrt{T}$ for the cases $\{p = 1 \ \text{and} \ \zeta \in (2,4)\}$ and $\{p = 2 \ \text{and} \ \zeta \in (4,8)\}$, see \cite{davis:mikosch:SV:2009} for additional details. 
%\textbf{As discussed in \cite{DM1998AS}, \cite{MS2000AS} and , among others, the above non-Gaussian stable limiting variables are complex and hard to describe.  
\end{example}

%{\color{blue}{RI: Is formulation of the remark okay?}}{\color{red} RSP: Made af few changes}

 The formulas for the covariance matrices of the limiting Gaussian variables in Lemmas \ref{NormalAsymp} and \ref{NormalAsymp1} are provided in the Online Appendix for the case of covariances and correlations, $\Cov(f(R_t), g(R_{t-h}))$ and $\Corr(f(R_t), g(R_{t-h}))$, for general functions $f$ and $g$. %\textcolor{red}{\textbf{RSP: Perhaps delete the following sentence.}} The limiting distributions are not particularly useful in practice, due to the complicated structure of the asymptotic covariance matrices. For instance, 
 The asymptotic covariance matrices have a complicated structure: For instance, for the case of (\ref{Autocov1}), the asymptotic covariance matrix depends on autocovariances of any order of the time series of the products $(|R_t|^p |R_{t-h}|^p)_{h=0,\ldots,m}.$ Under suitable conditions, including more restrictive moment conditions, the limiting covariance matrices may be estimated by HAC-type estimators, as discussed in the introduction. %(see \citeauthor{Newey}, \citeyear{Newey}, \citeauthor{Andrews}, \citeyear{Andrews}). 
 For instance, following \citet[Theorem 2]{Newey}, one has to assume that $E[|R_t|^{2p(4+\epsilon)}]<\infty$ for some $\epsilon >0$ for Newey-West-type standard errors to be applicable. Such conditions are restrictive for financial applications as, e.g., letting $p=1$ requires that $R_t$ has finite eighth-order moments, i.e., $\zeta>8$ in terms of the tail index (see also Remark \ref{tHAC1} on self-normalization and HAR approaches to time series inference that typically may be used under more relaxed moment conditions as compared to HAC). Importantly,  HAC-based inference methods often have poor finite sample properties, even in rather standard inference problems; see the discussion and references in the introduction.

 %\textcolor{red}{\textbf{RSP: Should we mention that HAC works OK in some cases in the simulations?}}

%{\color{blue}{RI: Are the conditions for autocorrelations or autocovariances? Do the above conditions become even more restrictive in the case of autocorrelations?}}

%{\color{red}\textbf{Mention general GARCH or okay as is?}}

%\begin{rem} \label{stableRem} 
%\end{rem}

%{\color{red}\textbf{Changed the text of the remark. Okay? Need to refer to Rasmus' results - in the intro?}}

In the next section, we propose robust approaches to inference
 on covariances and correlations of the form \eqref{eq:autocov:power}-\eqref{eq:autocrosscorr:power} using $t-$statistics in estimates of these quantities computed over groups of time series observations. The main advantage of these approaches is that no estimation of limiting covariance matrices is needed. We establish asymptotic validity of the robust inference approaches by relying on the large-sample results in Lemmas \ref{NormalAsymp} and \ref{NormalAsymp1} and new results on asymptotic independence of the group-based estimators.
 
%{\color{red}\textbf{Made changes in the above paragraph for clarification.}}
\subsection{Robust inference on market (non-)efficiency, volatility clustering and nonlinear dependence} \label{robust}

Following IM (2010, 2016), we consider robust $t-$statistic inference on a parameter $\beta$ of a general stationary process $(R_t)_{t\in \mathbb{Z}}$. In the following, the parameter $\beta$ of interest may be the population covariance $\beta=\gamma_{|R|^p}(h), \gamma'_{R,|R|^s \sign(R)}(h)$ or correlation $\beta=\rho_{|R|^p}(h),\rho'_{R,|R|^s \sign(R)}(h)$. %For the ease of exposition, we focus on the case of covariances, but emphasize that the results also apply to correlations.}
Let $(R_t)_{t=1,\ldots,T}$ be a sample of observations. Consider a partition of the sample into a fixed number $q\ge 2$ of (approximately) equal sized groups of consecutive observations, i.e. the observations in group $j=1,\dots,q$ have time indexes $(j-1)\lfloor T/q \rfloor <t\le j \lfloor T/q \rfloor$, where $\lfloor x \rfloor$ is the integer part of $x\in \mathbb{R}$. The robust $t-$statistic inference on $\beta$ is conducted using its group estimators, $(\hat{\beta}_j)_{j=1,\ldots,q}$, given by the sample covariances/correlations in (\ref{eq:autocov:power})-(\ref{eq:autocrosscorr:power}) based on the observations in group $j$, e.g., $\hat{\beta}_j$ may equal
\begin{eqnarray} \label{Groupj} \hat{\gamma}_{j, |R|^p}(h)=\frac{1} {\lfloor T/q \rfloor} \sum_{t=(j-1)\lfloor T/q \rfloor+h+1}^{j\lfloor T/q \rfloor} (|R_t|^p-\hat{\mu}_{j, |R|^p}) (|R_{t-h}|^p-\hat{\mu}_{j, |R|^p}), \end{eqnarray}
or
\begin{eqnarray} \label{Groupj1}  \hat{\gamma'}_{j, R, |R|^s \sign(R)}(h)=\frac{1} {\lfloor T/q \rfloor} \sum_{t=(j-1)\lfloor T/q \rfloor+h+1}^{j\lfloor T/q \rfloor} (R_t-\hat{\mu}_{j, R}) (|R_{t-h}|^s \sign(R_{t-h})-\hat{\mu}_{j,|R|^s \sign(R)}),\end{eqnarray}
where $\hat{\mu}_{j,|R|^p}$ is the group-based version of $\hat{\mu}_{|R|^p}$ in \eqref{eq:sample_means} based on the observations in group $j,$ and similar for $\hat{\mu}_{j, R}$ and $\hat{\mu}_{j,|R|^s \sign(R)}.$
%\textcolor{red}{RSP: Is the following not obvious from the context? (the group sample correlations $\hat{\rho}_{j, |R|^p}(h),$ $\hat{\rho'}_{j, R, |R|^s \sign(R)}(h)$ are defined in a similar way). Here $ \hat{\mu}_{j, |R|^p}=\lfloor T/q \rfloor^{-1} \sum_{t=(j-1)\lfloor T/q \rfloor+1}^{j\lfloor T/q \rfloor}  |R_t|^p$ denotes the sample mean of the $p$-th powers $|R_t|^p$ of absolute values of observations $R_t$ in group $j$, and, similarly,   $ \hat{\mu}_{j, R}=$ \\$\lfloor T/q \rfloor^{-1} \sum_{t=(j-1)\lfloor T/q \rfloor+1}^{j\lfloor T/q \rfloor}  R_t$ and  $ \hat{\mu}_{j, |R|^s sign(R)}=$ $ \lfloor T/q \rfloor^{-1} \sum_{t=(j-1)\lfloor T/q \rfloor+1}^{j\lfloor T/q \rfloor}  |R_t|^s \sign(R_t)$.}

%{\color{red}Should we change the index $k$ to $t$ in the above}

Suppose that one seeks to test the null hypothesis $H_0:\beta = \beta_0$, e.g. that the covariance $\gamma'_{R, |R|^s \sign(R)}(h)=0$, against the two-sided alternative $H_a: \beta \neq \beta_0$. Let $t_\beta$ denote the $t$-statistic in the group estimators, $(\hat{\beta}_j)_{j=1,\ldots,q}$, i.e.
\begin{eqnarray} \label{eq:def:tstat}
t_{\beta}=\sqrt{q}\frac{\overline{\hat{\beta}}-\beta_0}{s_{\hat{\beta}}},
\end{eqnarray}
with $\overline{\hat{\beta}}=q^{-1}\sum_{j=1}^q \hat{\beta}_j$ and $s^2_{\hat{\beta}}=(q-1)^{-1} \sum_{j=1}^q (\hat{\beta}_j-\overline{\hat{\beta}})^2$.
The robust $t$-statistic approaches rely on rejecting the null hypothesis $H_0$ in favor of the two-sided alternative $H_a$ at level $\tilde{\alpha}\le 0.083\dots$, if the absolute value of the $t-$statistic, $|t_{\beta}|$, exceeds the $(1-\tilde{\alpha}/2)$ quantile of a Student's $t$-distribution with $q-1$ degrees of freedom. The test of $H_0$ against $H_a$ at level $\tilde{\alpha}\le 0.1$ is conducted in the same way if $2\le q\le 14.$ Using the results in \cite{Bakirov2005student} and IM (2010), one can further calculate the $p-$values of the above $t-$statistic robust tests in the case of an arbitrary number $q$ of groups thus enabling conducting $t-$statistic robust tests of an arbitrary level (see Theorem \ref{thm:sizecontrol} below and the empirical applications in Section \ref{sec:empirical}).%\footnote{\textbf{One-sided tests are conducted in a similar way; one may note that percentiles of Student-$t$ distributions with $q-1$ degrees of freedom can also be used in one-sided tests of level $\alpha\le 0.1$ if $q\in \{2, 3\}.$}}

%\textcolor{red}{RI: Added the footnote and further text on an arbitrary level. Maybe better to delete. Percentiles or quantiles throughout?}

According to Theorem \ref{thm:sizecontrol} below, the $t$-statistic approaches to inference on properties \ref{Autocorr2} and \ref{AbsPower} are asymptotically valid and have asymptotically correct size. The theorem follows from, firstly, asymptotic normality and asymptotic independence of the group estimators  $(\hat{\beta}_j)_{j=1,\ldots,q},$ implied by Lemmas \ref{NormalAsymp} and \ref{NormalAsymp1} in the previous section and Lemma \ref{lemma} below, and, secondly, a small sample result on the conservativeness property of the $t$-statistic in heterogeneous normal r.v.'s originally proved by \cite{Bakirov2005student}; see also IM (2010, Thm. 1). %The theorem follows from, firstly, verifying that the group estimators  $(\hat{\beta}_j)_{j=1,\ldots,q}$ are asymptotically normal (Lemmas \ref{NormalAsymp} and \ref{NormalAsymp1} in the previous section) and asymptotically independent (Lemma \ref{lemma} below), and, secondly, applying a small sample result on the conservativeness property of the $t$-statistic in heterogeneous normal r.v.'s originally proved by \cite{Bakirov2005student}; see also IM (2010, Thm. 1). 

As discussed in IM (2010), asymptotic validity of the robust $t-$statistic approaches further implies that the confidence intervals 
\begin{eqnarray} \label{CI} \overline{\hat{\beta}}\pm \textrm{cv} s_{\hat{\beta}},
\end{eqnarray} 
where $\textrm{cv}$ is the usual $(1+C)/2\times 100$ percentile of the Student-$t$ distribution with $q-1$ degrees of freedom, have asymptotic coverage of at least $C$ for all $C\ge 0.917\dots.$ Further, for $2\le q\le 14,$ confidence intervals (\ref{CI}) with $\textrm{cv}$ being the usual 95\% percentile of the Student-$t$ distribution with $q-1$ degrees of freedom, have asymptotic coverage of at least $0.9.$%\footnote{Moreover, for $C\ge 0,$ let $\tilde{\textrm{cv}}=\max_{R(t_{\beta})<k\le q} \sqrt{(q-1)k\textrm{cv}_k^2/(q(k-1)+(q-k)\textrm{cv}_k^2)},$ $\textrm{cv}_k$ is the usual \textbf{$(1+C)/2$ percentile} of the Student-$t$ distribution with $k-1$ degrees of freedom. Then the interval
%$ \overline{\hat{\beta}}\pm \tilde{\textrm{cv}} s_{\hat{\beta}}$ has asymptotic coverage of at least $(1+C)/2.$}

%{\color{red}{Maybe emphasize further the difficulties in establishing asymptotic independence and also generality of the results?}}
%From the general results discussed in Appendix \ref{robust} it follows that the above $t-$statistic approach to robust inference on autocovariances $\gamma_{|R|^p}(h),$ $\gamma'_{R,|R|^p sign(R)}(h)$ and robust tests of properties (\ref{Autocorr2New}), (\ref{AbsPowerNew}) is asymptotically valid under the asymptotic normality and asymptotic independence of the group sample autocovariances $\hat{\gamma}_{j, |R|^p}(h)$ and $\hat{\gamma'}_{j, R |R|^p sign(R)}(h)$ (group estimators $\hat{\beta}_j$ in this context). Asymptotic normality of the above group sample autocovariances hold as long it holds for the full sample autocovariances $\hat{\gamma}_{|R|^p}(h)$ and $\hat{\gamma'}_{R,|R|^s sign(R)}(h)$ (see Theorems \ref{NormalAsymp} and  \ref{NormalAsymp1}).

According to the following lemma, the group estimators, $(\hat{\beta}_j)_{j=1,\ldots,q}$, are asymptotically independent under the assumptions in Lemmas \ref{NormalAsymp} and  \ref{NormalAsymp1}. %This, together with Lemmas \ref{NormalAsymp} and  \ref{NormalAsymp1} implying asymptotic normality of the group estimators, justifies asymptotic validity 
%The results in the lemma thus complete the justification of the applicability and \textbf{asymptotic validity} of the robust $t-$statistic approaches in the settings considered.

%{\color{red}{Maybe should refer to IM that asymptotic validity holds under asymptotic normality and asymptotic independence of group estimators.}}

\begin{lemma}  \label{lemma} Suppose that $(R_t)_{t\in \mathbb{Z}}$ is stationary and $\beta$-mixing. Under the assumptions of Lemma \ref{NormalAsymp},  the centered and scaled group sample covariances $\sqrt{\lfloor T/q \rfloor}(\hat{\gamma}_{i, |R|^p}(h) - \gamma_{|R|^p}(h))$ and $\sqrt{\lfloor T/q \rfloor}(\hat{\gamma}_{j, |R|^p}(h) - \gamma_{|R|^p}(h))$, are asymptotically independent for $i, j=1, 2, ..., q,$ with $i\neq j$. Likewise, under the conditions of Lemma \ref{NormalAsymp1}, the group sample covariances $\sqrt{\lfloor T/q \rfloor}(\hat{\gamma'}_{i, R, |R|^s sign(R)}(h)-\gamma'_{R,|R|^s \sign(R)})$ and $\sqrt{\lfloor T/q \rfloor}(\hat{\gamma'}_{j, R, |R|^s sign(R)}(h)-\gamma'_{R,|R|^s sign(R)})$  are asymptotically independent for $i, j=1, 2, ..., q,$ with $i\neq j$. The asymptotic independence property also holds for the group sample correlations $\hat{\rho}_{j, |R|^p}(h)$ and $\hat{\rho'}_{j, R, |R|^s sign(R)}$,  $j=1, 2, ..., q$, under the assumptions in Lemmas \ref{NormalAsymp} and \ref{NormalAsymp1}, respectively.
\end{lemma}

%{\color{red}{Used 'auto(cross)', changed a couple of words.}}

\begin{rem}\label{rem:coupling}
The proof of Lemma \ref{lemma} in the case of covariances and correlations of general functions of the process $(R_t)$ is given in the Online Appendix and relies on exact coupling properties for $\beta$-mixing processes. A similar argument was recently used in \cite{Pedersen} in the context of inference on the autoregressive coefficient in linear autoregressive models in the presence of heavy-tailed GARCH-type errors under symmetry. The proof is general also in the sense that the arguments hold for arbitrary limiting distributions of the group estimators. The limiting distributions may be non-Gaussian, e.g., stable (with the index of stability less than 2), as discussed in Example \ref{ex:garch:appl:limit}.
%Note that an alternative approach to establishing asymptotic independence of group estimators under Gaussian limits is to show that the joint Gaussian limiting distribution of the group-based estimators has a block-diagonal covariance matrix.  
\end{rem}

The following main result follows by Lemmas \ref{NormalAsymp}-\ref{lemma} and IM (2010, Thm. 1).

\begin{thm}\label{thm:sizecontrol}
Consider, as above, testing the hypothesis
$H_0: \beta=\beta_0$ against $H_a: \beta\neq \beta_0$ for covariances/correlations $\beta=\gamma_{|R|^p}(h),$ $\gamma'_{R,|R|^s \sign(R)}(h),$ $\rho_{|R|^p}(h),$ $\rho'_{R,|R|^s \sign(R)}(h)$. Suppose that the assumptions of Lemma \ref{lemma} are satisfied. Let $T_{q-1}$ denote a r.v. with a Student's $t$-distribution with $q-1$ degrees of freedom, and let $\mathrm{cv}(q, \tilde{\alpha})$ satisfy $P(T_{q-1}>\mathrm{cv}(q, \tilde{\alpha}))=\tilde{\alpha}/2$. With $t_\beta$ defined in (\ref{eq:def:tstat}) using group estimators, $\hat{\beta}_j$, of $\beta$ one has, under the null hypothesis $H_0,$
 \begin{eqnarray} \label{IMTStat}
 \limsup_{T\to \infty}P(|t_\beta| > \mathrm{cv}(q, \tilde{\alpha})|H_0) \le \tilde{\alpha},
 \end{eqnarray}
 for any $\tilde{\alpha}\le 2\Phi(-\sqrt{3})=0.08326\dots$, where $\Phi(\cdot)$ is the standard normal cdf.
Inequality (\ref{IMTStat}) also holds for $2\le q\le 14$ if $\tilde{\alpha}\le 0.1.$ Moreover, for any $x\ge 0$ and $q\ge 2,$
  \begin{eqnarray} \label{IMTStat1}
 \limsup_{T\to \infty}P(|t_\beta| > x|H_0) \le \max_{R<k\le q} P\Big(|T_{k-1}|>\sqrt{\frac{R(k-1)}{k-R}}\Big),
 \end{eqnarray}
 where $R=R(x)=\frac{q x^2}{x^2+q-1}.$
 %and for $q\in \{2, 3\}$ if $\alpha\le 0.2.$}
\end{thm}

%{\color{red}{Define $T_{k}$ as Student-$t$ r.v. before?}}

%By Theorem \ref{NormalAsymp}, the limits of the group estimators $\hat{\gamma}_{j, |R|^p}(h)$ and $\hat{\rho}_{j, |R|^p}(h)$  are normal if $p<\zeta/4.$ By Theorem \ref{NormalAsymp1}, the limits of group $\hat{\gamma'}_{j, R, |R|^s}(h)$  are normal if $0<s<\zeta/2-1.$ Under zero population autocorrelations ${\rho'}_{ R, |R|^s sign(R)}(h))_{h=0, 1, ..., m}=0$ and all $0<s<\zeta/2-1,$ aymptototic normality also holds for group estimators $\hat{\rho'}_{j, R, |R|^s}(h)$ (the limits become stable for group sample autocovariances and ratios of stable r.v.'s for group sample autocorrelations if these inequalities are not satisfied, see Remark \ref{stableRem}). 
%\textbf{Are the inequalities between $p$ and $\zeta$ in the lemma correct? or need to formulate separately the case $p<\zeta/4$ and $p>\zeta/4$ for instance}

%\begin{lemma}  \label{lemma} The group sample autocorrelations $\hat{\rho}_{i, |R|^p}(h)$ and $\hat{\rho}_{j, |R|^p}(h)$ are asymptotically independent for $i, j=1, 2, ..., q,$ with $i\neq j,$ if $0<p<\zeta/2.$ The group sample autocovariances $\hat{\gamma}_{i, |R|^s sign(R)}(h)$ and $\hat{\gamma}_{j, |R|^s sign(R)}(h)$ are asymptotically independent for $i, j=1, 2, ..., q,$ with $i\neq j,$ if $0<s<\zeta-1.$ 
%\end{lemma}

Theorem \ref{thm:sizecontrol} demonstrates the asymptotic validity of robust $t-$statistic approaches for inference on the properties \ref{Autocorr2}-\ref{AbsPower} under appropriate conditions on the degree of heavy-tailedness (the tail index $\zeta$) for the return process $(R_t)$ and powers $p$ and $s$ in \ref{Autocorr2}-\ref{AbsPower} like $p<\zeta/4,$ or $s<\zeta/2-1.$ % (see Example \ref{ex:garch:appl:limit} in the previous section and Example \ref{ex:garch:independence} below).} %($4p<\zeta$ in Lemma \ref{NormalAsymp} and $2(1+s)<\zeta$ in Lemma \ref{NormalAsymp1}, see Example \ref{ex:garch:appl:limit} in the previous section and Example \ref{ex:garch:independence} below).} %given in Lemmas \ref{NormalAsymp} and \ref{NormalAsymp1}. 
%Following the approaches, for instance, the null hypothesis $H_0: \beta = \beta_0$ on the auto(cross)covariance or auto(cross)correlation parameter $\beta$ is rejected in favor of the two-sided alternative $H_a: \beta\neq \beta_0$ at significance level $\alpha\le 0.05$ if the absolute value $|t_{\beta}|$ of the $t-$statistic in $q$ group estimates of $\beta$ given by group sample auto(cross)covariances/auto(cross)correlations exceeds the $(1-\alpha/2)$ percentile of the Student's $t$-distribution with $q-1$ degrees of freedom.
In contrast to HAC-based inference approaches, with estimates of limiting covariance matrices that depend on the particular values of powers $p$ and $s$ appearing in $\gamma_{|R|^p}(h)$  and $\gamma'_{R, |R|^s \sign(R)}(h)$, the robust $t-$statistic approaches are applicable irrespective of %the particular values 
the values of $p$ and $s$ %in \textbf{the measures in \ref{Autocorr2}-\ref{AbsPower}, 
provided they satisfy the above conditions/inequalities.

%\textcolor{red}{RSP: We may want to combine Remark 2.3 and Example 2.3 with the above paragraph.}

 \begin{rem}\label{rem:scalemix}
As pointed out by IM (2010), the $t$-statistics approaches are applicable whenever the group-based estimators weakly converge to scale mixtures of normal distributions and are asymptotically independent (Example \ref{ex:garch:appl:limit} and Remark \ref{rem:coupling}). Hence, the limiting distributions are, in principle, not required to be Gaussian, as further discussed in \cite{Pedersen} in the symmetric case as well as in Example \ref{ex:garch:independence} below. 
\end{rem}
\begin{example}\label{ex:garch:independence}
  As discussed in Section \ref{Sample}, under suitable conditions, the GARCH(1,1) process is $\beta$-mixing (with geometric rate) and, hence, satisfies the assumptions of Lemmas \ref{NormalAsymp}-\ref{lemma} under the conditions on the tail index $\zeta$ and powers $p, s$ discussed in Example \ref{ex:garch:appl:limit}. Hence, Theorem \ref{thm:sizecontrol} holds. Thus, for instance, the (centered and scaled) group estimators $\hat{\gamma}_{j, |R|^p}(h)$ and $\hat{\rho}_{j, |R|^p}(h)$  are asymptotically independent and normal if $4p<\zeta$. Likewise, the group estimators $\hat{\gamma'}_{j, R, |R|^s \sign(R)}(h)$ are asymptotically independent and normal if $2(1+s)<\zeta$. Similar conclusions hold for the group estimators $\hat{\rho'}_{j, R, |R|^s \sign(R)}(h)$. Suppose instead that $2(1+s)> \zeta > 1+s$. Then the group estimators $\hat{\gamma'}_{j, R, |R|^s \sign(R)}(h)$ are asymptotically independent and stable; see also Remark \ref{rem:coupling}. If the innovations, $Z_t$, are symmetric, then the limiting stable distributions are symmetric, and given by scale mixtures of normal distributions. In such situations, as pointed out in Remark \ref{rem:scalemix}, the $t-$statistic approaches are applicable for inference on property \ref{Autocorr2}. These aspects of the $t$-statistic approaches are further investigated in the simulation experiments in the next section.
\end{example}

%{\color{red}\textbf{The remark refers to the condition $\zeta>2(1+s)$ - is this because of mds we have equality to zero for autocovariances and autocorrelations? Should we discuss that the results in the paper hold in the asymmetric case further}}

%{\color{blue}\textbf{Simulations, applications of this?}}
\begin{rem} \label{tHAC11} Inference using $t-$statistics, essentially in any context, is related to using self-normalized sums and statistics. E.g., as is well-known, for the $t-$statistic $t_q=\sqrt{q} \overline{X}_q/s_q$ in any r.v.'s $X_1, ..., X_q,$ $q\ge 2$ (e.g., group estimators $\hat{\beta}_j,$ $j=1, ..., q,$ of a parameter of interest $\beta,$ as in robust $t-$statistic inference approaches proposed in IM (\citeyear{IM2010}, \citeyear{IM2016}) and this paper), where, as usual, $\overline{X}_q=\frac{1}{q} \sum_{j=1}^q X_j,$ $s_q^2=\frac{1}{q-1} \sum_{j=1}^q (X_j-\overline{X}_q)^2,$ and the  self-normalized sum $SN_q=\frac{\sum_{j=1}^q X_j}{\sqrt{\sum_{j=1}^q X_j^2}}$ of $X_j's$ one has $SN_q=t_q/\sqrt{1+(t_q^2-1)/q}$ (see, among others, \citeauthor{Efron}, \citeyear{Efron}, \citeauthor{Edelman}, \citeyear{Edelman}, \citeauthor{Pinelis}, \citeyear{Pinelis}, \citeauthor{DufourHallin}, \citeyear{DufourHallin}, \citeauthor{selfnorm}, \citeyear{selfnorm}, and references therein). Therefore, for any $x>0,$ \begin{eqnarray} \label{tSN} P(|t_q|>x)=P(|SN_q|>x/\sqrt{1+(x^2-1)/q}).\end{eqnarray} As discussed in the above and other works in the literature (see, among other, \citeauthor{PLS}, \citeyear{PLS}, and references therein), self-normalization allows one to conduct inference under heavy tails and relaxed moment conditions. For instance, in the case of an arbitrary sample size $q\ge 2,$ and independent r.v.'s $X_1, ..., X_q$ symmetric about a common median $\mu,$ the test of $H_0: \mu=0$ against $H_a: \mu\neq 0$ can be based on bounds for the tail probabilities of linear combinations of i.i.d. symmetric Bernoulli r.v.'s and the implied bounds for $t-$statistics $t_q$ and self-normalized sums $SN_q$ of $X_j'$s in \eqref{tSN} established in the above papers. In particular, from \cite{Edelman} it follows that, under $H_0,$ for any $x>0,$
\begin{eqnarray}  &&P(|t_q|>x)=P(|SN_q|>x/\sqrt{1+(x^2-1)/q})< \notag \\ 
&&1-\Phi\Big[\frac{x}{\sqrt{1+(x^2-1)/q}}-1.5\frac{\sqrt{1+(x^2-1)/q}}{x}\Big]=G(x).\label{Edel}
\end{eqnarray}
Thus, $H_0: \mu=\mu_0$ is rejected in favor of $H_a: \mu\neq \mu_0$ at level $\tilde{\alpha}$ if $G(|t_q|)\le \tilde{\alpha}.$
%1-\Phi\Big[\frac{|t_q|}{\sqrt{1+(t_q^2-1)/q}}-1.5\frac{\sqrt{1+(t_q^2-1)/q}}{|t_q|}\Big]<\alpha.$$
\cite{DufourHallin} show that similar bounds can be used in tests on regression and autocorrelation coefficients under independence and symmetry in observations, and \cite{selfnorm} consider the case of random polynomials and generalized cross-correlations. In the context of robust $t-$statistic inference approaches using group estimators $\hat{\beta}_j,$ $j=1, ..., q,$ of a parameter of interest $\beta,$ i.e. the autocovariances/autocorrelations of functions of a GARCH time series as considered in this paper, the bounds on the $t-$statistics $t_q$ and self-normalized sums $SN$ in (\ref{tSN}) under independence and symmetry can be used to construct tests on $\beta$ that are valid under asymptotically independent group estimators $\hat{\beta}_j,$ $j=1, ..., q,$ with \emph{any} (not necessarily identical) limiting distributions that are symmetric about $\beta.$ E.g., under these conditions, using bounds \eqref{Edel} in \cite{Edelman}, the hypothesis $H_0: \beta=\beta_0$ is rejected in favor of $H_a:\beta\neq \beta_0$ at level $\tilde{\alpha}$ if the $t-$statistic $t_{\beta}$ in group estimators $\hat{\beta}_j,$ $j=1, ..., q,$ in \eqref{eq:def:tstat} if  $G(|t_{\beta}|)\le \tilde{\alpha}.$ The small-sample results in \cite{Bakirov2005student} and $t-$statistic robust inference approaches in IM (\citeyear{IM2010},\citeyear{IM2016}) and this paper may be viewed as improvements on bounds \eqref{Edel} for independent and symmetric observations and the implied (conservative) testing procedures using the observations or asymptotically independent and symmetric group estimators of a parameter of interest. The improvements hold under the assumptions that the group estimators are asymptotically independent and have asymptotic distributions that are scale mixtures of normal ones. \end{rem}
%$$1-\Phi\Big[\frac{|t|}{\sqrt{1+(t^2-1)/q}}-1.5\frac{\sqrt{1+(t^2-1)/q}}{|t_q|}\Big]<\alpha.$$

\begin{rem}\label{tHAC1} Several works in the literature have focused on HAR inference using self-normalized statistics with non-standard asymptotics, including testing uncorrelatedness of a dependent time series (see \citeauthor{Lobato}, \citeyear{Lobato}, and the review in \citeauthor{Shao}, \citeyear{Shao}). \cite{Politis} (see also the working paper version, \citeauthor{Politis1}, \citeyear{Politis1}) has proposed a new class of higher-order accurate large-sample covariance and spectral density matrix estimators based on flat-top kernels. The above inference approaches may be used under more relaxed moment assumptions as compared to HAC methods. %based on flat-top kernels that may be used in HAC/HAR inference \textbf{on parameters of a stationary parameters under a potential lack of finite fourth moments of the time series used in their estimation.} 
%The above approaches may be useful in inference on the measures of market (non-)efficiency, volatility clustering and nonlinear dependence dealt with in this paper. 
However, direct application of HAC/HAR and self-normalization approaches, including the aforementioned ones, in the context of inference on, e.g., linear autocovariances and autocorrelations of GARCH-type time series, % and their squares, requires %like linear autocorrelations and autocorrelations of squares of the time 
 requires %finite variances and 
asymptotic normality of respective sample autocovariances and autocorrelations. %numerators of corresponding $t-$statistics \textbf{or self-normalized statistics}  \textbf{and finite variances of sample covariances } 
On the other hand, according to Lemmas \ref{NormalAsymp} and \ref{NormalAsymp1} (see also the discussion in the introduction), %finiteness of variances and the 
asymptotic normality of the above sample linear autocovariances and autocorrelations requires finite fourth moments of the GARCH processes (we note, in particular, that GARCH processes dealt with in the numerical analysis of inference on linear autocorrelations in \citeauthor{Lobato}, \citeyear{Lobato}, have coefficients implying finite fourth moments).

\end{rem}

%\textcolor{red}{On robustness to heavy tails, we should discuss in the remark further that $t-$statistic approaches are valid under convergence to symmetric heavy-tailed stable distributions.}
\section{Finite-sample properties}\label{sec:simulations}

%{\color{red}\textbf{Maybe have the title as 'Inference on measures of market (non-)efficiency, volatility clustering and nonlinear dependence: Finite sample properties'? Or retain as is to minimize the changes?}}

In this section, we present numerical results on finite-sample properties of the $t$-statistic approaches and compare them with those of HAC-based approaches in inference on properties \ref{Autocorr2}-\ref{AbsPower}. We consider the AR-ARCH DGPs given by
\begin{eqnarray} \label{ARGARCH}
R_t&=&\phi R_{t-1}+\varepsilon_t,\quad t=1,\dots,T, \label{DGP1} \\
\varepsilon_t&=&\sigma_t Z_t, \\
%\end{eqnarray}
%Models \ref{ARCH1}-\ref{ARCH1AsyH} below assume ARCH(1) dynamics for the volatility process  $(\sigma_t^2):$ 
%\begin{eqnarray} 
%\varepsilon_t&=&\sigma_t Z_t, \; t=2,\dots,T,\\
\sigma_t^2 &=&\omega  + \alpha \varepsilon_{t-1}^2, \label{DGP1.1}
 \end{eqnarray}
where $\omega=0.1,$ $0<\alpha<1,$ $0\le \phi<1,$ and $(Z_t)_{t=1,\ldots,T}$ are i.i.d. r.v.'s with $E(Z_t)=0$ and $\Var(Z_t)=1$. In terms of the distribution of the innovations $Z_t$, we consider:

\begin{enumerate}[label=(\alph*)]
\item \label{ARCH1} \textit{Symmetric light-tailed distribution}:
$Z_t$ is standard normal, $Z_t\sim \mathcal{N}(0, 1).$ 
%\begin{eqnarray} \label{DGP1.1}
%\varepsilon_t&=&\sigma_t Z_t, \; t=2,\dots,T,\\
%\sigma_t^2&=&0.1 + (\pi^{1/3}/2) \varepsilon_{t-1}^2,
% \end{eqnarray}
%and $(Z_t)$ is a sequence of i.i.d. standard normal $N(0,1)$ r.v.'s.

\item \label{ARCH1Asy} \textit{Asymmetric light-tailed distribution}: 
$Z_t$ has an asymmetric (skewed) $t$-distribution with 50 degrees of freedom and the skewness parameter of 0.5: $Z_t\sim t(50, 0.5)$.
%\begin{eqnarray} \label{DGP1.2}
%\sigma_t^2&=&0.1 + (\pi^{1/3}/2) R_{t-1}^2,
% \end{eqnarray}

\item \label{ARCH1AsyH} \textit{Asymmetric heavy-tailed distribution}: $Z_t$ has an asymmetric (skewed) $t$-distribution with 3 degrees of freedom and the skewness parameter of 0.5: $Z_t\sim t(3, 0.5).$
%\begin{eqnarray} \label{DGP1.2}
%\sigma_t^2&=&0.1 + (\pi^{1/3}/2) R_{t-1}^2,
% \end{eqnarray}
%and $(Z_t)$ is a sequence of i.i.d. r.v.'s that have a $t$ distribution with 3 degree of freedom and the skewness parameter of 0.5 (see Section 6.7 in \cite{ChrisRM}).

%\item \label{GJRGARCH} \textit{\color{red}{GJR-GARCH(1, 1, 1)} with normal innovations}: \textbf{The dynamics of the volatility} $\sigma_t^2$ is given by
%\begin{eqnarray} \label{DGP1.3}
%\sigma_t^2 =\omega+\alpha(|\varepsilon_{t-1}|-\gamma\varepsilon_{t-1})^2+\beta\sigma_{t-1}^2= \nonumber \\  \omega+2\alpha(1-\gamma)^2 R_{t-1}^2+2\alpha \gamma R_{t-1}^2I(R_{t-1}<0),
 %\end{eqnarray}
%where $I_{t-1}=1$ for $R_{t-1}<0$ and $I_{t-1}=0$ otherwise; $\omega=0.1,$ $\alpha=0.04545,$ $\beta=0.9,$ $\gamma=1,$ and the distribution of the innovations 
% \begin{eqnarray} \label{DGP1.3}
%\sigma_t^2&=&0.1+\frac{0.1818}{4}(|\varepsilon_{t-1}|-\varepsilon_{t-1})^2+0.9\sigma_{t-1}^2
 %\end{eqnarray}
 %$(Z_t)$ is a standard normal $N(0,1).$

\end{enumerate}
%{\color{blue}\textbf{RI: There are several definition of asymmetric t-distributions, e.g. at {\url{https://en.wikipedia.org/wiki/Skewed_generalized_t_distribution#Skewed_t_distribution}} and Christoffersen's book. Which one do we use, should provide a ref. Maybe Johnson and Kotz's books? - or it's Christoffersen's definition with the paramegters chosen appropriately - $d_1$ and $d_2$?}}

The densities of the asymmetric $t-$distributions in \ref{ARCH1Asy} and \ref{ARCH1AsyH} are given in (10)-(13) in \cite{hansen1994} with $\lambda=0.5$ and $\eta=50, 3,$ respectively.

The conclusions from the numerical results are similar for other AR-GARCH-type processes, including processes with asymmetric GJR-GARCH dynamics. %; see the working paper \cite{IPSWP}. 
All tests considered have a 5\% nominal level. We use a sample size of 5,000 observations and 10,000 Monte Carlo replications. % \textbf{for the rejection frequencies.} 
All computations were done in MATLAB (v$.$ 2020a).

%\textcolor{red}{Include version number of MATLAB??}

 %{\color{red}{Include refs? GJR, Christoffersen, Bollerslev, Ding et al. in the references' list. Can $\gamma$ be equal to zero so that the coefficient at $R_{t-1}^2$ is zero?}}

\subsection{Testing for linear (in)dependence and market (non-)efficiency} \label{ind}
%\ref{Autocorr2}: $\rho'_{R, |R|^s \sign(R)}(h)=0$,
 
%{\color{red}{} \textbf{Maybe specify the process considered in the numerical analysis as
% \begin{eqnarray}\nonumber \label{DGP1}
%R_t&=&\phi R_{t-1}+\varepsilon_t,\\
%\varepsilon_t&=&\sigma_t Z_t, \; t=2,\dots,T,
%\end{eqnarray}
%where the GARCH-type process $(\varepsilon_t)$ with $\alpha$ and then specify $\phi, \alpha$ and $Z_t'$ in size and power analysis?).}}

%{\color{red}{What does (1, 1, 1) stand for in GJR-GARCH(1, 1, 1). Skip?}}

For investigating the finite-sample properties of inference methods, we consider the processes in (\ref{DGP1})-(\ref{DGP1.1}) with $\alpha=\pi^{1/3}/2$ and innovations $Z_t$ with distributions in \ref{ARCH1}-\ref{ARCH1AsyH}. By Kesten's equation (\ref{KestenEq}), when $Z_t\sim \mathcal{N}(0, 1)$ as in \ref{ARCH1}, the processes $R_t$ and $\varepsilon_t$ in (\ref{ARGARCH})-(\ref{DGP1.1}) have heavy-tailed power-law distributions as in \ref{Heavy}, with the tail index $\zeta=3$. Likewise, solving Kesten's equation (\ref{KestenEq}) numerically, we obtain that $R_t$ and $\varepsilon_t$ have heavy-tailed power-law distributions with the tail indices $\zeta \approx 2.89$ and $\zeta \approx 2.24$, respectively, in cases \ref{ARCH1Asy} and \ref{ARCH1AsyH}. Therefore, under distributions \ref{ARCH1}-\ref{ARCH1AsyH} for innovations $Z_t,$ the tail indices $\zeta$ of the processes $R_t$ and $\varepsilon_t$ lie in the interval $(2, 4)$, as is typically the case for financial returns in developed markets (see the discussion in the introduction). Further, as $\zeta\in (2, 3],$ the absolute third moments of the processes are infinite.

%\textcolor{red}{Maybe tail index discussion should go in the previous section as also referred in the section on finite sample properties of tests for nonlinear dependence - RI}

%\textcolor{red}{(2, 4) not discussed in intro! Add?}
%with 50 and 0.5 we have the tail index   urther, the $t$ distributions $t(50, 0.5)$ and $t(3, 0.5)$ in \ref{ARCH1Asy}-\ref{ARCH1AsyH} are more heavy-tailed as compared to the normal case in (\ref{ARCH1}). Therefore, from Kesten's equation (\ref{KestenEq}) it further follows the considered (AR-)GARCH processes $(R_t)$ and $(\varepsilon_t)$ with $\alpha=\alpha=\pi^{1/3}/2$ and $t$ distributed innovations $Z_t:$ $Z_t\sim t(50, 0.5),$ $Z_t\sim t(43 0.5)$ in \ref{ARCH1Asy}-\ref{ARCH1AsyH} heavy-tailed power law distributions \ref{Heavy1} with tail indices $\zeta$ that satisfy $2<\zeta<3.$
%\textcolor{red}{Removed the MATLAB part. OK? /RSP Ok, usually I write about the coding, but this may be omitted if footnotes are not allowable. / AS}
%All simulations in the numerical analysis in this section were done %in MATLAB. simulations of GARCH-type processes considered were %based on the MFE\_Toolbox from Kevin Sheppard's website, see %https://www.kevinsheppard.com/MFE\_Toolbox for details.

We consider the finite-sample size and power properties of tests of the null hypotheses $H_0: \beta=0$ against the two-sided alternative $H_a: \beta\neq 0$ for $\beta=\rho'_{R, |R|^s \sign(R)}(1)=\Corr(R_t, |R_{t-1}|^s \sign(R_{t-1}))$ in \ref{Autocorr2} with the lag $h=1$ and different powers $s>0.$ Note that, under $H_0$, the autoregressive coefficient in (\ref{DGP1}) is zero, $\phi=0$. For $s=1$, $H_0$ corresponds to the standard property of absence of linear autocorrelations, $\Corr(R_t, R_{t-1})=0$, as in \ref{Autocorr}. In simulations below, we consider the powers $s= 0.1, 0.25, 0.5$. 

Based on the stated values of the tail index, $\zeta=4, 2.89, 2.24$, Example \ref{ex:garch:appl:limit} implies asymptotic normality of the full-sample estimator $\hat{\beta}=\hat{\rho}'_{R, |R|^s \sign(R)}(1)$ of $\beta=\rho'_{R, |R|^s \sign(R)}(1)$ in \eqref{eq:conv:autocrosscorr:nullity} whenever $s<0.5$, $s<0.445$, and $s<0.12,$ respectively, for the cases \ref{ARCH1}, \ref{ARCH1Asy} and  \ref{ARCH1AsyH}. Likewise, from Example \ref{ex:garch:independence} it follows that, under $H_0$ and the same conditions on powers $s,$  asymptotic normality and asymptotic independence hold for the group estimators $\hat{\beta}_j=\hat{\rho}'_{j, R, |R|^s \sign(R)}(1)$, implying, by Theorem \ref{thm:sizecontrol}, asymptotic validity of robust $t-$statistic inference approaches.% using the group estimators $\hat{\beta}_j.$

%\textbf{\color{red}{Okay to refer to Examples not Lemmas and Theorems? - RI}}
%, and based on the values of the tail index $\zeta$, we have that asymptotic normality of the estimators of $\rho'_{R, |R|^s \sign(R)}(1)$ applies whenever $s<0.5$, $s<0.445$, and $s<0.12$, respectively,
The first class of tests we consider is based on the HAC-based $t-$statistic of (full-)sample correlations $\hat{\beta}=\hat{\rho'}_{R,|R|^{s} \sign(R)}(1)$, with a long-run variance estimator based on a QS kernel with automatic bandwidth selection (\citeauthor{Andrews}, \citeyear{Andrews}; see also Section \ref{AsNorm}), and the critical values based on the standard normal distribution. %. \textcolor{red}{, with critical values based on the standard normal distribution.} \textcolor{red}{SHOULD WE MENTION THAT CV IS BASED ON $N(0,1)$??} 
The second class of tests is based on $t-$statistics in the group estimators for $q=4, 8, 12$ and $16$ groups.

\subsubsection{Size properties}\label{sec:size:noneff}

%{\color{red}{Include 'finite-sample' in the sections' titles?}}

%\textbf{\color{red}{Maybe skip the text on Matlab instead of making the foonote on this the main text?}}
%The second and third ARCH processes have the same $\alpha=\pi^{1/3}/2$ and asymmetric Student-$t-$distributed innovations that are more heavy-tailed than normal. By Kesten's equation (\ref{KestenEq}), the tail index $\zeta$ in \ref{Heavy} for these processes satisfies $2<\zeta<3$. For the choice of the parameters of the GJR-GARCH(1,1,1) model, the tail $\zeta$ equals to 3 as in the case of ARCH with normal innovations. The parameters of the GJR-GARCH process that correspond to the tail index $\zeta=3$ were found using simulations from the analogue of Kesten's equation (\ref{KestenEq}) for the stochastic recurrence (difference) equation defining the process (see, among others, \cite{MS2000AS}). All simulations were done in MATLAB. The generation of GARCH type processes were done via MFE\_Toolbox from Kevin Sheppard's website, see https://www.kevinsheppard.com/MFE\_Toolbox for details. %\textbf{Do we need note in appendix for this?} 

%(with standard normal innovations) the value $\alpha=\pi^{1/3}/2$ corresponds to the tail index $\zeta=3$ in \ref{Heavy}.

%\textcolor{red}{Table \ref{tab1}: Typo in column 2 ($\rho$). Row with GJR-GARCH should be removed. /RSP}

The results on size properties of HAC-based and robust $t-$statistic approaches for testing $H_0: \beta=\rho'_{R, |R|^s \sign(R)}(1)=0$, as in property \ref{Autocorr2}, against $H_a: \beta\neq 0$ are provided in Table \ref{tab1}.
%\textcolor{red}{RI: Important: Need to add 'HAC' in the first four columns of Table 1 and '$t-$statistic tests' or similar in the last big four columns of Table 1.}
%We present the results for tests of the null hypotheses $H_0: \rho'_{R, |R|^s \sign(R)}(1)=\Corr(R_t, |R_{t-1}|^s \sign(R_{t-1}))=0$ in property (\ref{Autocorr2}) for the lag $h=1$ and the power values $s=1$ (corresponding to the standard property on absence of linear autocorrelations: $\rho_{R}(1)=\Corr(R_t, R_{t-1})=0$ as in \ref{Autocorr} with $h=1$) and $s= 0.5, 0.25, 0.1$ against the two-sided alternatives $H_a: \rho'_{R, |R|^s \sign(R)}(1)\neq 0.$ The first class of tests considered is based on the HAC $t-$statistics of (full-sample) sample autocorrelations  $\hat{\rho}_{R}(1),$ $\hat{\rho'}_{R,|R|^{0.5} \sign(R)}(1)$, $\hat{\rho'}_{R,|R|^{0.25} sign(R)}(1)$, $\hat{\rho'}_{R,|R|^{0.1} \sign(R)}(1).$ The second class of tests considered are robust tests based on $t-$statistics in the group counterparts of the above sample autocorrelations for the number of groups $q=4, 8, 12$ and 16. 
%\textbf{In this section, we analyse finite sample size properties of the tests 
%of the null hypotheses $H_0: \rho'_{R, |R|^s \sign(R)}(1)=0$ in (\ref{Autocorr2}) with $h=1$ and $s= 0.1, 0.25, 0.5, 1$ against the two-sided alternatives $H_a: \rho'_{R, |R|^s \sign(R)}(1)\neq 0$ described in the previous section by setting  $\phi=0$ in \eqref{DGP1}. }
We note that the standard HAC-based tests are oversized. In particular, in the case of asymmetric heavy-tailed innovations $Z_t$ (case \ref{ARCH1AsyH}), the size distortions are severe in the case where $s=1$, i.e., when one carries out the usual test for the absence of  linear autocorrelations as in property \ref{Autocorr}. For the robust $t-$statistic approaches, the size control is good even for the case $s=1$, except for the case of asymmetric heavy-tailed innovations. For the latter case, the limit of the full-sample and group estimators $\hat{\beta}=\hat{\rho'}_{R,|R|^{s} \sign(R)}(1)$ and $\hat{\beta}_j=\hat{\rho'}_{j, R,|R|^{s} \sign(R)}(1)$ is asymmetric stable, invalidating the use of both the HAC and robust $t-$statistic approaches (Example \ref{ex:garch:appl:limit}). In contrast, in the case where $s=1$ and the distribution of $Z_t$ is standard normal (case \ref{ARCH1}), one has that the group estimators $\hat{\beta}_j=\hat{\rho'}_{j, R,|R|^{s} \sign(R)}(1)$ are asymptotically independent and \textit{symmetric} stable (Example  \ref{ex:garch:independence}), implying that the $t-$statistic robust tests of $H_0$ are asymptotically valid. This is reflected in the attractive size properties of the robust $t-$statistic approaches, e.g., in contrast to those of the standard HAC-approaches. The same conclusions hold in the case $s=0.5$. %\textbf{The same is the case} for both HAC and robust $t-$statistic approaches to testing the hypothesis $H_0$ with $s=0.5$ based on (full-sample and group) sample auto(cross)correlations $\hat{\rho'}_{R, |R|^{0.5} \sign(R)}(1)$ (as indicated above, one needs $s$ to be smaller than 0.5 for asymptotic normality of sample analogues of these autocorrelations). 
The robust $t-$statistic approaches to testing the hypotheses $H_0$ under \ref{ARCH1}, \ref{ARCH1Asy} with $s=0.1, 0.25$ and under \ref{ARCH1AsyH} with $s=0.1$ -- where estimators are asymptotically normal -- are slightly over-sized, with reasonable size control for $q=4$ and $q=8$.  Quite remarkably, the tests with the most desirable size properties are the ones for testing $H_0$ with $s=0.1$ based on the $t-$statistic approaches with $q=4$ or $q=8$ number of groups. The reason might be that the asymptotic (Gaussian) distributions in \eqref{eq:conv:autocrosscorr:nullity} in Lemma \ref{NormalAsymp1} provide relatively good approximations to the distributions of the group estimators for small choices of powers $s$, whereas the quality of the approximations worsens for larger values of $s$.

\begin{rem}\label{rem:results_HAC_self}
We note that for some of the DGPs, and for some powers $s$, the size distortions of the HAC-based approach are not overly severe, although the approach is not theoretically justified, in contrast to $t-$statistic inference. The reason may be two-fold. Firstly, finite sample distributions of the HAC-based $t$-statistics may be fairly well approximated by a standard Gaussian distribution, if the tails of the DGPs are not too heavy. Secondly, kindly pointed out by the Associate Editor, for any fixed bandwidth, the HAC-based $t$-statistic may be viewed as having a self-normalized structure. Further, as discussed in Remarks \ref{tHAC11} and Remark \ref{tHAC1}, time series inference approaches based on self-normalization typically exhibit robustness to infinite higher-order moments. %, e.g., as compared to HAC methods. %; see also the discussion in Remark \ref{tHAC1}.
\end{rem}

%\textcolor{red}{Note that under (c), asymptotic normality holds only with $s=0.1.$ Tried to accommondate this above - Okay?}

%\textcolor{red}{ The choice of smaller $s$ leads to better size properties. The reason is the following. If, e.g.,  the tail index $\zeta=3$, then the power $s$ for $\rho'_{R,|R|^{s} \sign(R)}(1)$ should be smaller than 0.5. I.e., $s=0.5-\epsilon$ with some small $\epsilon$ should give asymptotic Normality for $\rho'_{R,|R|^{s} \sign(R)}(1)$. But in finite sample, the asymptotic approximation is worse for $s$ very close to 0.5, and it becomes better if $s$ decreases. Obviously, $s=0$ make no sense.}

%{\color{red}\textbf{Should we refer to the convergence to non-standard limits - ratios of stables as conjecture or a fact. Should we use $H_0$ as is now, should we change 'autocorrelations' to 'auto(cross)correlations as above. $s$ depends on the tail index of asymmetric heavy-tailed GARCH processes to guarantee asymptotic normality, maybe derived the tail index by simulations?}}

%\textcolor{red}{Note: In table 1, the second row ($t(50,0.5)$) has to be "moved to the right". /RSP}

\begin{center}
    [Table \ref{tab1} about here.]
\end{center}
%\textcolor{red}{RSP: In Table \ref{tab1}: some of the horizontal lines should be extended, so that they go through all columns.}
%\textcolor{red}{Removed (for now) the comments about the small powers. Perhaps include some additional comments... /RSP }

%\textbf{AS: We need to differentiate the q as number of groups and q as power.}

%textbf{AS: I think, this is mirroring the models with near unit root. The normal approximation is better for smaller $\rho$. So, the q=0.1 gives  better normal approximation than q=0.25 which is 'local to stable distribution'.}

\subsubsection{Power properties}

%{\color{red}{Maybe delete 'size-adjusted in the title?}}

\begin{figure}[h]%
\begin{center}%
\subfigure[$N(0,1)$ noise]{
\includegraphics[width=0.45\linewidth]{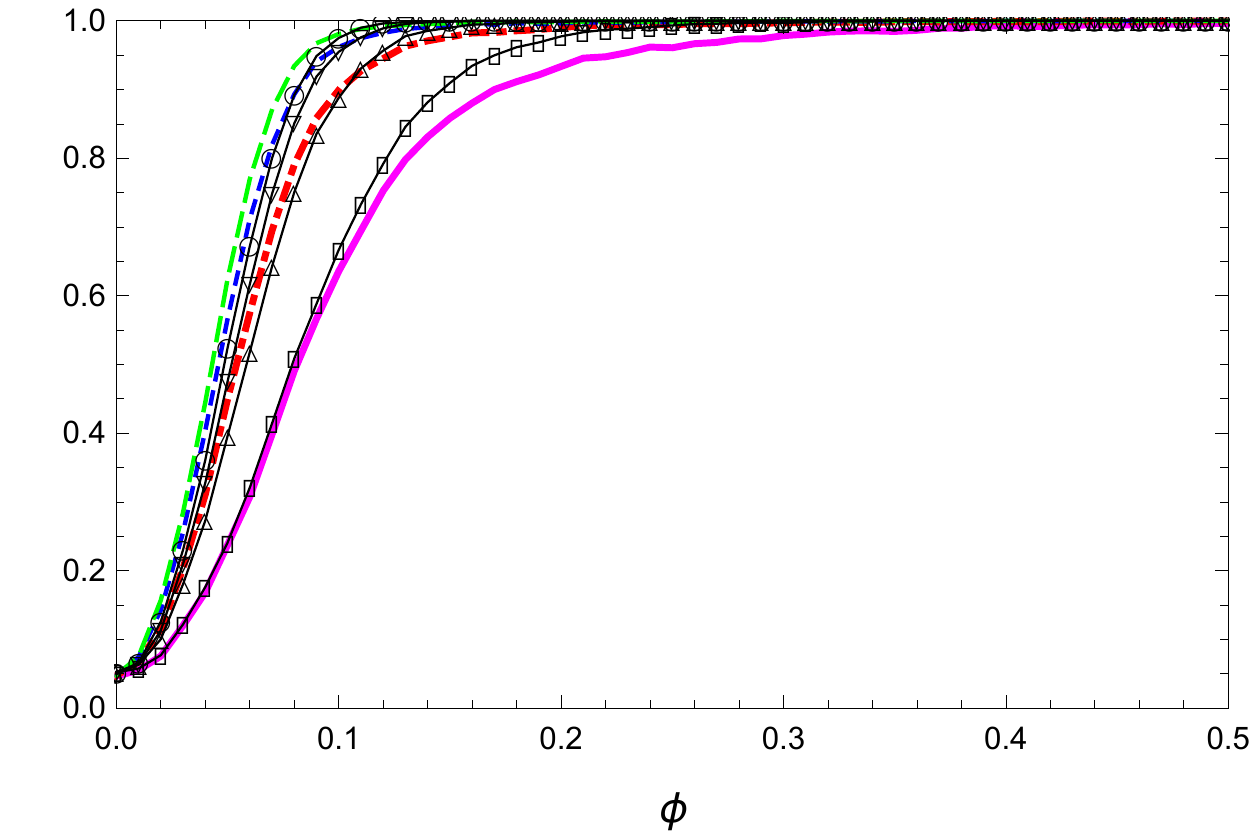}\label{fig:1:1}
}
\subfigure[$t(3,0.5)$ noise]{
\includegraphics[width=0.45\linewidth]{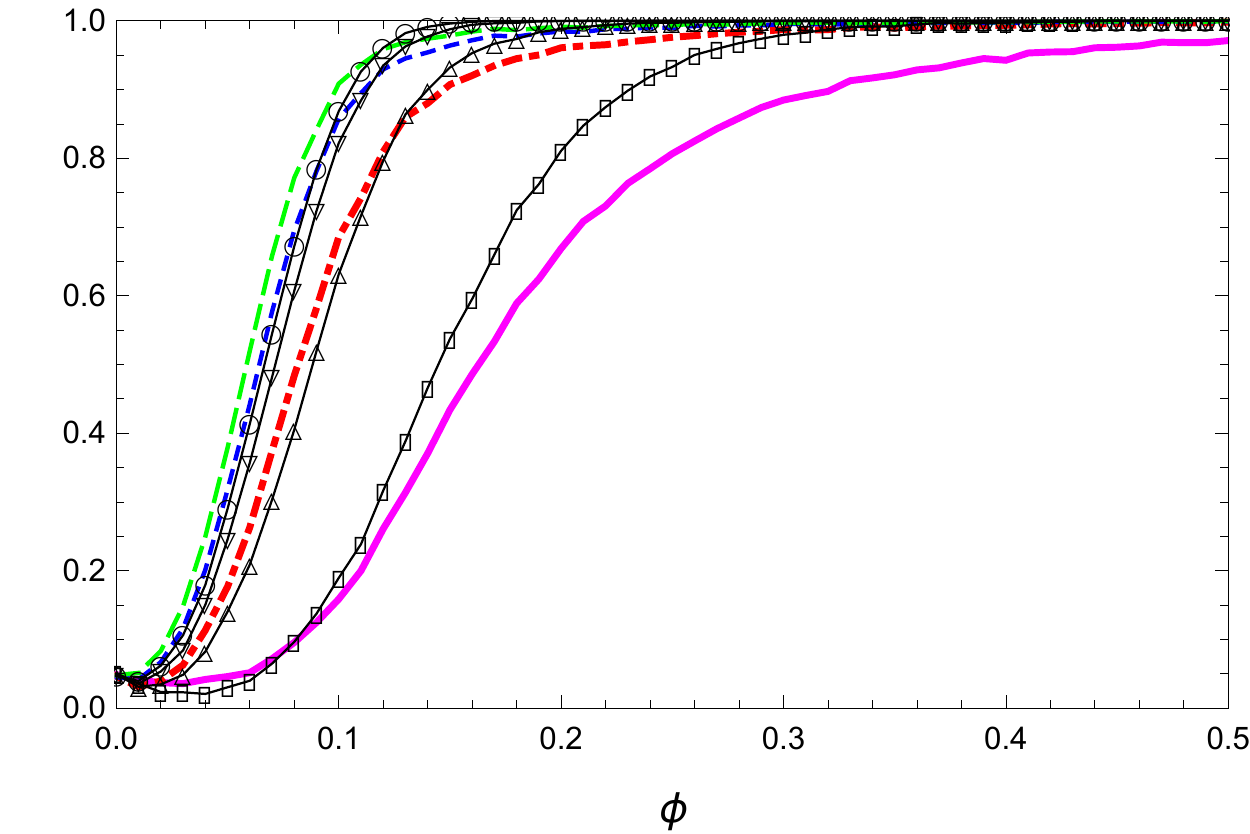}\label{fig:1:2}
}\\
%\subfigure[$N(0,1)$ noise]{
%\includegraphics[width=0.45\linewidth]{Monte-Carlo/Figures/fig1_1_1_SA_p1.pdf}\label{fig:1:113}
%}
\end{center}%
\caption{Size-adjusted power}
\label{fig1}
\centering
${\rho'}_{R, R}(h), \ \text{HAC} \ $:$\textcolor{magenta}{\rule[0.25em]{2em}{1.6pt}\ }$,
${\rho'}_{R, |R|^{0.5} sign(R)}(h), \ \text{HAC} \ $:$\textcolor{red}{\rule[0.25em]{0.6em}{1.7pt} \ \mathbf{\cdot} \ \rule[0.25em]{0.6em}{1.7pt} \ }$,
%$A^*(t_{HLT},s_\alpha):\rule[0.25em]{2em}{0.5pt}\ $
${\rho'}_{R,|R|^{0.25} sign(R)}(h), \ \text{HAC} \
$:$\textcolor{blue}{\rule[0.25em]{0.4em}{1.6pt} \ \rule[0.25em]{0.4em}{1.6pt}\ }$,
${\rho'}_{R, |R|^{0.1} sign(R)}(h), \ \text{HAC} \ $:$\textcolor{green}{\rule[0.25em]{0.8em}{1.6pt} \ \rule[0.25em]{0.8em}{1.6pt}\ }$,
${\rho'}_{R, R}(h)$, $q=8: \rule[0.25em]{2em}{0.5pt} \!\!\!\!\!\!\!\!\! \square \;\;\;\;$,
${\rho'}_{R, |R|^{0.5} sign(R)}(h)$, $q=8:\rule[0.25em]{2em}{0.5pt} \!\!\!\!\!\!\!\!\! \bigtriangleup \;\;\;\;$,
${\rho'}_{R, |R|^{0.25} sign(R)}(h)$, $q=8: \rule[0.25em]{2em}{0.5pt} \!\!\!\!\!\!\!\!\! \bigtriangledown \;\;\;\;$,
${\rho'}_{R, |R|^{0.1} sign(R)}(h)$, $q=8: \rule[0.25em]{2em}{0.5pt} \!\!\!\!\!\!\!\!\! \bigcirc \;\;\;\;\;$
\end{figure}

To investigate the power properties of the HAC-based and robust $t-$statistics tests of $H_0: \beta=0$ against $H_a: \beta\neq 0$ for $\beta=\rho'_{R, |R|^s \sign(R)}(1)=\Corr(R_t, |R_{t-1}|^s \sign(R_{t-1}))$ as in \ref{Autocorr2}, we consider the alternatives where the autoregressive parameter $\phi$ in \eqref{ARGARCH} ranges from 0 to 0.5. Figure \ref{fig1} provides the size-adjusted rejection frequencies for the HAC-based and robust $t-$statistic (with $q=8$ groups) tests of $H_0$ for the case of normal innovations (case \ref{ARCH1}) and asymmetric heavy-tailed innovations (case \ref{ARCH1AsyH}). When $Z_t$ are standard normal, the power curves for the HAC and the robust $t-$statistic tests are very close to each other. We note that the rejection frequencies are generally lower when $s=1$, i.e., when testing the classical hypothesis of no linear dependence/absence of linear autocorrelations as in  \ref{Autocorr}. As in the previous section, we note that, under $s=1$ and standard normal innovations, the robust $t-$statistic tests remain asymptotically valid, in contrast to the HAC-based tests. Similar conclusions hold in the case of asymmetric heavy-tailed innovations (case \ref{ARCH1AsyH}), with the only exception that the robust $t-$statistic tests have much better power properties than the HAC-based tests for the case of $s=1$, although, as discussed in the previous section, the use of HAC-based and the robust $t-$statistic tests are not theoretically justified for this case. For the sake of brevity, we do not report the results for the case of light-tailed asymmetric innovations (case \ref{ARCH1Asy}), where the power properties of the tests are similar to those for the Gaussian case in \ref{ARCH1}. Overall, the tests with $s=0.1, 0.25$ are the most powerful.

%\textcolor{red}{Is the notation ${\rho'}_{R, R}(h)$ in Figure 1 okay? Or we used ${\rho'}_{R}(h)$ before?}

%\textcolor{red}{Note that there is a typo in Figure \ref{fig2_sa}: Check the correlation for $s=0.25$. ALSO: We should change $\rho_R(h)$ to $\rho_{R,R}(h)$ or $\rho_{R,|R|\sign(R)}(h)$. I think that it might be a good idea to also report Figure \ref{fig4}, as this provides the pattern of rejection frequencies for any choice of $s$ and distributions of $Z_t$. /RSP  }

%{\color{red}\textbf{changed somewhat. Always similar, but rather poor power in the case $s=1,$ should we mention that power is large for the remaining cases}}

%In the asymmetric case with heavy-tailed innovations (Model  \ref{ARCH1AsyH}), the conclusions are similar, with the most powerful tests of $H_0$ obtained for $s=0.1, 0.25$ and the sample auto(cross)correlations $\hat{\rho'}_{R, |R|^{0.1} \sign(R)}(1)$  and $\hat{\rho'}_{R, |R|^{0.25} \sign(R)}(1)$ (and former is slightly more powerful than the later).

%{\color{red}{Include 'market efficiency' in relation to the hypotheses through this section? Is this for both HAC and $t-$statistics on the most powerful tests.}}
%Similar conclusions also hold in the GJR-GARCH case (Model 4).
\begin{center}
    [Figure \ref{fig1} about here.]
\end{center}
%{\color{red}{The results for Model 2 and 4 are virtually the same and provided in Appendix \ref{app:powercurves}, Figures \ref{fig2_sa_new}-\ref{fig3_sa}. }}

%\textbf{We note that, similar to size properties and based on the same reasons, the size-adjusted power is higher for smaller values of $s$.}

\begin{figure}[h!]
\begin{center}%
\includegraphics[width=0.45\linewidth]{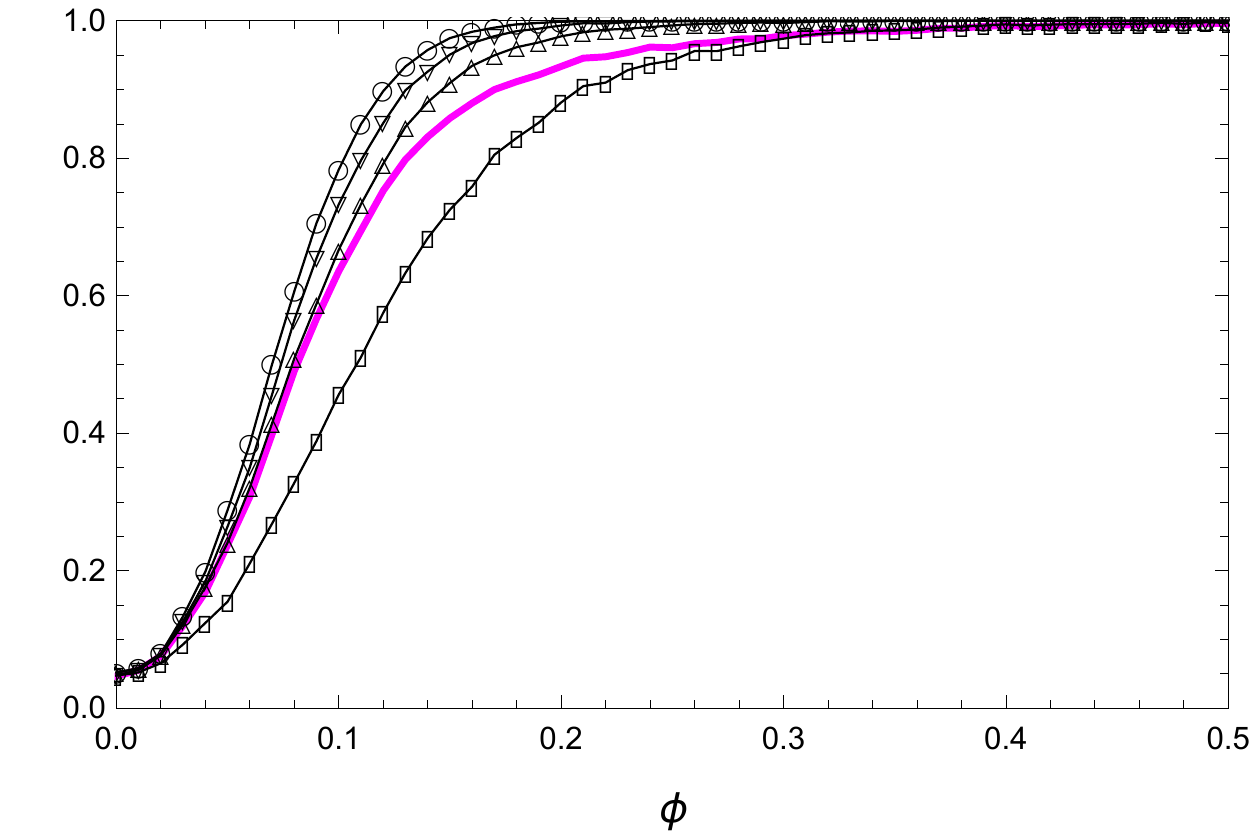}
\end{center}%
\caption{Size-adjusted power for ARCH(1) with $N(0,1)$ noise, ${\rho'}_{R,R}(h)$.}
\label{fig2}
\centering
HAC:$\textcolor{magenta}{\rule[0.25em]{2em}{1.6pt}\ }$,
 $q=4: \rule[0.25em]{2em}{0.5pt} \!\!\!\!\!\!\!\!\! \square \;\;\;\;$,
$q=8:\rule[0.25em]{2em}{0.5pt} \!\!\!\!\!\!\!\!\! \bigtriangleup \;\;\;\;$,
$q=12: \rule[0.25em]{2em}{0.5pt} \!\!\!\!\!\!\!\!\! \bigtriangledown \;\;\;\;$,
$q=16: \rule[0.25em]{2em}{0.5pt} \!\!\!\!\!\!\!\!\! \bigcirc \;\;\;\;\;$
\end{figure}

Figure \ref{fig2} contains rejection frequencies for the tests under standard normal innovations (case \ref{ARCH1}) for $s=1$ and different number $q$ of groups in robust $t-$statistic approaches. In this case, as mentioned, the robust $t-$statistic approaches are asymptotically valid, in contrast to HAC-based tests.  Note that the rejection frequencies are increasing in $q$, and that the tests based on $q=8, 12, 16$ groups appear to be more powerful than the HAC-based tests. Similar conclusions hold for other powers, $s$, and other distributions of the innovations $Z_t$ (see Online Appendix).

%\textcolor{red}{Mention that more powerful in terms of size-adjusted power? So, q=16 is oversized.}

%\textcolor{red}{Mention that ``for finitesamples the Gaussian approximation might be poor for values ofsclose to 0.5, andthe  approximation  is  better  for  smaller  values  ofs.   This  might  explain  why  theproposed inference approaches seem to work better for small positive powerssandp.'' Answer for Comment 4 of referee.}

%{\color{red}{RI: Is it okay on validity or take off. What explains performance under $s=1.$}. RSP: Symmetric stable limits?}
%According to Figures \ref{fig4}-\ref{fig7_3} presented in Appendix \ref{app:powercurves}, the power of robust $t-$statistic tests of $H_0$ increases as the number of groups $q$ increases, but the differences between the power curves become negligible as the power $s$ in the auto(cross)correlations $\hat{\rho'}_{R, |R|^{s} \sign(R)}(1)$ decreases { \color{red}{(except for the test with $q=4$ groups).}}
\begin{center}
    [Figure \ref{fig2} about here.]
\end{center}

%{\color{red}{RI: Minor: Mention 'available on request' where unreported results are mentioned here and below?}}

To conclude, the numerical results indicate that in order to conduct reliable inference on property \ref{Autocorr2}, one may choose $s>0$ small and rely on the robust $t$-statistic approaches with $q=4$ or $q=8$ groups. This ensures very reasonable size control as well as quite attractive power properties, e.g., in comparison with the widely used HAC-based approaches. The robust $t-$statistic approaches to inference may thus be viewed as useful complements to the traditional HAC-based methods.

%\textbf{for HAC tests considered in the previous section and robust $t-$statistic approaches based on  with $q=8.$ }
%\textbf{AS: Actually we need only figures with powers 0.25 and 0.1, and the tests with q=8,12,16 are equally powerful.}

%\ref{fig2_sa_new} and \ref{fig3_sa} present the results on finite size-adjusted power for Model 1 (ARCH with normal innovations), Model 2 (ARCH with asymmetric innovations), Model 3 (ARCH with asymmetric heavy-tailed innovations) and Model 4 (GJR-GARCH with normal innovations)

\subsection{Testing for nonlinear dependence and volatility clustering}

Next, we turn to the finite sample properties of the HAC-based and robust $t-$statistic approaches to inference on property \ref{AbsPower} with the lag order of $h=1$ and powers $p\in \{0.1, 0.25, 1, 2 \}$. The DGPs considered follow (\ref{DGP1})-(\ref{DGP1.1}) with $\phi=0$ and the ARCH parameter $\alpha\in (0, 1).$ %$\alpha$ ranges from 0 to 1. 
For the sake of brevity, we present the results for the case of standard normal innovations $Z_t$ in \ref{ARCH1}. The results in the Online Appendix for asymmetric/heavy-tailed cases \ref{ARCH1Asy} and  \ref{ARCH1AsyH} reveal the same qualitative conclusions.

We investigate the relative performance of the $t$-statistic and HAC-based approaches by comparing the coverage levels of the corresponding $t$-statistic and HAC-based confidence intervals for the unknown population correlation $\beta=\rho_{|R|^p}(1)$. Specifically, the confidence intervals based on the $t$-statistic approaches are constructed as in (\ref{CI}), and the HAC-based confidence intervals are computed by standard methods, relying on asymptotic normality of the full-sample estimator $\hat{\beta}=\hat{\rho}_{|R|^p}(1)$.
%\textcolor{red}{; see Lemma \ref{NormalAsymp} and the subsequent discussion.}  
%We provide the numerical results on finite-sample properties of HAC and robust $t-$statistic approaches to the analysis of the property $\rho_{|R|^{p}}(1)=\Corr(|R_t|^p, |R_{t-1}|^p)>>0$ as in \ref{AbsPower} with the lag $h=1$ and powers $p=2, 1, 0.5, 0.25, 0.1$ of absolute returns $R_t.$ The HAC inference approaches are based on estimators $\hat{\rho}_{R^2}(1)$ $\hat{\rho}_{|R|}(1)$, $\hat{\rho}_{|R|^{0.5}}(1)$, $\hat{\rho}_{|R|^{0.25} }(1)$, $\hat{\rho}_{|R|^{0.1}}(1)$ and, for robust $t-$statistic inference, on $t-$statistics in group counterparts of the estimators with the number of groups $q=4, 8, 12$ and 16. The comparisons are based on the coverage level of the corresponding HAC and $t-$statistic based (\ref{CI}) confidence intervals for the unknown population \textbf{autocorrelations} $\rho_{R^2}(1),$ $\rho_{|R|}(1)$, $\rho_{|R|^{0.5}}(1)$, $\rho_{|R|^{0.25} }(1)$, $\rho_{|R|^{0.1}}(1).$ 
Note that with $\alpha\in (0,1)$, the processes $R_t$ and $\varepsilon_t$ have heavy-tailed power-law distributions as in \ref{Heavy}, with the tail index $\zeta > 2$. % (and $\zeta = 2$ if $\alpha=1$). %, and $\zeta  = 2$ if only if $\alpha =1$.
From Example \ref{ex:garch:appl:limit}, it follows that the sample correlation $\hat{\beta}=\hat{\rho}_{|R|^{p} }(1)$, is asymptotically normal if $\zeta>4p$, and hence whenever $p\le 0.5$.  %\textbf{(and also for $p=0.5$, if $\alpha<1$ so that $\zeta>2$).} 
From Example \ref{ex:garch:independence}, under the same conditions on powers $p$, it holds that the group-based estimators are asymptotically normal and asymptotically independent, implying asymptotic validity of the robust $t-$statistic inference approaches. %\textcolor{red}{(Theorem \ref{thm:sizecontrol})}. %for different values of the ARCH parameter $\alpha$ in \ref{DGP2a} ranging from 0 to 1 ($\alpha=1$ implies infinite second moment and $\zeta=2$ in power law distributions (\ref{Heavy1}) for ARCH processes considered so that, by Lemma \ref{NormalAsymp}, the power $p$ in \ref{AbsPower} should be smaller than 0.5 for asymptotic normality of the sample autocorrelations $\rho_{|R|^{p} }(1)$ for all $0<p\le 1$).

\begin{figure}[h]\label{fig:power2}%
\begin{center}%
\subfigure[${\rho}_{R^2}(h)$]{
\includegraphics[width=0.45\linewidth]{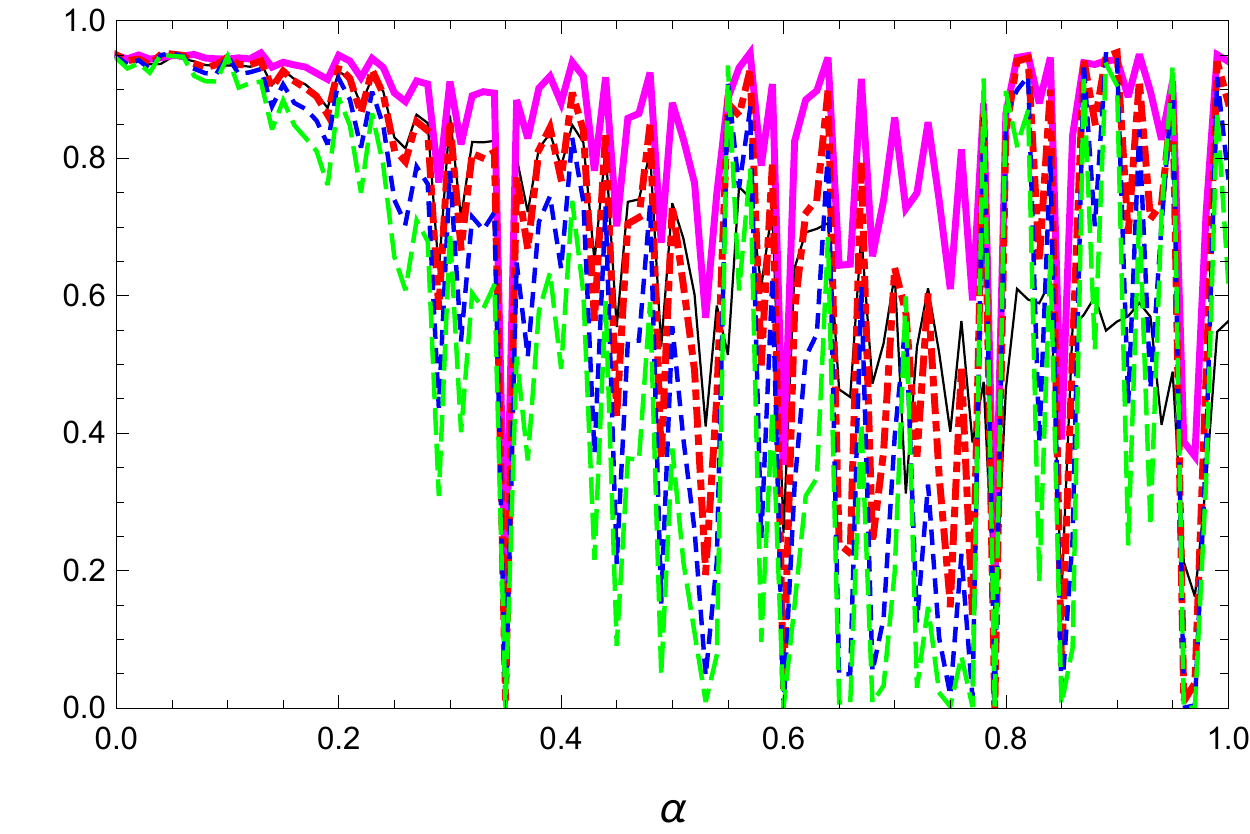}\label{fig:3:1}
}
\subfigure[${\rho}_{|R|}(h)$]{
\includegraphics[width=0.45\linewidth]{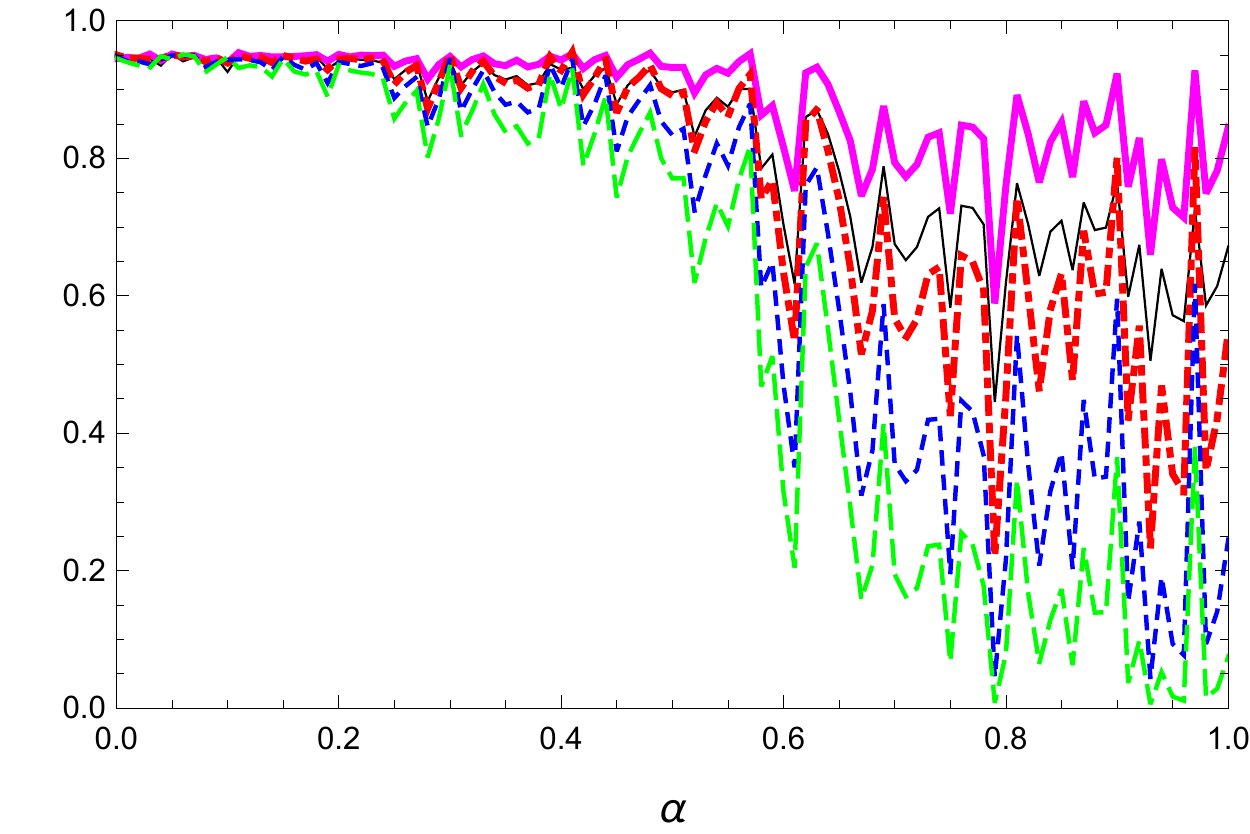}\label{fig:3:2}
}\\
\subfigure[${\rho}_{|R|^{0.5}}(h)$]{
\includegraphics[width=0.45\linewidth]{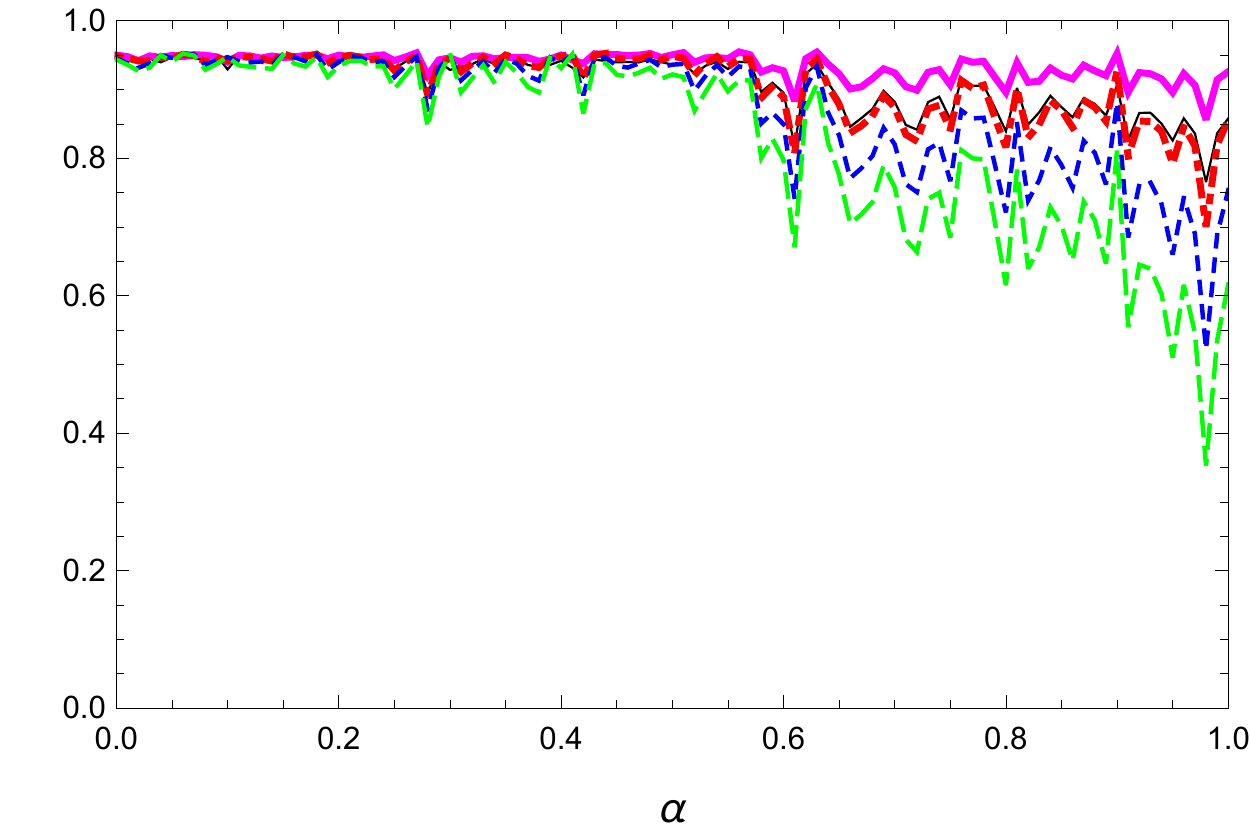}\label{fig:3:3}
}
\subfigure[${\rho}_{|R|^{0.25}}(h)$]{
\includegraphics[width=0.45\linewidth]{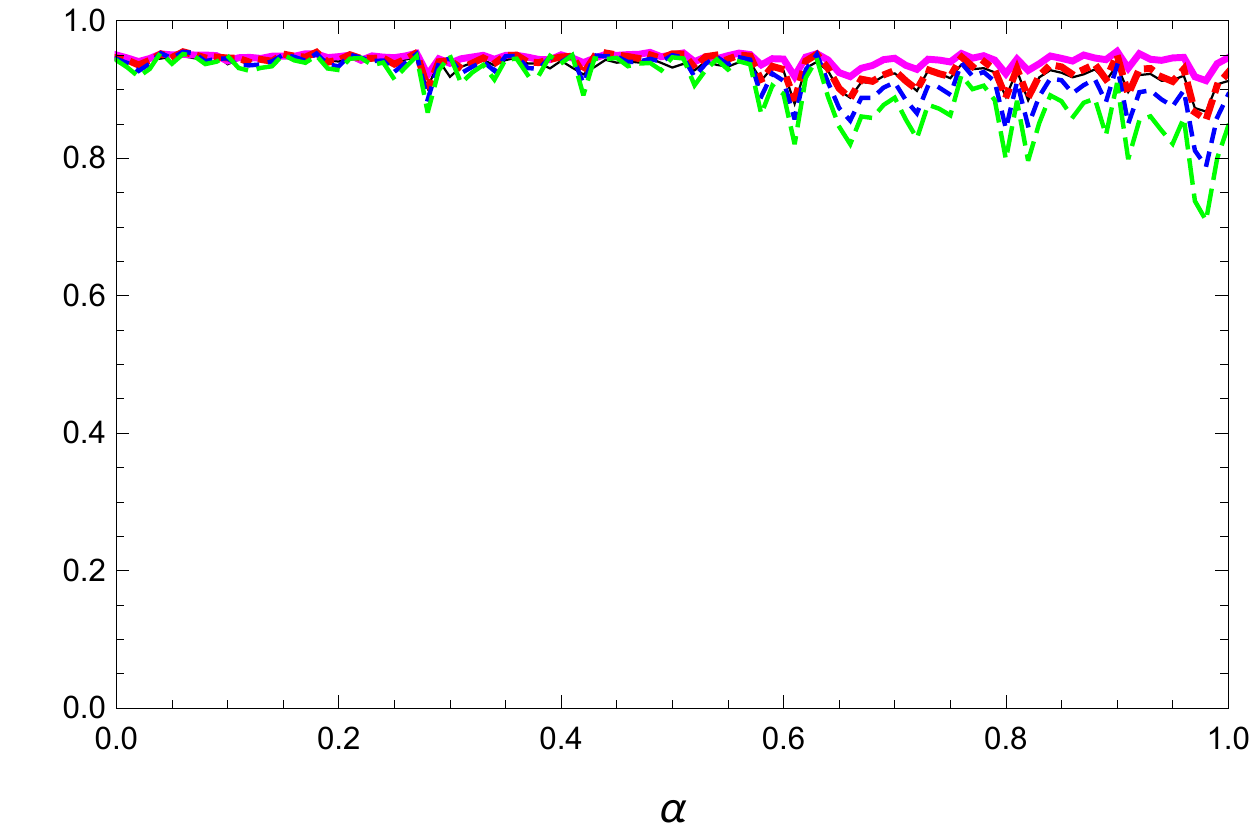}\label{fig:3:4}
}\\
\subfigure[${\rho}_{|R|^{0.1}}(h)$]{
\includegraphics[width=0.45\linewidth]{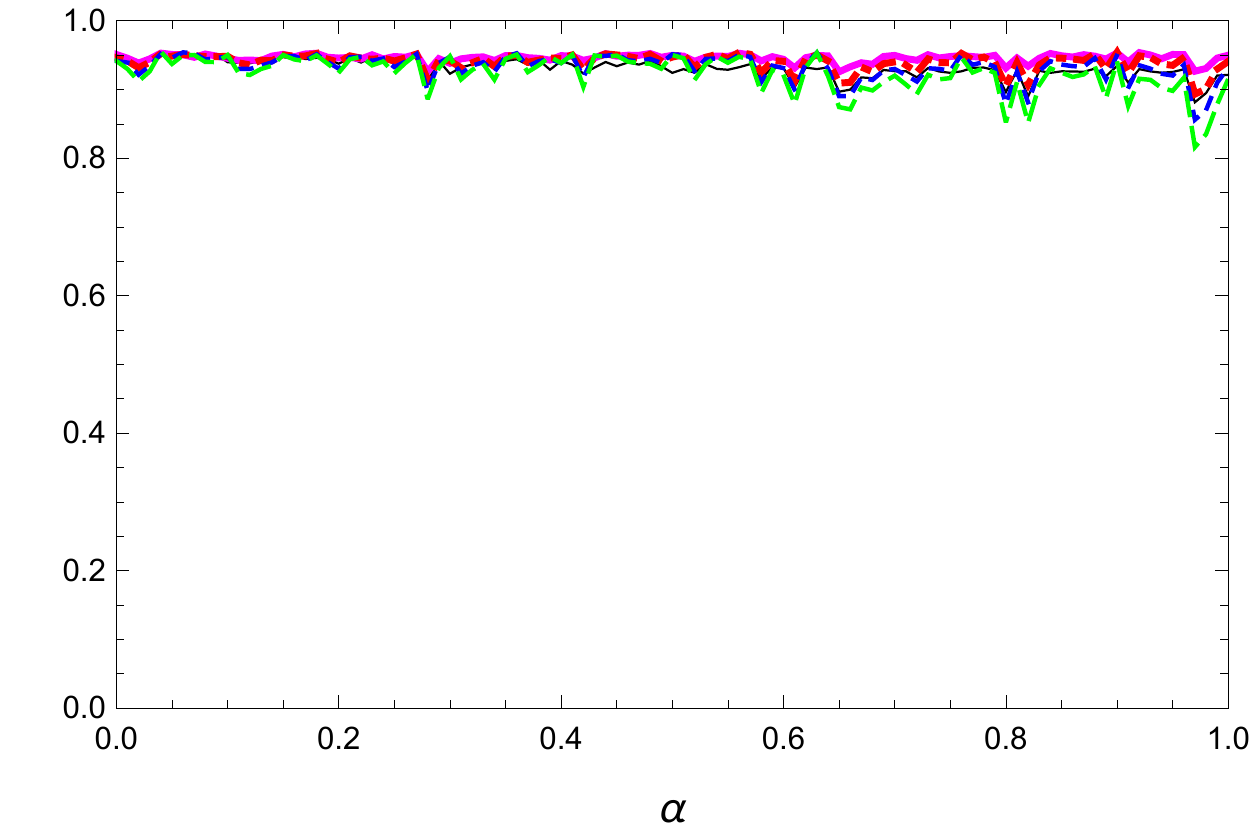}\label{fig:3:5}
}
\end{center}%
\caption{Coverage level for ARCH(1) with $N(0,1)$ noise}
\label{fig3}
\centering
HAC:$\textcolor{black}{\rule[0.25em]{2em}{1.6pt}\ }$,
$q=4$:$\textcolor{magenta}{\rule[0.25em]{2em}{1.6pt}\ }$,
$q=8$:$\textcolor{red}{\rule[0.25em]{0.6em}{1.7pt} \ \mathbf{\cdot} \ \rule[0.25em]{0.6em}{1.7pt} \ }$
%$A^*(t_{HLT},s_\alpha):\rule[0.25em]{2em}{0.5pt}\ $
$q=12$:$\textcolor{blue}{\rule[0.25em]{0.4em}{1.6pt} \ \rule[0.25em]{0.4em}{1.6pt}\ }$, $q=16$:$\textcolor{green}{\rule[0.25em]{0.8em}{1.6pt} \ \rule[0.25em]{0.8em}{1.6pt}\ }$
\end{figure}

Figure \ref{fig3} contains coverage levels of the confidence intervals for the HAC-based and $t-$statistic inference approaches for different choices of powers $p$ and different number of groups $q$ in the $t$-statistic approaches. 
%present the numerical results on the coverage of confidence intervals in the normal ARCH case for the tests based on $\hat{\rho}_{R^2}(1)$ (Figure \ref{fig8}),  $\hat{\rho}_{|R|}(1)$ (Figure \ref{fig9}), $\hat{\rho}_{|R|^{0.5}}(1)$ (Figure \ref{fig10}), $\hat{\rho}_{|R|^{0.25} }(1)$ (Figure \ref{fig11}) and $\hat{\rho}_{|R|^{0.1}}(1)$ (Figure \ref{fig12}).
As expected, the coverage is very unstable for the tests based on the powers $p=2$ and  $p=1$, due to the loss of asymptotic normality for the full-sample and group estimators $\hat{\beta}=\hat{\rho}_{|R|^{p}}$ and $\hat{\beta}_j=\hat{\rho}_{j, |R|^{p}}$ and their convergence to functions of r.v.'s with asymmetric stable distributions for sufficiently large values of $\alpha$ (Example \ref{ex:garch:appl:limit}). Specifically, this holds %the estimators are asymptotically asymmetric stable  
if $\zeta \le 4p $, and, hence, by Kesten's equation (\ref{KestenEq}), whenever $\alpha \ge 3^{-1/2} \approx 0.5574$ ($\alpha \ge 105^{-1/4}\approx 0.3124$) for $p=1$ ($p=2$). 

%{\color{red}\textbf{Or refer to non-Gaussianity, it's functions of stables for correlations - RI}}

The coverage improves for smaller powers, in particular for $p=0.1, 0.25$. The best coverage across all values of $\alpha$ is observed for the robust $t-$statistic tests with $p=0.1$ and $q=4, 8$ number of groups. For $p\le 0.5,$ the coverage levels of the HAC-based approaches are comparable to those of the $t-$statistic approaches with $q=8$ groups, although the coverage for $q=4$ groups is always better and closer to the correct 95$\%$. % whenever the moment conditions for asymptotic normality of the estimators are satisfied. 

%Further, in the case $p=0.5,$ the HAC approach has very similar coverage to the robust $t-$statistic approaches with $q = 8$ groups, although the coverage for $q=4$ groups is always larger and closer to the correct 95$\%$.  
We emphasize that for the case where $p=0.5$, the HAC-based methods are not theoretically justified due to infinite moments (as this requires $\zeta > 8p$). This is in contrast to the robust $t-$statistic approaches that are asymptotically valid under $p=0.5$. Again, the reasonable performance of the HAC-based approach may be due to self-normalized structure of the HAC-based $t$-statistic, as discussed in Remark \ref{rem:results_HAC_self}.% and thus their application is justified.}
%It is important to emphasize that, in the cases like those with $p=0.5$, moment  despite similar finite-sample performance, HAC methods lack asymptotic validity and are thus not theoretically justified due to infinite moments while asymptotic validity holds for the robust $t-$statistic approaches whose application is thus justified in contrast to HAC inference procedures.}  

Similar to the findings in Section \ref{ind}, to make reliable inference on property \ref{AbsPower}, one may choose $p>0$ small in the robust $t$-statistic approaches with $q=4$ or $q=8$ groups. Similar to the discussion in Section \ref{ind}, the latter approaches may be viewed as useful complements to HAC-based inference methods in the analysis of nonlinear dependence/volatility clustering property \ref{AbsPower}.
\begin{center}
    [Figure \ref{fig3} about here.]
\end{center}
%\textbf{The conclusions} are similar for the case of $t(50,0.5)$-case (Figures \ref{fig8_2_new}-\ref{fig12_2_new} in Appendix \ref{app:coverage}) and the $t(3,0.5)$-case (Figures \ref{fig8_2}-\ref{fig12_2} in Appendix \ref{app:coverage}).

\section{Illustration: Revisiting \cite{baltussen2019}}\label{sec:empirical}

In this section, we revisit a recent study by \cite{baltussen2019} that (among other contributions) tests for linear dependence in returns on the world's major stock market indices. Specifically, relying on HAC-based inference applied to the first-order autocorrelations,  $\rho'_{R,R}(1)$, \cite{baltussen2019} state that serial dependence in daily returns on 20 major market indices covering 15 countries in North America, Europe, and Asia was significantly positive until the end of the 1990s, and switched to being significantly negative since the early 2000s. In light of the discussion in the introduction and Examples \ref{ex:garch:appl:limit} and \ref{ex:garch:independence} with $s=1$, asymptotic normality of sample linear autocorrelations requires finite fourth-order moments of the (GARCH-type) return process. In the case where such moment conditions are not satisfied, the sample linear autocorrelations weakly converge (under suitable conditions) to functions of non-Gaussian stable variables, invalidating the HAC-based inference approaches based on asymptotic normality.
%limiting distribution is (under suitable conditions) a function of non-Gaussian stable variables, invalidating the HAC-based inference approach based on asymptotic normality.

%{\color{red}\textbf{Minor: tests for linear dependence, must be okay - RI}}

We consider daily percentage returns on the major stock indexes from March 3, 1999 to December 31, 2016 as in  \cite{baltussen2019}.
%January 14, 2020.
%\footnote{The data series were retrieved from Bloomberg. We choose the same starting date (March 3, 1999) as \cite{baltussen2019}, but extend the sample until January 14, 2020 (Baltussen et al. use data series ending on December 31, 2016).}  
The second and third columns of Table \ref{tab2} provide, respectively, the (bias-corrected) log-log rank-size regression estimates of the tail indices for the return time series and their 95\% confidence intervals (see \citeauthor{GI}, \citeyear{GI}). Importantly, for 19 out of 20 series, the estimates of the tail index are smaller than 4. The left end-points of the confidence intervals vary from 2.49 to 3.41 across the return series, and several of the intervals lie to the left of 4. This indicates that standard HAC-based inference on linear autocorrelations, $\rho'_{R,R}(1)$, is \emph{invalid} for several of the data series.

%\textcolor{red}{Note: Typo in column 6 of Table 2 for the correlation coefficient. /RSP}
%\textcolor{red}{RI: Minor: 'suggests' or 'indicates'.}
%\textcolor{red}{Would it make sense to report HAC-based 90\% CIs in column 5 (or in a new column), so that we may compare directly to the findings in Baltussen et al.? /RSP}

%\textcolor{red}{Could we make the analysis much more clear, by reducing the content of Table 2? Since the t-statistic approaches are not theoretically justified for $s=1$ (unless we have symmetry!), we may just remove column 6. In column 5, we could state 90\% CIs so that we have the results from Baltussen et al. We could remove columns 9-10 and simply mention that if we let $s=0.1$ for all series, we find qualitatively the same conclusions as when choosing $s$ based on the estimated tail index. I do not think that anything important is left out by doing this. /RSP}

%\textcolor{red}{Perhaps, we could consider a robustified MAC(5) index, where we take the average of $\rho_{R,|R|^s \sign(R)}(h)$ for $h=1,\dots,5$?, and then state the robustified CI for this instead of the last column of Table 2? \textbf{If we implement the suggestions above, we should edit the following paragraph. But that should be fairly quick to do. /RSP}}
%\textcolor{red}{\textbf{RSP: Would it be possible to have two digits on all tail index estimates?}}

Columns 4 and 7 in Table \ref{tab2} contain full-sample estimates $\hat{\rho}'_{R, |R|^{s} \sign(R)}$ of the correlations $\rho'_{R, |R|^{s} \sign(R)}(1)$ (as in \ref{Autocorr2}) for different values of $s$, and column 10 contains estimates of the multi-period (auto)correlation MAC(5) (a weighted sum of the correlation coefficients $\rho'_{R, |R|^{s} \sign(R)}(h)$ of order $h=1, \dots, 5$)  used in \cite{baltussen2019}. Column 5 provides, similar to \cite{baltussen2019}, HAC-based $t$-statistics for the nullity of the linear autocorrelation $\rho'_{R, R}(1)$ (as in \ref{Autocorr}), whereas column 6 contains the $t$-statistics in (\ref{eq:def:tstat}) based on $q=8$ group estimates of $\rho'_{R, R}(1)$.
In column 7, for each return time series, the power $s$ is chosen based on the left end-points of the confidence intervals for the tail index of the time series (column 3 of Table \ref{tab2}). Specifically, following Examples \ref{ex:garch:appl:limit} and \ref{ex:garch:independence}, if the left end-point exceeds 3, we set $s=0.5;$ if the end-point lies between 2.5 and 3.0, we set $s=0.25;$ and if the end-point lies between 2.2 and 2.5, we set $s=0.1.$ Under the above values of powers $s$ (and assuming that the true tail index $\zeta$ belongs to the reported confidence intervals), the moment conditions for asymptotic normality of $\hat{\rho'}_{R,|R|^s \sign(R)}(1)$ and asymptotic independence and normality of the group estimators  (under the hypothesis $H_0: \rho'_{R,|R|^s \sign(R)}(1)=0$ in consideration) in Lemmas \ref{NormalAsymp1} and \ref{lemma} are satisfied. Consequently, by Theorem \ref{thm:sizecontrol}, the robust $t-$statistic approaches for testing $H_0: \rho'_{R,|R|^s \sign(R)}(1)=0$ are asymptotically valid.

%\textcolor{red}{Below note that t-stat approaches are justified under symmetry but this doesn't hold for returns, right - RI}

The reported HAC $t$-statistics and $t-$statistics in group estimates as in (\ref{eq:def:tstat}) in columns 5 and 6 suggest that the hypothesis of zero linear autocorrelation, $H_0: \rho'_{R,R}(1)=0$, cannot be rejected for most of the series at conventional significance levels. However, the HAC and robust $t-$statistic approaches are not theoretically justified in the case $s=1.$

Further, based on the theoretically justified %robust $t$-statistics approaches with 
$t-$statistics in group estimates (as in (\ref{eq:def:tstat})) in column 9, the hypothesis $H_0: \rho'_{R,|R|^s \sign(R)}(1)=0$ is rejected only for six of the series. Hence, based on the theoretically justified robust $t-$statistic approaches, we find evidence of zero correlations in most of the series in contrast to the conclusions in \cite{baltussen2019}. On the other hand, similar to \cite{baltussen2019}, we find somewhat stronger evidence for non-zero weighted autocorrelations (MAC(5)),  based on the robust $t$-statistics in group estimates reported in column 12, where the null of $\text{MAC(5)}=0$ is rejected for 11 out of 20 series.

\begin{center}
    [Table \ref{tab2} about here.]
\end{center}

Table \ref{tab5} contains the results on testing for nonlinear dependence and volatility clustering in the return time series considered using the 95\% confidence intervals constructed on the base of robust $t-$statistic approaches applied to inference on the autocorrelations ${\rho}_{|R|^{p}}(5)$ with $p=0.1, 0.5, 1, 2$. %${\rho}_{R^2}(5),$ ${\rho}_{|R|}(5)$ and ${\rho}_{|R|^{p}}(5)$ 
%with $p=0.1, 0.5$.
The table also presents the results on tail index estimation and HAC-based confidence intervals for the dependence measures considered. %In light of the estimates of the tail indices and their confidence intervals in Table \ref{tab5}, 
The $t$-statistic approaches are theoretically justified for the powers $p = 0.1, 0.5$ (but not for $p=1, 2$) provided that the true tail indices belong to the confidence intervals reported in the table. One should further note that, for the above tail indices, %provided that the true tail indices belong to the confidence intervals reported in the table, 
the HAC approaches are theoretically
justified only under $p=0.1.$ Overall, the results in the table confirm the presence of nonlinear dependence and volatility clustering in the returns on the most of the financial indices. Exceptions are the ASX 200 and Russell 2000 indices, where $t-$statistic approaches with $q=8$ applied to $\rho_{|R|^{0.1}}(5)$ indicate, somewhat surprisingly, absence of volatility clustering.

\begin{center}
    [Table \ref{tab5} about here.]
\end{center}

%{\color{red}\textbf{TheMaybe note that $p<0.5</\zeta/4$ and in accordance with the left-end points of the intervals, so that asymptotic normality is satisfies.}}

%\textcolor{red}{Need to say that justified when $p=0.1, 0.5$ but not for $p=1, 2$? - RI}

\section{Conclusion and suggestions for further research}\label{sec:conclusion}
%\textcolor{red}{\textbf{RSP: Perhaps rephrase? Focus on developing new methods instead of just justifying the applicability of the IM approach?}}

The paper proposes new approaches to inference on measures of market (non-)efficiency, volatility clustering and nonlinear dependence in the case of general heavy-tailed dependent time series, including GARCH-type processes. We provide the results that motivate the use of measures of market (non-)efficiency and volatility clustering based on (small) powers of absolute returns and their signed versions.

The inference approaches dealt with in the paper are based on robust $t-$statistic tests and several new results on their applicability in the settings considered. Theoretical and numerical results and empirical applications in the paper confirm validity, appealing finite sample properties, and wide applicability of the proposed inference approaches. %under heterogeneity and dependence assumptions satisfied in real-world financial markets. %, their advantagesand the wide applicability of the proposed ahpproaches. %We provide the results that justify validity  robust inference under heterogeneity and dependence assumptions satisfied inreal-world financial markets.

% In the approaches, parameter  estimates  (e.g.,  estimates  of  measures  of  nonlinear  dependence)  are  computed for groups of data and the inference is based ont?statistics in the resulting group estimates.  Thisresults in valid robust inference under heterogeneity and dependence assumptions satisfied inreal-world financial markets.  Numerical results and empirical applications confirm advantagesand the wide applicability of the proposed ahpproaches.

%The focus of this paper was on the case of testing of stylized facts of financial markets in form \ref{Autocorr}-\ref{AutocorrSq}, as well as \ref{Autocorr2}-\ref{AbsPower}, with application of inference on one parameter of interest (an autocovariance or autocorrelation based on powers of absolute returns). 

The future research may focus on the development of the two-sample analogues of the approaches to robust inference on market (non-)efficiency and volatility clustering dealt with in the paper using the results in \cite{IM2016}. The results in this direction may be used in testing for structural breaks in the dynamics of key economic and financial time series, including financial returns and foreign exchange rates, and comparisons of the properties of the dynamics of different economic and financial markets. The research on the above inference problems is currently under way by the authors, and will be presented elsewhere.

%. Such inference may be based on two-sample $t-$statistics in group estimators of parameters of interest %(e.g., measures of market inefficiency and analogues of the approaches for the two-sample case in \cite{IM2016}
  %  provide robust tests for equality of two parameters of interest, e.g., $\beta_1=\beta_2,$ using two-sample $t-$statistics in group estimators of these parameters.

%in the case of testing of owever, analogues of the approaches for the two-sample case in \cite{IM2016}
 %   provide robust tests for equality of two parameters of interest, e.g., $\beta_1=\beta_2,$ using two-sample $t-$statistics in group estimators of these parameters. These results that may be used in testing for structural breaks and comparisons of economic and financial models and time series. Using the arguments similar to those in this paper, one can show asymptotic validity of the robust two-sample $t-$statistic approaches in the context of testing the hypothesis on equality of measures of market efficiency and nonlinear dependence dealt with in the paper.

%may be used The results in \cite{IM2016} on robust inference on equality of two parameters of interest can be further used for robust tests of structural breaks in the measures of market (non-)efficiency, volatility clustering and nonlinear dependence considered in this work and differences across the markets. These applications of the robust inference approaches are currently under way by the authors and will be presented elsewhere.

\section*{Acknowledgements}
%We thank Christian Hansen (editor), an Associate Editor, and two anonymous referees for very helpful comments that have led to a much improved paper. In particular, we thank the Associate Editor for kindly pointing out relevant references on self-normalization-based inference and for references on HAC/HAR inference under relaxed moment conditions. 
We thank W. Distaso, J.-M. Dufour, L. Giraitis, P. Kattuman, A. Min, T. Mikosch, U. K. M\"uller, J. L. M. Olea, A. Prokhorov, A. Rahbek, G. Sz{\' e}kely, L. Trapani and Y. Zu for comments and discussions. Research of R. Ibragimov and A. Skrobotov  was supported in part by a grant from the Russian Foundation for Basic Research (RFBR, Project No. 20-010-00960). %Russian Science Foundation (Project No. 16-18-10432). 
R. S. Pedersen was supported by a grant from the Independent Research Fund Denmark (FSE 0133-00162B). Pedersen is a research fellow at the Danish Finance Institute.

\bibliography{IbragimovBrown}
    \bibliographystyle{agsm}
%\newpage
\bigskip

\bigskip

%\section{Additional simulations results}

%\subsection{Size-adjusted powers for $\rho'_{R,|R|^ssign(R)}(h)$}\label{app:powercurves}
%\begin{center}
%    [Figures \ref{fig2_sa_new}-\ref{fig7_3} about here.]
%\end{center}

%\subsection{Coverage levels for $\rho_{|R|^p}(h)$} \label{app:coverage}
%\begin{center}
%    [Figures \ref{fig8_2_new}-\ref{fig12_2} about here.]
%\end{center}

\newcolumntype{C}{>{\centering\arraybackslash}X}

\newpage 

\begin{landscape}

\begin{table}[h!]
\centering
\footnotesize
\caption{Size of tests for absence of autocorrelations, $h=1$\label{tab1}}

\begin{tabularx}{1.35\textwidth}{ccccccccccccccccccccc} \toprule
&\multicolumn{4}{c}{HAC} & \multicolumn{16}{c}{$t-$statistic approach}\\
\cmidrule(r){2-5}\cmidrule(r){6-21}
&\rotatebox[origin=c]{90}{${\rho'}_{R,R}(h)$}&\rotatebox[origin=c]{90}{${\rho'}_{R, |R|^{0.5} sign(R)}(h)$}&\rotatebox[origin=c]{90}{${\rho'}_{R, |R|^{0.25} sign(R)}(h)$}&\rotatebox[origin=c]{90}{${\rho'}_{R, |R|^{0.1} sign(R)}(h)$}&\multicolumn{4}{c}{${\rho'}_{R,R}(h)$}&\multicolumn{4}{c}{${\rho'}_{R, |R|^{0.5} sign(R)}(h)$}&\multicolumn{4}{c}{${\rho'}_{R, |R|^{0.25} sign(R)}(h)$}&\multicolumn{4}{c}{${\rho'}_{R, |R|^{0.1} sign(R)}(h)$}\\
\cmidrule(r){2-5}\cmidrule(r){6-9}\cmidrule(r){10-13}\cmidrule(r){14-17}\cmidrule(r){18-21}
$q$&&&&&4&8&12&16&4&8&12&16&4&8&12&16&4&8&12&16\\\hline
 ARCH(1), $N(0,1)$
  &7.9&6.1&6.2&6.1&4.6&4.5&4.8&4.8&4.9&4.9&5.1&5.2&5.3&5.1&5.2&5.5&5.4&5.0&5.2&5.4\\
 ARCH(1), $t(50,0.5)$ &8.6&5.9&5.8&5.8&4.4&5.1&5.6&6.2&4.6&5.1&5.3&6.1&4.9&5.4&5.5&5.8&5.0&5.3&5.5&5.7\\
 ARCH(1), $t(3,0.5)$
  &17.3&9.7&8.2&7.6&6.6&9.5&11.5&13.6&5.8&7.3&8.2&9.3&5.5&6.5&7.1&8.0&5.2&5.9&6.3&7.3\\
%GJR-GARCH, $N(0,1)$&5.8&5.6&5.7&5.6&4.6&5.2&5.8&5.9&4.6&5.2&5.6&5.8&4.8&5.1&5.5&5.6&4.9&5.1&5.4&5.6\\
 \bottomrule
\end{tabularx}
\end{table}
\end{landscape}

\newpage

\newpage

\begin{landscape}

\begin{table}[h!]
\centering
\begin{threeparttable}
%\footnotesize
\scriptsize
\caption{Empirical results, testing for efficient market hypothesis, $h=1$\label{tab2}}

\begin{tabularx}{1.11\textwidth}{ccccccccccccccc} \toprule
Series&
\rotatebox[origin=c]{90}{Estimate of tail index}&
\rotatebox[origin=c]{90}{CI for tail index}&
\rotatebox[origin=c]{90}{$\hat{\rho'}_{R,R}(h)$}& 
\rotatebox[origin=c]{90}{t-statistic for $\rho'_{R,R}(h)=0$ (HAC)}& %\rotatebox[origin=c]{90}{Reject null of zero correlation?}&
\rotatebox[origin=c]{90}{$t_\beta$ for $\rho_{R,R}(h)=0$ ($q=8$)}&
\rotatebox[origin=c]{90}{$\hat{\rho'}_{R, |R|^{s} sign(R)}(h)$}& 
\rotatebox[origin=c]{90}{t-statistic for $\rho'_{R, |R|^{s} sign(R)}(h)=0$ (HAC)}&
\rotatebox[origin=c]{90}{$t_\beta$ for $\rho'_{R, |R|^{s} sign(R)}(h)=0$ ($q=8$)}&
\rotatebox[origin=c]{90}{Estimate of MAC(5)}&
\rotatebox[origin=c]{90}{t-statistic for $\text{MAC(5)}=0$ (HAC)}&
\rotatebox[origin=c]{90}{$t_\beta$ for $\text{MAC(5)}=0$ ($q=8$)}\\\hline
S\&P 500&3.26&[2.66,3.86] &-0.074&-3.44$^{***}$&-3.12$^{***}$&-0.056&-3.74$^{***}$&-3.32$^{***}$&-0.085&-4.41$^{***}$&-5.02$^{***}$\\
FTSE 100&3.48&[2.84,4.12] &-0.040&-2.24$^{***}$&-1.75 &-0.029&-1.96$^{**}$&-1.69 &-0.075&-4.31$^{***}$&-3.12$^{***}$\\
DJESI 50&3.73&[3.05,4.42] &-0.029&-1.66$^{**}$&-2.1$^{**}$&-0.027&-1.71$^{**}$&-2.03$^{**}$&-0.057&-3.59$^{***}$&-3.31$^{***}$\\
TOPIX&3.31&[2.69,3.93] &0.014&0.66 &1.63 &0.036&2.26$^{***}$&2.05$^{**}$&-0.014&-0.87 &1.42\\ 
ASX 200&3.41&[2.78,4.04] &-0.034&-1.77$^{**}$&-1.23 &-0.018&-1.16 &-0.71 &-0.038&-1.86$^{**}$&-0.96 \\
TSE 60&3.13&[2.55,3.70] &-0.039&-1.7$^{**}$&-0.15 &-0.003&-0.2 &0.06 &-0.058&-2.93$^{***}$&-1.55 \\
CAC 40&3.64&[2.97,4.30] &-0.027&-1.51 &-1.77 &-0.025&-1.65 &-1.72 &-0.061&-3.57$^{***}$&-2.76$^{***}$\\
DAX&3.62&[2.96,4.29] &-0.013&-0.8 &-0.64 &-0.008&-0.53 &-0.6 &-0.026&-1.54 &-1.48 \\
IBEX 35&3.77&[3.07,4.46] &0.004&0.23 &-0.09 &0.005&0.29 &-0.04 &-0.036&-2.12$^{***}$&-1.91$^{**}$\\
%&3.17&[2.4,3.94] &-0.029&-1.48 &-2.45$^{***}$&-0.026&-1.46 &-2.54$^{***}$&-0.021&-0.84 &-2.54$^{***}$\\
MIB&3.75&[3.06,4.44] &-0.024&-1.48 &-1.91$^{**}$&-0.027&-1.78$^{**}$&-1.91$^{**}$&-0.031&-1.79$^{**}$&-2.93$^{***}$\\
AEX Index&3.32&[2.71,3.93] &-0.007&-0.37 &-0.19 &0.002&0.12 &0.14 &-0.026&-1.58 &-1.14 \\
OMX Stockholm&4.18&[3.41,4.95] &0.001&0.06 &-0.86 &0.013&0.83 &-0.03 &-0.030&-2.07$^{***}$&-1.59 \\
SMI&3.19&[2.60,3.78] &0.029&1.49 &0.89 &0.017&1.15 &0.69 &-0.016&-0.78 &-1.24 \\
Nikkei 225&3.32&[2.70,3.95] &-0.042&-1.95$^{**}$&-2.05$^{**}$&-0.024&-1.48 &-1.54 &-0.046&-2.96$^{***}$&-2.26$^{**}$\\
HSI&3.41&[2.78,4.05] &-0.004&-0.2 &0.51 &0.026&1.69$^{**}$&1.88 &-0.014&-0.9 &0.76 \\
Nasdaq 100&3.64&[2.97,4.32] &-0.059&-3.17$^{***}$&-2.18$^{**}$&-0.036&-2.6$^{***}$&-2.59$^{***}$&-0.087&-4.72$^{***}$&-3.36$^{***}$\\
NYSE&3.05&[2.49,3.62] &-0.057&-2.57$^{***}$&-1.56 &-0.025&-1.77$^{**}$&-1.23 &-0.067&-3.54$^{***}$&-2.42$^{***}$\\
Russell 2000&3.51&[2.86,4.16] &-0.058&-2.62$^{***}$&-1.41 &-0.026&-1.58 &-1.24 &-0.060&-3.48$^{***}$&-2.37$^{***}$\\
S\&P 400&3.38&[2.76,4.01] &-0.029&-1.34 &-0.4 &0.005&0.3 &0.59 &-0.053&-2.93$^{***}$&-2.4$^{***}$\\
%&3.87&[3.15,4.59] &0.030&1.54 &3.27$^{***}$&0.039&2.31$^{***}$&3.25$^{***}$&-0.004&-0.23 &1.28 \\
KOSPI 200&3.88&[3.16,4.61] &0.023&1.23 &2.05$^{**}$&0.031&1.88$^{**}$&2.71$^{***}$&-0.014&-0.96 &0.17 \\
 \bottomrule
\end{tabularx}
 \begin{tablenotes}
      \small
      \item \textit{Note}: $^{*}$, $^{**}$ and $^{***}$ denote the significance at 10, 5 and 1\% levels respectively.
    \end{tablenotes}
\end{threeparttable}
\end{table}

\end{landscape}

\newpage

\begin{landscape}

\begin{table}[h!]
\centering
\begin{threeparttable}
%\footnotesize
\scriptsize
\caption{Empirical results, testing for non-linearity, $h=5$\label{tab5}}

\begin{tabularx}{1.43\textwidth}{ccccccccccccc} \toprule
Series&%\rotatebox[origin=c]{90}{Estimate of tail index}&
%\rotatebox[origin=c]{90}{CI for tail index}&
\rotatebox[origin=c]{90}{$\hat{\rho}_{R^2}(h)$}& 
\rotatebox[origin=c]{90}{CI for $\rho_{R^2}(h)$ (HAC)}& %\rotatebox[origin=c]{90}{Reject null of zero correlation?}&
\rotatebox[origin=c]{90}{CI for $\rho_{R^2}(h)$ ($q=8$)}&
\rotatebox[origin=c]{90}{$\hat{\rho}_{|R|}(h)$}& 
\rotatebox[origin=c]{90}{CI for $\rho_{|R|}(h)$ (HAC)}& %\rotatebox[origin=c]{90}{Reject null of zero correlation?}&
\rotatebox[origin=c]{90}{CI for $\rho_{|R|}(h)$ ($q=8$)}&
\rotatebox[origin=c]{90}{$\hat{\rho}_{|R|^{0.5} }(h)$}&\rotatebox[origin=c]{90}{CI for $\rho_{|R|^{0.5} }(h)$ (HAC)}& \rotatebox[origin=c]{90}{CI for $\rho_{|R|^{0.5} }(h)$ ($q=8$)}&
\rotatebox[origin=c]{90}{$\hat{\rho}_{|R|^{0.1} }(h)$}& \rotatebox[origin=c]{90}{CI for $\rho_{|R|^{0.1} }(h)$ (HAC)}&\rotatebox[origin=c]{90}{CI for $\rho_{|R|^{0.1} }(h)$ ($q=8$)}
%\rotatebox[origin=c]{90}{Reject null of zero correlation?}&
\\\hline
%\rotatebox[origin=c]{90}{Reject null of zero correlation?}\\\hline
S\&P 500&0.316&[0.17,0.47]$^*$ &[0.05,0.207]$^*$ &0.331&[0.26,0.4]$^*$ &[0.05,0.262]$^*$ &0.261&[0.21,0.31]$^*$ &[0.04,0.24]$^*$ &0.177&[0.15,0.21]$^*$ &[0.02,0.19]$^*$ \\
FTSE 100&0.330&[0.17,0.49]$^*$ &[0.09,0.243]$^*$ &0.302&[0.23,0.37]$^*$ &[0.11,0.256]$^*$ &0.240&[0.2,0.28]$^*$ &[0.09,0.21]$^*$ &0.143&[0.11,0.17]$^*$ &[0.03,0.12]$^*$ \\
DJESI 50&0.259&[0.16,0.35]$^*$ &[0.02,0.211]$^*$ &0.255&[0.2,0.31]$^*$ &[0.05,0.215]$^*$ &0.203&[0.16,0.24]$^*$ &[0.05,0.17]$^*$ &0.126&[0.09,0.16]$^*$ &[0.02,0.1]$^*$ \\
TOPIX&0.203&[0.11,0.3]$^*$ &[0.05,0.19]$^*$ &0.219&[0.14,0.29]$^*$ &[0.06,0.235]$^*$ &0.170&[0.12,0.22]$^*$ &[0.05,0.2]$^*$ &0.109&[0.08,0.14]$^*$ &[0.03,0.14]$^*$ \\
ASX 200&0.228&[0.12,0.34]$^*$ &[0.01,0.147]$^*$ &0.243&[0.18,0.3]$^*$ &[0.03,0.17]$^*$ &0.188&[0.14,0.23]$^*$ &[0.02,0.14]$^*$ &0.107&[0.08,0.14]$^*$ &[-0.01,0.09] \\
TSE 60&0.350&[0.22,0.48]$^*$ &[0.04,0.25]$^*$ &0.354&[0.26,0.44]$^*$ &[0.04,0.27]$^*$ &0.272&[0.22,0.33]$^*$ &[0.03,0.23]$^*$ &0.165&[0.13,0.2]$^*$ &[0.01,0.16]$^*$ \\
CAC 40&0.262&[0.15,0.37]$^*$ &[0.01,0.20]$^*$ &0.248&[0.19,0.3]$^*$ &[0.05,0.21]$^*$ &0.201&[0.16,0.24]$^*$ &[0.05,0.18]$^*$ &0.142&[0.11,0.17]$^*$ &[0.04,0.13]$^*$ \\
DAX&0.246&[0.16,0.34] &[0.05,0.203]$^*$ &0.274&[0.23,0.32]$^*$ &[0.08,0.226]$^*$ &0.234&[0.2,0.27]$^*$ &[0.08,0.19]$^*$ &0.169&[0.14,0.2]$^*$ &[0.05,0.14]$^*$ \\
IBEX 35&0.159&[0.05,0.26]$^*$ &[0.02,0.13]$^*$ &0.217&[0.16,0.27]$^*$ &[0.04,0.17]$^*$ &0.191&[0.15,0.23]$^*$ &[0.04,0.15]$^*$ &0.141&[0.11,0.17]$^*$ &[0.03,0.11]$^*$ \\
%MIB 30&0.302&[0.19,0.42] &[0.01,0.219] &0.319&[0.25,0.39] &[0.02,0.23] &0.267&[0.22,0.32] &[0.02,0.19] &0.172&[0.13,0.21] &[0.01,0.13] 
MIB&0.212&[0.12,0.3]$^*$ &[0.04,0.19]$^*$ &0.251&[0.2,0.3]$^*$ &[0.05,0.21]$^*$ &0.220&[0.18,0.26]$^*$ &[0.04,0.18]$^*$ &0.159&[0.13,0.19]$^*$ &[0.03,0.13]$^*$ \\
AEX Index&0.377&[0.25,0.51]$^*$ &[0.03,0.28]$^*$ &0.325&[0.26,0.39]$^*$ &[0.06,0.27]$^*$ &0.248&[0.2,0.29]$^*$ &[0.06,0.22]$^*$ &0.152&[0.12,0.18]$^*$ &[0.02,0.15]$^*$ \\
OMX Stockholm&0.232&[0.14,0.32]$^*$ &[0.06,0.188]$^*$ &0.240&[0.19,0.29]$^*$ &[0.09,0.194]$^*$ &0.208&[0.17,0.24]$^*$ &[0.07,0.16]$^*$ &0.150&[0.12,0.18]$^*$ &[0.03,0.13]$^*$ \\
SMI&0.283&[0.19,0.38]$^*$ &[0.07,0.246]$^*$ &0.302&[0.23,0.37]$^*$ &[0.1,0.28]$^*$ &0.244&[0.2,0.29]$^*$ &[0.09,0.23]$^*$ &0.163&[0.13,0.19]$^*$ &[0.05,0.16]$^*$ \\
Nikkei 225&0.214&[0.11,0.32]$^*$ &[0.03,0.18]$^*$ &0.219&[0.14,0.3]$^*$ &[0.05,0.22]$^*$ &0.174&[0.13,0.22]$^*$ &[0.05,0.2]$^*$ &0.126&[0.09,0.16]$^*$ &[0.05,0.15]$^*$ \\
HSI&0.177&[0.1,0.25]$^*$ &[0.02,0.22]$^*$ &0.235&[0.18,0.29]$^*$ &[0.03,0.21]$^*$ &0.195&[0.15,0.24]$^*$ &[0.01,0.17]$^*$ &0.143&[0.11,0.17]$^*$ &[0.003,0.13]$^*$ \\
Nasdaq 100&0.224&[0.11,0.34]$^*$ &[0.04,0.17]$^*$ &0.346&[0.3,0.39]$^*$ &[0.06,0.22]$^*$ &0.310&[0.27,0.35]$^*$ &[0.06,0.19]$^*$ &0.217&[0.19,0.25]$^*$ &[0.05,0.11]$^*$ \\
NYSE&0.366&[0.19,0.54]$^*$ &[0.04,0.23]$^*$ &0.347&[0.27,0.43]$^*$ &[0.05,0.27]$^*$ &0.261&[0.21,0.31]$^*$ &[0.05,0.24]$^*$ &0.178&[0.15,0.21]$^*$ &[0.03,0.19]$^*$ \\
Russell 2000&0.315&[0.22,0.41]$^*$ &[0.04,0.20]$^*$ &0.288&[0.22,0.36]$^*$ &[0.03,0.24]$^*$ &0.206&[0.15,0.26]$^*$ &[0.01,0.21]$^*$ &0.113&[0.06,0.16]$^*$ &[-0.001,0.16] \\
S\&P 400&0.338&[0.22,0.46]$^*$ &[0.06,0.23]$^*$ &0.326&[0.26,0.4]$^*$ &[0.06,0.25]$^*$ &0.245&[0.19,0.29]$^*$ &[0.05,0.22]$^*$ &0.161&[0.13,0.19]$^*$ &[0.04,0.16]$^*$ \\
%KOSPI&0.227&[0.13,0.32] &[0.07,0.269] &0.269&[0.22,0.32] &[0.08,0.244] &0.244&[0.21,0.28] &[0.05,0.19] &0.178&[0.15,0.21] &[0.01,0.13] 
KOSPI 200&0.218&[0.13,0.3]$^*$ &[0.07,0.256]$^*$ &0.263&[0.22,0.31]$^*$ &[0.07,0.23]$^*$ &0.238&[0.2,0.27]$^*$ &[0.05,0.19]$^*$ &0.144&[0.11,0.17]$^*$ &[0.002,0.12]$^*$ \\
 \bottomrule
\end{tabularx}
 \begin{tablenotes}
      \small
      \item \textit{Note}: $^{*}$ denote the significance at 5\% level.
    \end{tablenotes}
\end{threeparttable}
\end{table}
\end{landscape}

\newpage

\newpage

\newpage

\clearpage

\appendix

\setcounter{equation}{0}
\setcounter{figure}{0}
\renewcommand{\theequation}{S.\arabic{equation}}%
\renewcommand{\thelemma}{S.%
\arabic{lemma}}
\renewcommand{\thefigure}{S.\arabic{figure}}%

\begin{center}
\vspace*{2in}{\Huge Online Appendix}

\bigskip

{\Huge \bigskip}

{\Large to}

\bigskip

{\Large \bigskip}

{\Large New Approaches to Robust Inference on Market (Non-)Efficiency, Volatility Clustering and Nonlinear Dependence}

\end{center}

\bigskip

\setcounter{section}{0}
\def\thesection{\arabic{section}}%

\renewcommand{\thesection}{S.\arabic{section}}
\setcounter{section}{0}%

\renewcommand{\thetable}{S.\arabic{table}}
\setcounter{table}{0}%

\renewcommand{\thepage}{[S.\arabic{page}]}
\setcounter{page}{1}%

\newpage

\section{Asymptotic normality of general sample covariances and correlations} \label{Proofs}
The following Lemma \ref{thm:limit:cov_corr} provides the results on asymptotic normality of sample covariances and correlations for arbitrary functions of stationary processes (see also Francq and Zako\"ian, 2006, and Lindner, 2009, in relation to GARCH processes).  The result relies on applying a central limit theorem (CLT) for stationary $\alpha$-mixing (i.e$.$ strongly mixing) processes. We refer to Rio (2017) for a detailed treatment of mixing processes. We note that Lemmas \ref{NormalAsymp} and \ref{NormalAsymp1} follow immediately from general Lemma \ref{thm:limit:cov_corr}.

%In order to show that the sample auto(cross)covariances and -correlations have asymptotic Gaussian limits, we rely on the following central limit theorem (CLT) for strictly stationary $\alpha$-mixing, i.e. strongly mixing, processes. We refer to Appendix \ref{sec:mixing} for additional details about mixing processes. The theorem is given in \cite{IL1971} (see also Theorem 4.2 in \cite{rio2017}):
%\begin{thm}[Theorem 18.5.3 of \cite{IL1971}]\label{thm:IL}
%	Let $(R_t)_{t\in \mathbb{Z}}$ be an $\mathbb{R}$-valued mean zero stationary strongly mixing process with mixing coefficient $\alpha(n)$ and let $E[|R_t|^{2+\delta}]<\infty$ for some $\delta >0$. If
%	\begin{eqnarray}\label{eq:mixingcoef}
%	 \sum_{n=1}^{\infty}\alpha(n)^{\delta/(2+\delta)}<\infty,
%	\end{eqnarray}
%	then $\sigma^2\equiv E[R_0^2] +2\sum_{j=1}^{\infty}E[R_0R_j]<\infty$, and $T^{-1/2}\sum_{t=1}^TR_t \overset{d}{\to} N(0,\sigma^2)$.
%\end{thm}

%We have the following result.
\begin{lemma}\label{thm:limit:cov_corr}
Let $(R_t)_{t\in \mathbb{Z}}$ be an $\mathbb{R}$-valued stationary and strongly mixing process with mxing coefficients $\alpha(n), n\in\mathbb{Z}$. Let $f:\mathbb{R} \to \mathbb{R}$ and $g:\mathbb{R} \to \mathbb{R}$ be measurable functions. Consider the sample covariance of $f(R_t)$ and $g(R_{t-h)}$ for some fixed $h\ge0$,
\begin{eqnarray}
	\hat{\gamma}_{T,f(R),g(R)}(h)=\frac{1} {T} \sum_{t=1}^{T} f(R_t) g(R_{t-h}) - (\frac{1} {T} \sum_{t=1}^{T} f(R_t)) (\frac{1} {T} \sum_{t=1}^{T} g(R_{t-h})),
\end{eqnarray}
and its population equivalent,
\begin{eqnarray}
	\gamma_{f(R),g(R)}(h)=\mathrm{Cov}(f(R_t),g(R_{t-h})) = E[f(R_t )g(R_{t-h})]-E[f(R_t)]E[g(R_{t-h})].
\end{eqnarray}
Likewise, consider the sample correlation,
\begin{eqnarray}
	\hat{\rho}_{T,f(R),g(R)}(h)= \frac{\hat{\gamma}_{T,f(R),g(R)}(h)}{\sqrt{\hat{\gamma}_{T,f(R),f(R)}(0)\hat{\gamma}_{T,g(R),g(R)}(0)}}, \quad h\ge1,
\end{eqnarray}
and its population equivalent
\begin{eqnarray}
	\rho_{f(R),g(R)}(h)=\frac{{\gamma}_{f(R),g(R)}(h)}{\sqrt{{\gamma}_{f(R),f(R)}(0){\gamma}_{g(R),g(R)}(0)}}.
\end{eqnarray}
Suppose that there exists a $\delta > 0$ such that
\begin{eqnarray} \label{eq:thm:moment1}
\max\{ E[|f(R_t))|^{2+\delta}],E[|g(R_t))|^{2+\delta}] \}<\infty \  \text{and}\  \max_{h=0,\ldots,m}\{E[|f(R_t)g(R_{t-h})|^{2+\delta}]\}<\infty, 
\end{eqnarray}
and such that the mixing coefficients of $(R_t)_{t\in \mathbb{Z}}$ satisfy
\begin{eqnarray}\label{eq:mixingcoef}
	 \sum_{n=1}^{\infty}\alpha(n)^{\delta/(2+\delta)}<\infty.
\end{eqnarray}
Then
\begin{eqnarray} \label{eq:autocov:conv}
	\sqrt{T}(\hat{\gamma}_{T,f(R),g(R)}(h) - \gamma_{f(R),g(R)}(h) )_{h=0,\ldots,m} \overset{d}{\to} (G_{h,f(R),g(R)})_{h=0,\ldots,m},
\end{eqnarray}
where $(G_{h,f(R),g(R)})_{h=0,\ldots,m}$ is an $(m+1)$-dimensional Gaussian vector with zero mean and the covariance matrix given by
\begin{eqnarray}\label{eq:def:Gamma}
 \Gamma & = & \mathrm{Var}(Y_0) + 2 \sum_{k=1}^{\infty}\mathrm{Cov}(Y_0,Y_k),
\end{eqnarray}
where $Y_t = (Y_{t,h})_{h=0,\ldots,m}$, $Y_{t,h} = (f(R_t) - E[f(R_t)])(g(R_{t-h}) - E[g(R_{t-h})])-\gamma_{f(R),g(R)}(h)$.

If $\gamma_{f(R),g(R)}(h) )_{h=1,\ldots,m} =(0,\ldots,0)$, then
\begin{eqnarray} \label{eq:autocorr:conv0}
	\sqrt{T}(\hat{\rho}_{T,f(R),g(R)}(h))_{h=1,\ldots,m} \overset{d}{\to} \left( (\gamma_{f(R),f(R)}(0)\gamma_{g(R),g(R)}(0))^{-1/2} G_{h,f(R),g(R)} \right)_{h=1,\ldots,m}.
\end{eqnarray}

If there exists a $\delta >0$ such that
\begin{eqnarray}\label{eq:thm:moment2}
\max\{E[|f(R_t)|^{4+\delta}],E[|g(R_t)|^{4+\delta}]\}<\infty 
\end{eqnarray}
and such that \eqref{eq:mixingcoef} holds, then
\begin{eqnarray} \label{eq:autocorr:conv}
	\sqrt{T}(\hat{\rho}_{T,f(R),g(R)}(h) - \rho_{f(R),g(R)}(h) )_{h=1,\ldots,m} \overset{d}{\to} (\tilde{G}_{h,f(R),g(R)})_{h=1,\ldots,m},
\end{eqnarray}
where $(\tilde{G}_{h,f(R),g(R)})_{h=0,\ldots,m}$ is an $(m+1)-$dimensional Gaussian vector with  zero mean and the covariance matrix given by $A\Gamma^{\dagger}A'$, where $A$ is constant matrix defined in \eqref{eq:def:matrixA} and
\begin{eqnarray}\label{eq:def:Gammadagger}
 \Gamma^{\dagger} & = & \mathrm{Var}(Y_0^{\dagger}) + 2 \sum_{k=1}^{\infty}\mathrm{Cov}(Y_0^{\dagger},Y_k^{\dagger}),
\end{eqnarray}
with $Y_t^{\dagger} = (Y_t',V_{t,1},V_{t,2})'$, $V_{t,1} = (f(R_t) - E[f(R_t)])^2-\gamma_{f(R),f(R)}(0) $, $V_{t,2} = (g(R_t) - E[g(R_t)])^2-\gamma_{g(R),g(R)}(0)$.

\end{lemma}

{\bf Proof:} Firstly, note that strong mixing is a property about the $\sigma$-field generated by $(R_t)_{t\in \mathbb{Z}}$. Since $f$ and $g$ are measurable, it follows that the process $(V_t)_{t\in\mathbb{Z}}$, with  $V_t = (f(R_t),g(R_{t}), g(R_{t-1}),\ldots, g(R_{t-m}) )$, is stationary and strongly mixing with mixing coefficients satisfying \eqref{eq:mixingcoef} (see also Francq and Zako\"ian, 2006). Likewise, the same property applies to the process $(\tilde{V}_{t,h})_{t\in\mathbb{Z}}$ with $\tilde{V}_{t,h} = f(R_t)g(R_{t-h})$, $h=0,\ldots,m$. This allows us to apply Theorem 18.5.3 of Ibragimov and Linnik (1971) [the CLT, henceforth] to the processes $(V_t)_{t\in\mathbb{Z}}$ and $(\tilde{V}_{t,h})_{t\in\mathbb{Z}}$, under the moments conditions in the lemma.
%assumed integrability conditions. 
Next, note that
	\begin{eqnarray*}
		\hat{\gamma}_{T,f(R),g(R)}(h) - \gamma_{f(R),g(R)}(h) &=& \frac{1} {T} \sum_{t=1}^{T} (f(R_t)-E[f(R_t)])(g(R_{t-h}) - E[g(R_{t-h})])-\gamma_{f(R),g(R)}(h)  \\
		& & - (\frac{1} {T} \sum_{t=1}^{T} (f(R_t)-E[f(R_t)]))(\frac{1} {T} \sum_{t=1}^{T} (g(R_{t-h}) - E[g(R_{t-h})).
	\end{eqnarray*}
	Suppose that \eqref{eq:thm:moment1} holds. By the CLT, $\frac{1} {T} \sum_{t=1}^{T} f(R_t) - E[f(R_t)]$ and $\frac{1} {T} \sum_{t=1}^{T} g(R_{t-h}) - E[g(R_{t-h})]$ are $O_p(T^{-1/2})$, so that
		\begin{eqnarray*}
		\sqrt{T}(\hat{\gamma}_{T,f(R),g(R)}(h) - \gamma_{f(R),g(R)}(h)) &=& \frac{1} {\sqrt{T}} \sum_{t=1}^{T} ((f(R_t) - E[f(R_t)])(g(R_{t-h}) - E[g(R_{t-h})]) \\
		& & -\gamma_{f(R),g(R)}(h))  + o_p(1).
	\end{eqnarray*}
	Let $Y_{t,h} = (f(R_t) - E[f(R_t)])(g(R_{t-h}) - E[g(R_{t-h})])-\gamma_{f(R),g(R)}(h) $. Using \eqref{eq:thm:moment1}, there exists a $\delta >0$  such that, by H{\"o}lder's inequality, $E[|Y_{t,h}|^{2+\delta}]<\infty$. Since $E[Y_{t,h}] = 0 $ and $(Y_{t,h}:t\in \mathbb{Z})$ is strongly mixing satisfying \eqref{eq:mixingcoef}, $\sqrt{T}(\hat{\gamma}_{T,f(R),g(R)}(h) - \gamma_{f(R),g(R)}(h)) \overset{d}{\to} G_{h,f(R),g(R)} $ by the CLT. By similar arguments applied to linear combinations of \\$\sqrt{T}(\hat{\gamma}_{T,f(R),g(R)}(h) - \gamma_{f(R),g(R)}(h) )_{h=0,\ldots,m}$, and the Cram\'{e}r-Wold device, (\ref{eq:autocov:conv}) holds.

	Turning to the sample correlations, the limiting distribution for the case $\gamma_{f(R),g(R)}(h) )_{h=1,\ldots,m} =0$ is immediate, by noting that $(\hat{\gamma}_{T,f(R),f(R)}(0) - \gamma_{f(R),f(R)}(0))$ and $(\hat{\gamma}_{T,g(R),g(R)}(0) - \gamma_{g(R),g(R)}(0))$ are $o_p(1)$ and using Slutsky's theorem. Next, using \eqref{eq:thm:moment2} and arguments as above,
	\begin{eqnarray} \label{eq:autocov:conv:proof}
	\sqrt{T}\left[
               \begin{array}{c}
                 (\hat{\gamma}_{T,f(R),g(R)}(h) - \gamma_{f(R),g(R)}(h) )_{h=0,\ldots,m} \\
                 \hat{\gamma}_{T,f(R),f(R)}(0) - \gamma_{f(R),f(R)}(0)   \\
                 \hat{\gamma}_{T,g(R),g(R)}(0) - \gamma_{g(R),g(R)}(0)
               \end{array}
             \right]
            \overset{d}{\to}G^{\dagger},
	\end{eqnarray}
	where $G^{\dagger}$ is an $(m+3)$-dimensional Gaussian vector with zero mean and the covariance matrix given by $\Gamma^\dagger$. Let $x = (x_1, \ldots , x_{m+3})'\in \mathbb{R}^{m+3}$ and define the function $\tilde{g}: \mathbb{R}^{m+3} \to \mathbb{R}^{m+1}$ as $\tilde{g}(x)  =   ( \frac{x_{1}}{\sqrt{x_{m+2}x_{m+3}}}   ,\ldots, \frac{x_{m+1}}{\sqrt{x_{m+2}x_{m+3}}} )'$.
	%\begin{eqnarray*}
	%	\tilde{g}(x)  =   ( \frac{x_{1}}{\sqrt{x_{m+2}x_{m+3}}}   ,\ldots, \frac{x_{m+1}}{\sqrt{x_{m+2}x_{m+3}}} )'.
	%\end{eqnarray*}
	Define the matrix
	\begin{eqnarray}\label{eq:def:matrixA}
	A & = & \left. \frac{\partial \tilde{g}(x)}{\partial x'} \right|_{x = \gamma^\dagger}, \quad \gamma^\dagger = ((\gamma_{f(R),g(R)}(h) )_{h=0,\ldots,m}',\gamma_{f(R),f(R)}(0),\gamma_{g(R),g(R)}(0))'.
	\end{eqnarray}
	The convergence in \eqref{eq:autocorr:conv} is then obtained by an application of the delta method. $\square$

% \section{$\alpha$- and $\beta$-mixing}\label{sec:mixing}

% Let $(\Omega,\mathcal{F},P)$ be a probability space.

% The $\beta$-mixing coefficient between two $\sigma$-fields $\mathcal{A}$
% and $\mathcal{B}$, $\mathcal{A},\mathcal{B}\subset\mathcal{F}$,
% is given by
% \[
% \beta(\mathcal{A},\mathcal{B}):=\frac{1}{2}\sup\sum_{i=1}^{I}\sum_{j=1}^{J}|P(A_{i}\cap B_{j})-P(A_{i})P(B_{j})|,
% \]
% where the supremum is taken over all finite partitions $(A_{i}:i=1,\ldots,I)$
% and $(B_{j}:j=1,\ldots,J)$ of $\Omega$ with $A_{i}\in\mathcal{A}$
% and $B_{j}\in\mathcal{B}$.

% Let $(x_{t}:t\in\mathbb{Z})$ be a sequence of r.v.'s. Define
% the $\sigma$-fields
% \begin{align*}
% \mathcal{F}_{n} & :=\sigma(x_{t}:t\in\mathbb{Z},t\le n),\\
% \mathcal{G}_{n} & :=\sigma(x_{t}:t\in\mathbb{Z},t\ge n).
% \end{align*}
% The $\beta$-mixing coefficients of $(x_{t}:t\in\mathbb{Z})$ are given
% by
% \[
% \beta(k):=\sup_{n\in\mathbb{Z}}\beta(\mathcal{F}_{n},\mathcal{G}_{n+k}).
% \]
% Note that if $(x_{t}:t\in\mathbb{Z})$ is strictly stationary, $\beta(k)=\beta(\mathcal{F}_{0},\mathcal{G}_{k})$.
% The sequence $(x_{t}:t\in\mathbb{Z})$ is said to be $\beta$-mixing
% if $\beta(k)\to0$ as $k\to\infty$. Likewise, the strictly  stationary sequence $(x_{t}:t\in\mathbb{Z})$ has $\alpha$-mixing coefficients
% \[
% \alpha(k):= \sup_{A \in \mathcal{F}_{0}, B \in \mathcal{G}_{k} }|P(A\cap B) - P(A)P(B)|,
% \]
% and is said to be $\alpha$-mixing if $\alpha(k) \to 0$ as $k\to \infty$. Notice that for a strictly stationary process, $\beta(k)\ge\alpha(k)$, and hence a $\beta$-mixing process is also $\alpha$-mixing.

\section{Proof of Lemma \ref{lemma}}\label{app:coupling}
Suppose that $(R_{t})_{t\in\mathbb{Z}}$ is stationary and
$\beta$-mixing. For the sake of clarity we focus on the asymptotic independence of group-based estimators for covariances. The sample correlations are dealt with in a similar fashion. Let $f:\mathbb{R} \to \mathbb{R}$ and $g:\mathbb{R} \to \mathbb{R}$ be measurable functions, and define $y_t = (y_{t,1},y_{t,2})' = (f(R_t) , g(R_{t-1}))'$. Since $f$ and $g$ are measurable, $(y_{t})_{t\in\mathbb{Z}}$ is stationary and $\beta$-mixing. With $i,j\in\mathbb{Z}$, $0\le i<j$, let
\[
\hat{\phi}_{i,j}:=(j-i+1)^{-1}\sum_{t=i}^{j}y_{t,1}y_{t,2}-\left((j-i+1)^{-1}\sum_{t=i}^{j}y_{t,1}\right)\left((j-i+1)^{-1}\sum_{t=i}^{j}y_{t,2}\right)-\phi_{0},
\]
where $\phi_{0}=E[y_{t,1}y_{t,2}]-E[y_{t,1}]E[y_{t,2}]$. Suppose that for some deterministic sequence $(a_T)$, satisfying $a_T \to \infty$,
\begin{equation}
a_{T}\hat{\phi}_{1,T}\to_d Z,\label{eq:limit:phi},
\end{equation}
for some r.v. $Z$ (potentially non-Gaussian). Consider the case of two equi-sized groups, such that we have the group-based estimators $\hat{\phi}_{1,\lfloor T/2\rfloor}$ and $\hat{\phi}_{\lfloor T/2\rfloor+1,2\lfloor T/2\rfloor}$. (In the case of more groups, one has to repeat the coupling argument.) We seek to show that $a_{\lfloor T/2\rfloor}\hat{\phi}_{1,\lfloor T/2\rfloor}$ and $a_{\lfloor T/2\rfloor}\hat{\phi}_{\lfloor T/2\rfloor+1,2\lfloor T/2\rfloor}$ are asymptotically independent. By the Cram\'{e}r-Wold device, the asymptotic independence holds, if we show that for any constants $(k_{1},k_{2})\in\mathbb{R}^{2}$, $k_{1}a_{\lfloor T/2\rfloor}\hat{\phi}_{1,\lfloor T/2\rfloor}+k_{2}a_{\lfloor T/2\rfloor}\hat{\phi}_{\lfloor T/2\rfloor+1,2\lfloor T/2\rfloor}\to_d k_{1}Z^{(1)}+k_{2}Z^{(2)}$ where $Z^{(1)}$ and $Z^{(2)}$ are independent copies of $Z$. Let $\tilde{T}:=\tilde{T}(T) \to \infty$ be a sequence of positive integers satisfying $\tilde{T}=o(T)$ as $T\to\infty$. It holds that
\begin{align} \label{eq:decomp1}
a_{\lfloor T/2\rfloor}\hat{\phi}_{\lfloor T/2\rfloor+1,2\lfloor T/2\rfloor} & =a_{\lfloor T/2\rfloor}\hat{\phi}_{\lfloor T/2\rfloor+1,\lfloor T/2\rfloor+1+\tilde{T}}+a_{\lfloor T/2\rfloor}\hat{\phi}_{\lfloor T/2\rfloor+2+\tilde{T},2\lfloor T/2\rfloor}\nonumber \\
 & =:S_{T}^{(1)}+S_{T}^{(2)},
\end{align}
where it holds, due to (\ref{eq:limit:phi}), that
\begin{equation}
S_{T}^{(1)}=o_{p}(1).\label{eq:Sn1}
\end{equation}
Let $(y_{t}^{\star}:t\in\mathbb{Z})$ be a sequence of r.v.'s with the same distribution as that of $(y_{t}:t\in\mathbb{Z})$ and independent of $\mathcal{F}_{\lfloor T/2\rfloor} = \sigma(y_t : t \le \lfloor T/2 \rfloor)$.
By Theorem 5.1 of Rio (2017), and using that $(y_t:t\in\mathbb{Z})$ is  stationary,
\begin{equation}
P(y_{t}^{\star}\ne y_{t}\ \text{for some}\ t\ge\lfloor T/2\rfloor+k)=\beta(k)\label{eq:result:rio},
\end{equation}
where $(\beta(k):t\in\mathbb{Z})$ denotes the sequence of $\beta$-mixing coefficients. Let
\[
\hat{\phi}_{i,j}^{\star}:=(j-i+1)^{-1}\sum_{t=i}^{j}y_{t,1}^{\star}y_{t,2}^{\star}-\left((j-i+1)^{-1}\sum_{t=i}^{j}y_{t,1}^{\star}\right)\left((j-i+1)^{-1}\sum_{t=i}^{j}y_{t,2}^{\star}\right)-\phi_{0}.
\]
Note that $S_{T}^{(2)}=a_{\lfloor T/2\rfloor}\hat{\phi}_{\lfloor T/2\rfloor+2+\tilde{T},2\lfloor T/2\rfloor}^{\star}+(a_{\lfloor T/2\rfloor}\hat{\phi}_{\lfloor T/2\rfloor+2+\tilde{T},2\lfloor T/2\rfloor}-a_{\lfloor
T/2\rfloor}\hat{\phi}_{\lfloor T/2\rfloor+2+\tilde{T},2\lfloor T/2\rfloor}^{\star})$.

For any $\varepsilon>0$, using (\ref{eq:result:rio}) and that $(y_{t}:t\in\mathbb{Z})$ is stationary and $\beta$-mixing,
\begin{align*}
 & P\left[\left|a_{\lfloor T/2\rfloor}\hat{\phi}_{\lfloor T/2\rfloor+2+\tilde{T},2\lfloor T/2\rfloor}-a_{\lfloor T/2\rfloor}\hat{\phi}_{\lfloor T/2\rfloor+2+\tilde{T},2\lfloor
 T/2\rfloor}^{\star}\right|>\varepsilon\right]\\
 & \le P\left[y_{t}^{\star}\ne y_{t}\ \text{for some}\ t\ge\lfloor T/2\rfloor+2+\tilde{T}\right]
% & =\beta(\mathcal{F}_{\lfloor T/2\rfloor},\mathcal{G}_{\lfloor T/2\rfloor+2+\tilde{T}})\\
  =\beta(2+\tilde{T}) =o(1),
\end{align*}
so that
\begin{equation}
S_{n}^{(2)}=a_{\lfloor T/2\rfloor}\hat{\phi}_{\lfloor T/2\rfloor+2+\tilde{T},2\lfloor T/2\rfloor}^{\star}+o_{p}(1).\label{eq:Sn2}
\end{equation}
Hence, combining (\ref{eq:decomp1}), (\ref{eq:Sn1}), and (\ref{eq:Sn2}),
\begin{align*}
a_{\lfloor T/2\rfloor}\hat{\phi}_{\lfloor T/2\rfloor+1,2\lfloor T/2\rfloor} & =a_{\lfloor T/2\rfloor}\hat{\phi}_{\lfloor T/2\rfloor+2+\tilde{T},2\lfloor T/2\rfloor}^{\star}+o_{p}(1).
\end{align*}
Using (\ref{eq:limit:phi}), we then obtain that for any $(k_{1},k_{2})\in\mathbb{R}^{2}$,
\begin{align*}
k_{1}a_{\lfloor T/2\rfloor}\hat{\phi}_{1,\lfloor T/2\rfloor}+k_{2}a_{\lfloor T/2\rfloor}\hat{\phi}_{\lfloor T/2\rfloor+1,2\lfloor T/2\rfloor} & =k_{1}a_{\lfloor T/2\rfloor}\hat{\phi}_{1,\lfloor T/2\rfloor}+k_{2}a_{\lfloor
T/2\rfloor}\hat{\phi}_{\lfloor T/2\rfloor+2+\tilde{T},2\lfloor T/2\rfloor}^{\star}+o_{p}(1)\\
 & \overset{w}{\to}k_{1}Z^{(1)}+k_{2}Z^{(2)},
\end{align*}
where $Z^{(1)}$ and $Z^{(2)}$ are copies of $Z$, and $Z^{(1)}$
and $Z^{(2)}$ are independent since $\text{\ensuremath{a_{\lfloor T/2\rfloor}\hat{\phi}_{\lfloor T/2\rfloor+2+\tilde{T},2\lfloor T/2\rfloor}^{\star}}}$
is independent of $\mathcal{F}_{\lfloor T/2\rfloor}$.

\newpage

\section*{References}

\noindent Francq, C. and Zako\"ian, J.-M. (2006), `Mixing properties of a general class of GARCH(1,1) models without moment  assumptions  on  the  observed process', \emph{Econometric  Theory} \textbf{22}, 815--834.

\noindent Ibragimov,  I.  A.  and  Linnik,  Y.  V.  (1971), \emph{Independent  and  Stationary  Sequences  ofRandom  Variables},  Wolters-Noordhoff Series of Monographs and Textbooks on Pure and Applied Mathematics, Wolters-Noordhoff.

\noindent Lindner,  A.  (2009),  'Stationarity,  mixing,  distributional  properties  and  moments  of GARCH($p,q$) processes', in T. G. Andersen, R. A. Davis, J.-P. Kreiss and T. Mikosch, eds, \emph{Handbook of Financial Time Series}, Springer, 43--69.

\noindent Pedersen, R. S. (2020), `Robust  inference  in  conditionally  heteroskedastic autoregressions', \emph{Econometric Reviews} \textbf{39}, 244--259.

\noindent Rio,  E.  (2017), \emph{Asymptotic  Theory  of  Weakly  Dependent  Random  Processes},  Springer-Verlag, Berlin.

\bigskip

\newpage 

\section{Additional figures}

\begin{figure}[h!]%
\begin{center}%
\subfigure[${\rho'}_{R,R}(h)$]{
\includegraphics[width=0.45\linewidth]{Monte-Carlo/Figures/fig1_1_1_SA_p1.pdf}\label{fig:5:1}
}
\subfigure[${\rho'}_{R, |R|^{0.5} sign(R)}(h)$]{
\includegraphics[width=0.45\linewidth]{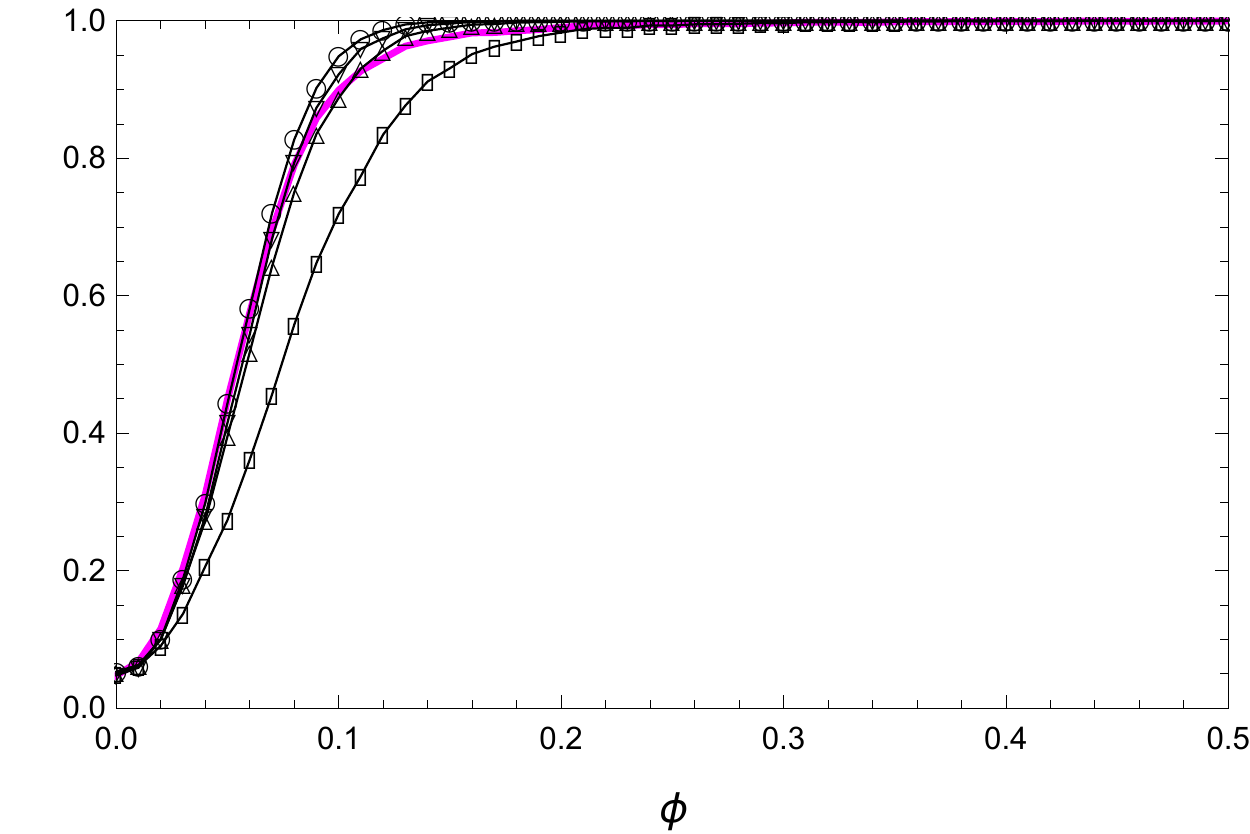}\label{fig:5:2}
}\\
\subfigure[${\rho'}_{R,|R|^{0.25} sign(R)}(h)$]{
\includegraphics[width=0.45\linewidth]{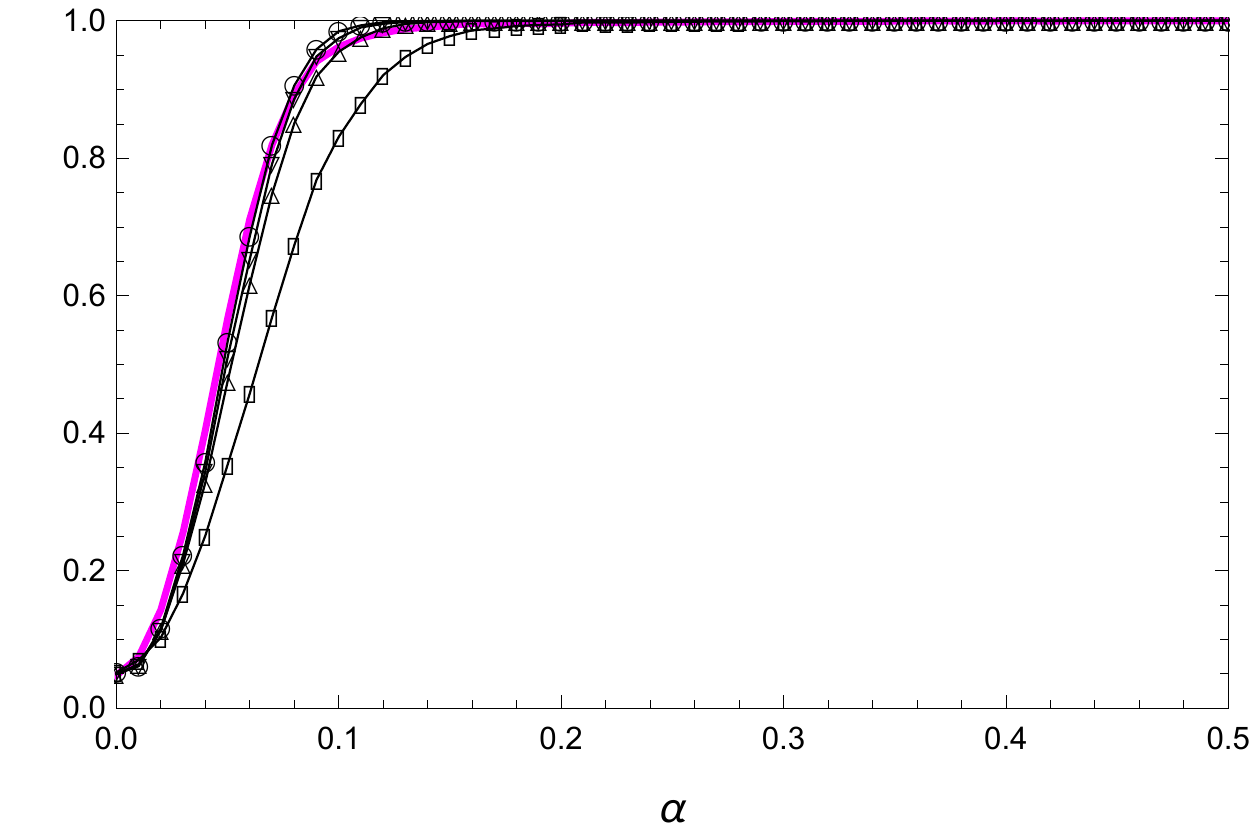}\label{fig:5:3}
}
\subfigure[${\rho'}_{R,|R|^{0.1} sign(R)}(h)$]{
\includegraphics[width=0.45\linewidth]{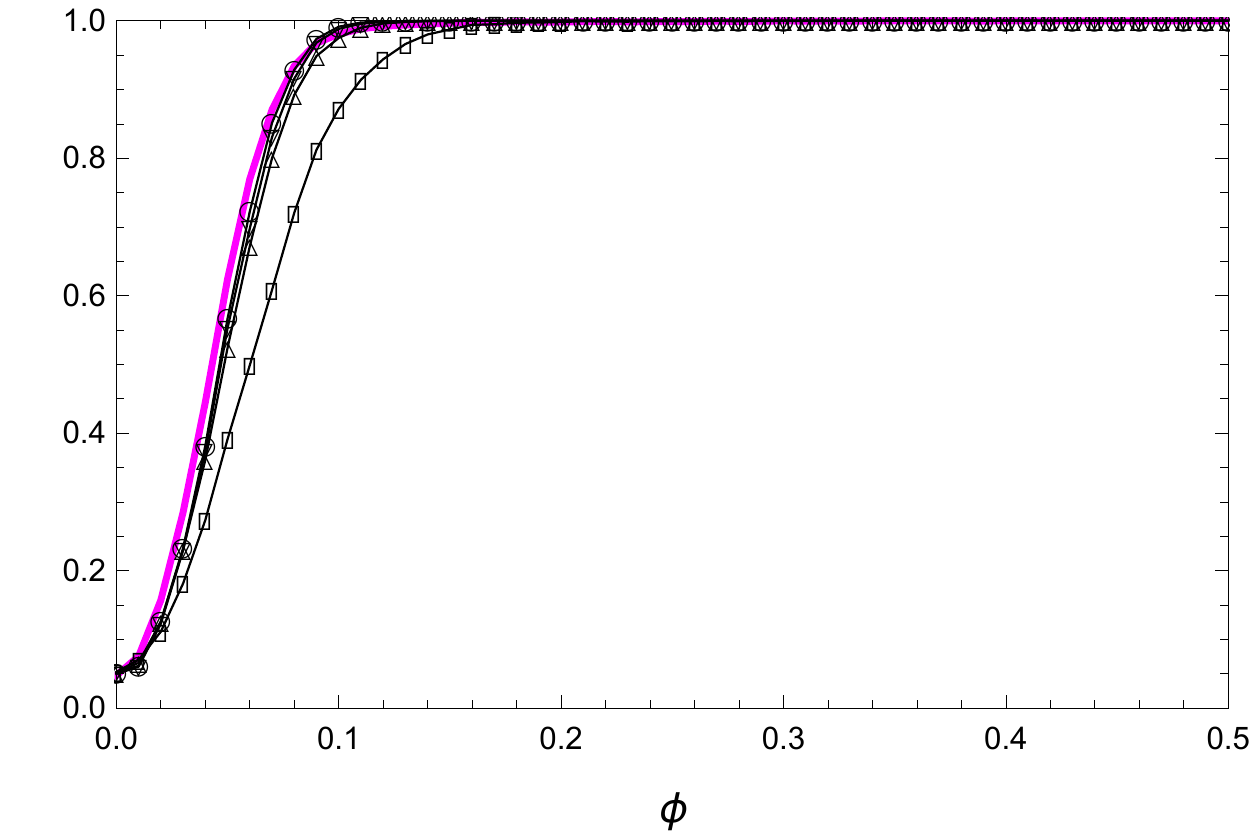}\label{fig:5:4}
}\\
\end{center}%
\caption{Size-adjusted power for ARCH(1) with $N(0,1)$ noise}
\label{fig5}
\centering
HAC:$\textcolor{magenta}{\rule[0.25em]{2em}{1.6pt}\ }$,
 $q=4: \rule[0.25em]{2em}{0.5pt} \!\!\!\!\!\!\!\!\! \square \;\;\;\;$,
$q=8:\rule[0.25em]{2em}{0.5pt} \!\!\!\!\!\!\!\!\! \bigtriangleup \;\;\;\;$,
$q=12: \rule[0.25em]{2em}{0.5pt} \!\!\!\!\!\!\!\!\! \bigtriangledown \;\;\;\;$,
$q=16: \rule[0.25em]{2em}{0.5pt} \!\!\!\!\!\!\!\!\! \bigcirc \;\;\;\;\;$
\end{figure}

\newpage

\begin{figure}[h!]%
\begin{center}%
\subfigure[${\rho'}_{R,R}(h)$]{
\includegraphics[width=0.45\linewidth]{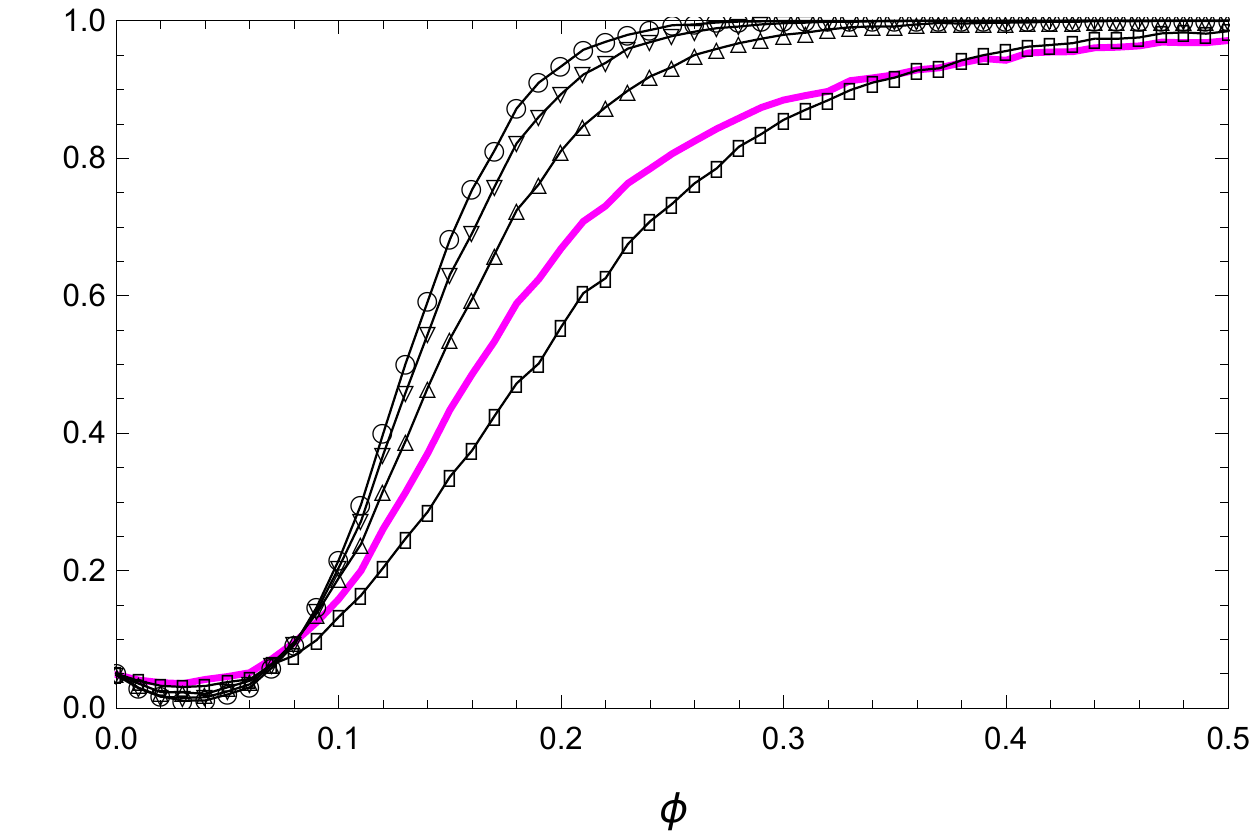}\label{fig:6:1}
}
\subfigure[${\rho'}_{R, |R|^{0.5} sign(R)}(h)$]{
\includegraphics[width=0.45\linewidth]{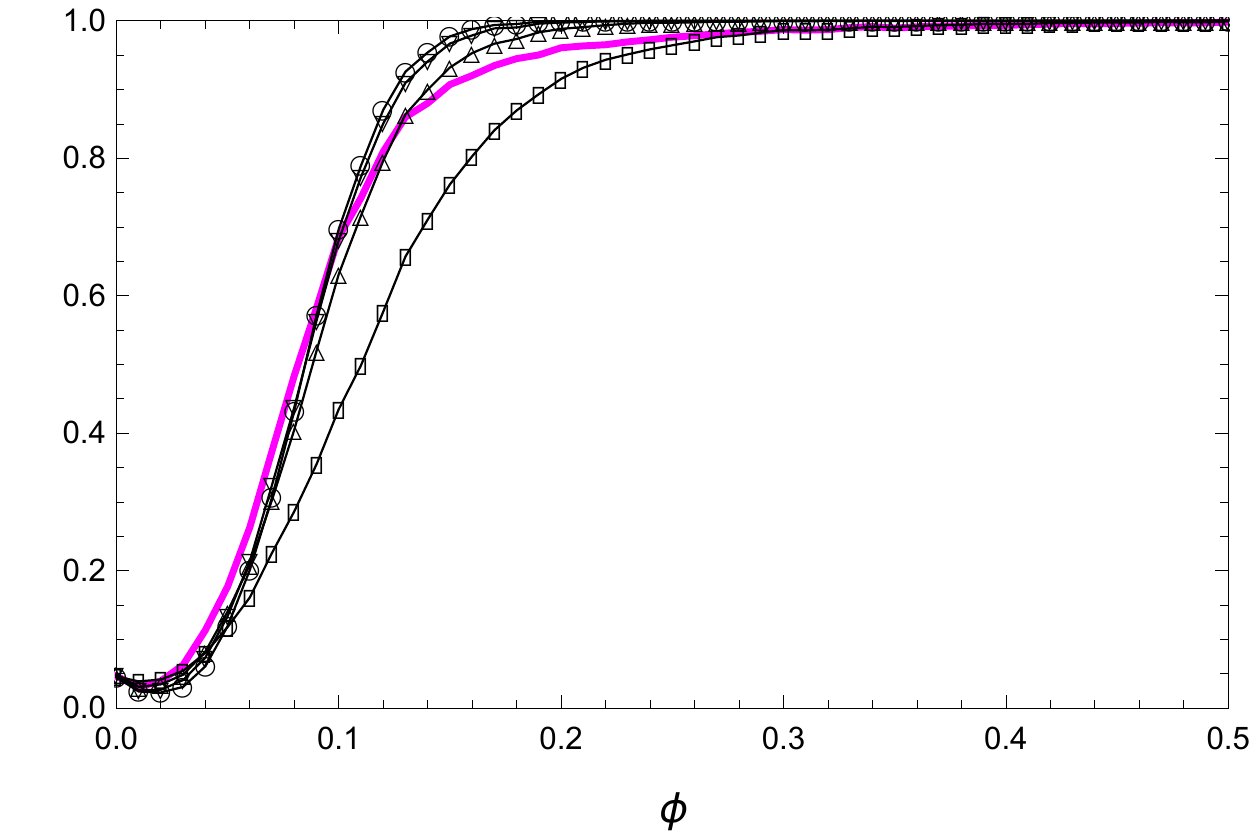}\label{fig:6:2}
}\\
\subfigure[${\rho'}_{R,|R|^{0.25} sign(R)}(h)$]{
\includegraphics[width=0.45\linewidth]{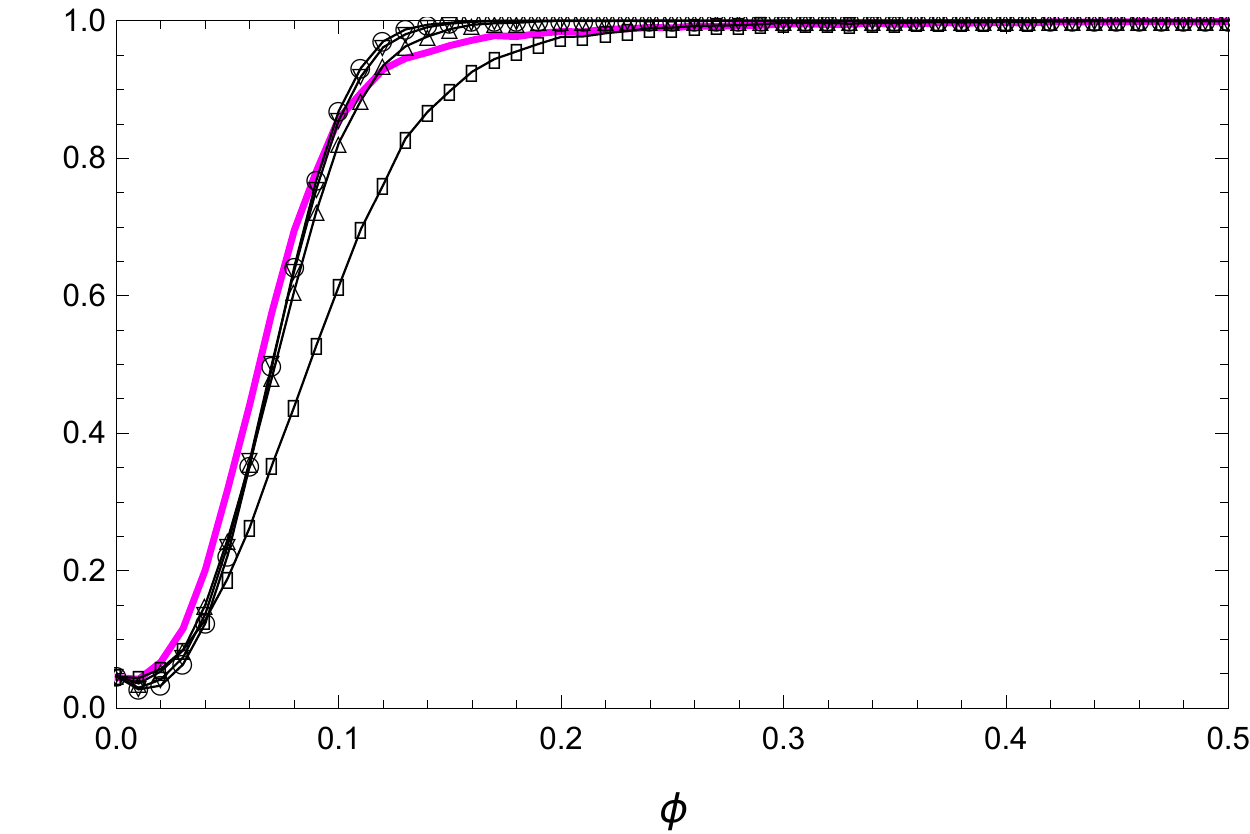}\label{fig:6:3}
}
\subfigure[${\rho'}_{R,|R|^{0.1} sign(R)}(h)$]{
\includegraphics[width=0.45\linewidth]{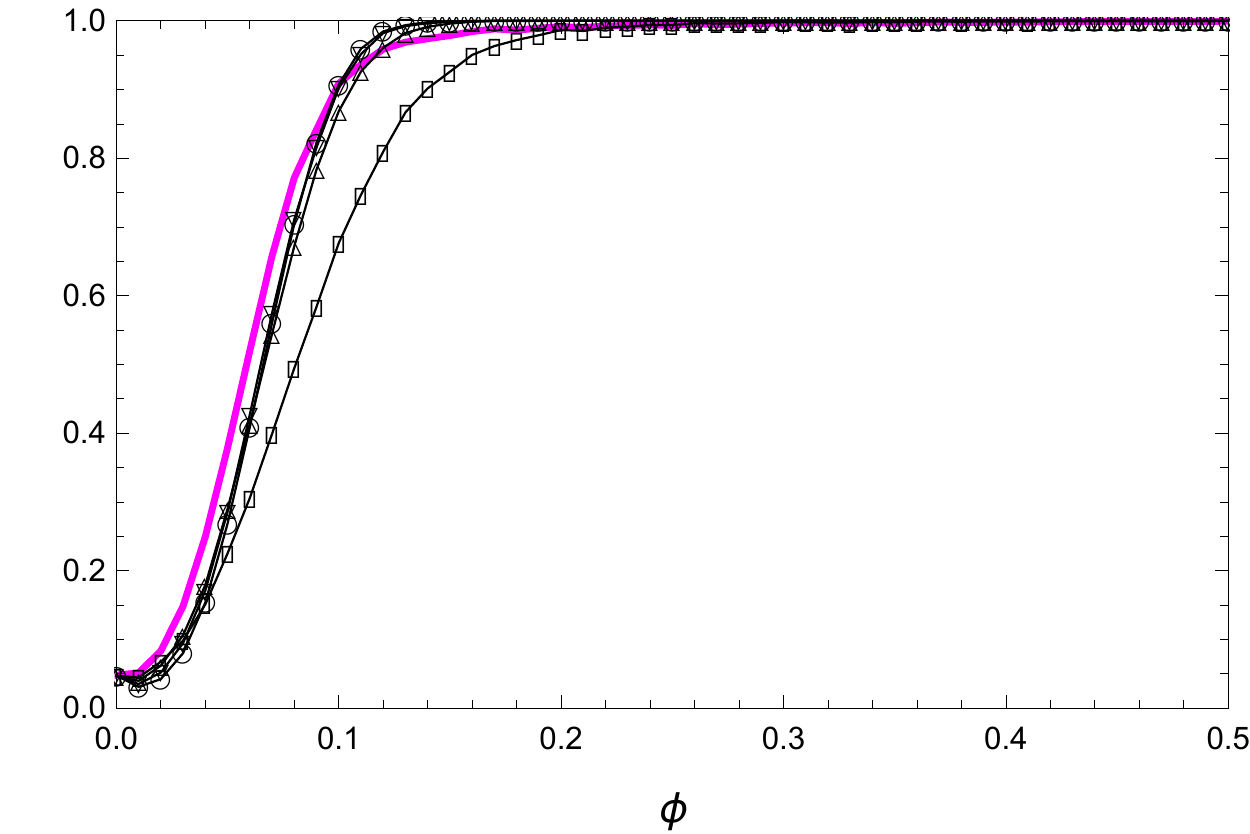}\label{fig:6:4}
}\\
\end{center}%
\caption{Size-adjusted power for ARCH(1) with $t(3,0.5)$ noise noise}
\label{fig6}
\centering
HAC:$\textcolor{magenta}{\rule[0.25em]{2em}{1.6pt}\ }$,
 $q=4: \rule[0.25em]{2em}{0.5pt} \!\!\!\!\!\!\!\!\! \square \;\;\;\;$,
$q=8:\rule[0.25em]{2em}{0.5pt} \!\!\!\!\!\!\!\!\! \bigtriangleup \;\;\;\;$,
$q=12: \rule[0.25em]{2em}{0.5pt} \!\!\!\!\!\!\!\!\! \bigtriangledown \;\;\;\;$,
$q=16: \rule[0.25em]{2em}{0.5pt} \!\!\!\!\!\!\!\!\! \bigcirc \;\;\;\;\;$
\end{figure}

\newpage

\begin{figure}[h!]%
\begin{center}%
\subfigure[${\rho'}_{R,R}(h)$]{
\includegraphics[width=0.45\linewidth]{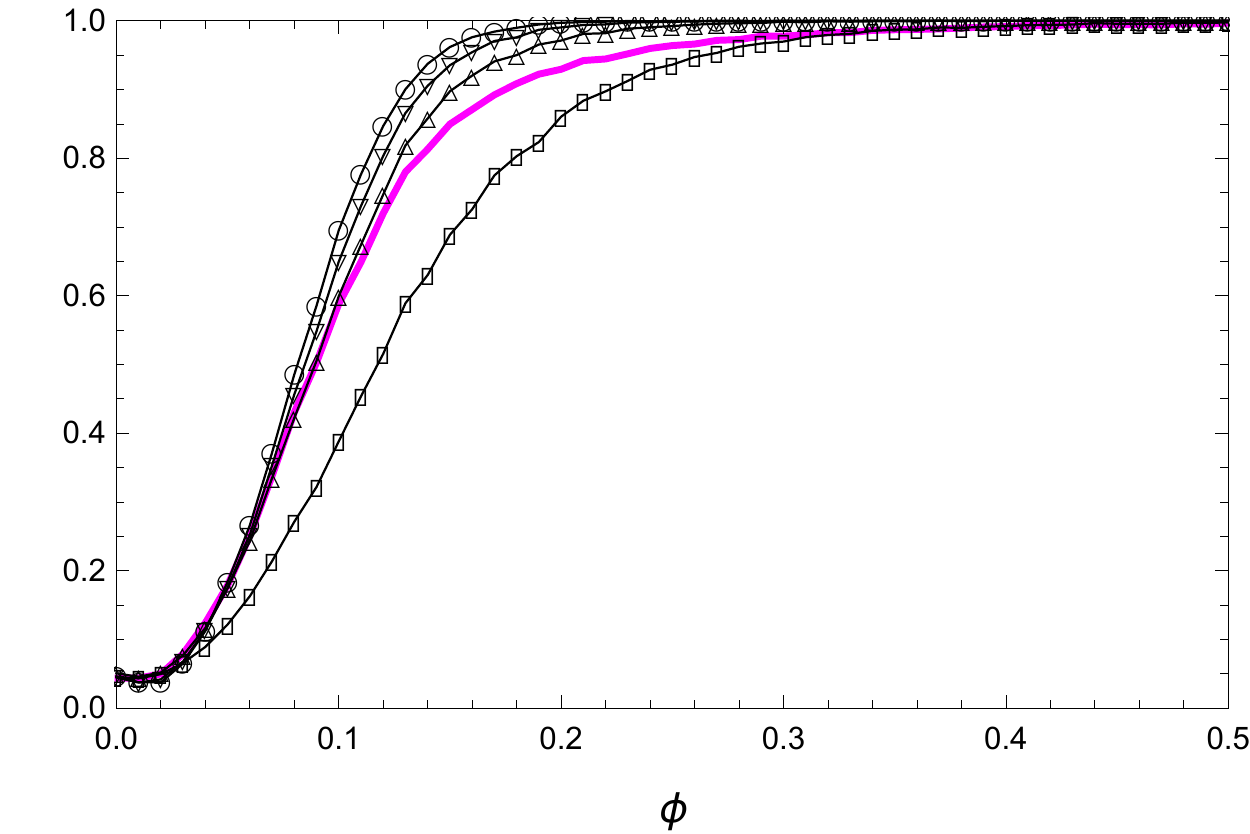}\label{fig:7:1}
}
\subfigure[${\rho'}_{R, |R|^{0.5} sign(R)}(h)$]{
\includegraphics[width=0.45\linewidth]{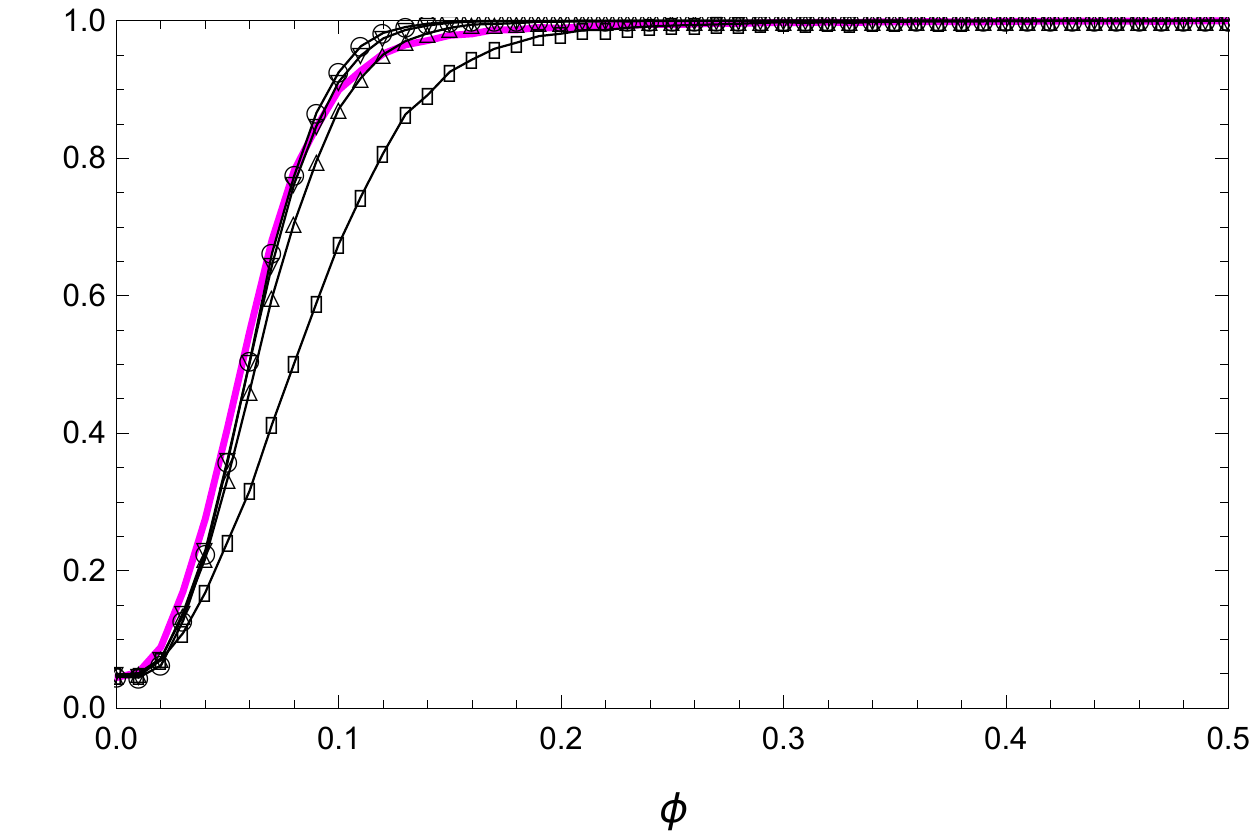}\label{fig:7:2}
}\\
\subfigure[${\rho'}_{R,|R|^{0.25} sign(R)}(h)$]{
\includegraphics[width=0.45\linewidth]{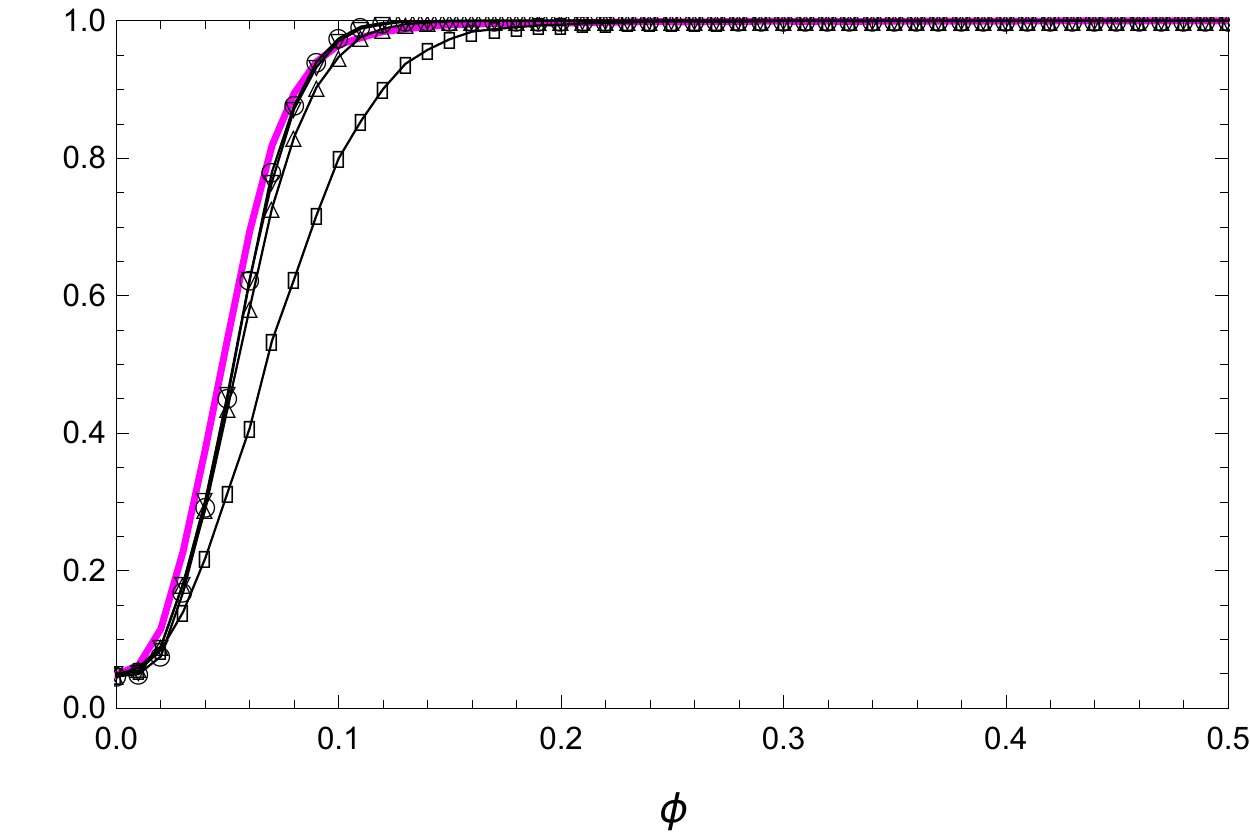}\label{fig:7:3}
}
\subfigure[${\rho'}_{R,|R|^{0.1} sign(R)}(h)$]{
\includegraphics[width=0.45\linewidth]{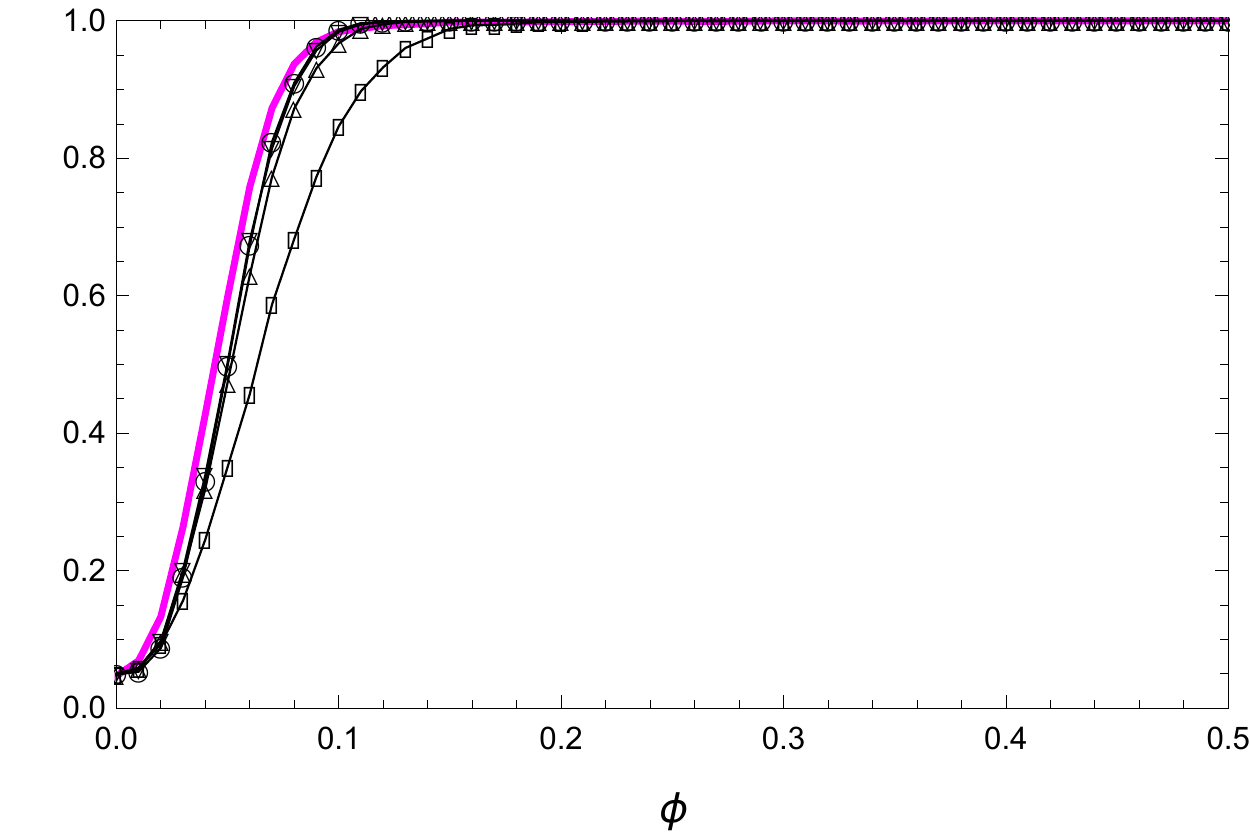}\label{fig:7:4}
}\\
\end{center}%
\caption{Size-adjusted power for ARCH(1) with $t(50,0.5)$ noise}
\label{fig7}
\centering
HAC:$\textcolor{magenta}{\rule[0.25em]{2em}{1.6pt}\ }$,
 $q=4: \rule[0.25em]{2em}{0.5pt} \!\!\!\!\!\!\!\!\! \square \;\;\;\;$,
$q=8:\rule[0.25em]{2em}{0.5pt} \!\!\!\!\!\!\!\!\! \bigtriangleup \;\;\;\;$,
$q=12: \rule[0.25em]{2em}{0.5pt} \!\!\!\!\!\!\!\!\! \bigtriangledown \;\;\;\;$,
$q=16: \rule[0.25em]{2em}{0.5pt} \!\!\!\!\!\!\!\!\! \bigcirc \;\;\;\;\;$
\end{figure}

\newpage

\begin{figure}[h]%
\begin{center}%
\subfigure[${\rho}_{R^2}(h)$]{
\includegraphics[width=0.45\linewidth]{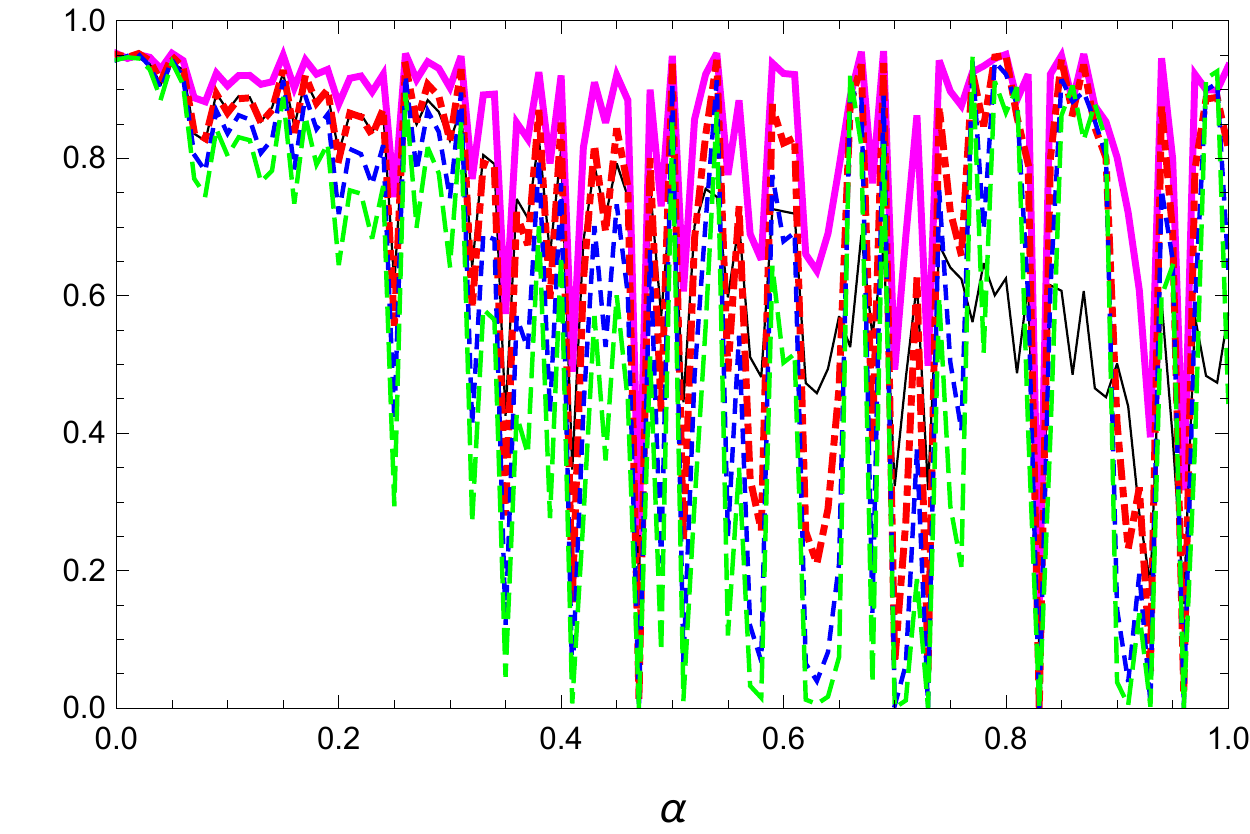}\label{fig:11:1}
}
\subfigure[${\rho}_{|R|}(h)$]{
\includegraphics[width=0.45\linewidth]{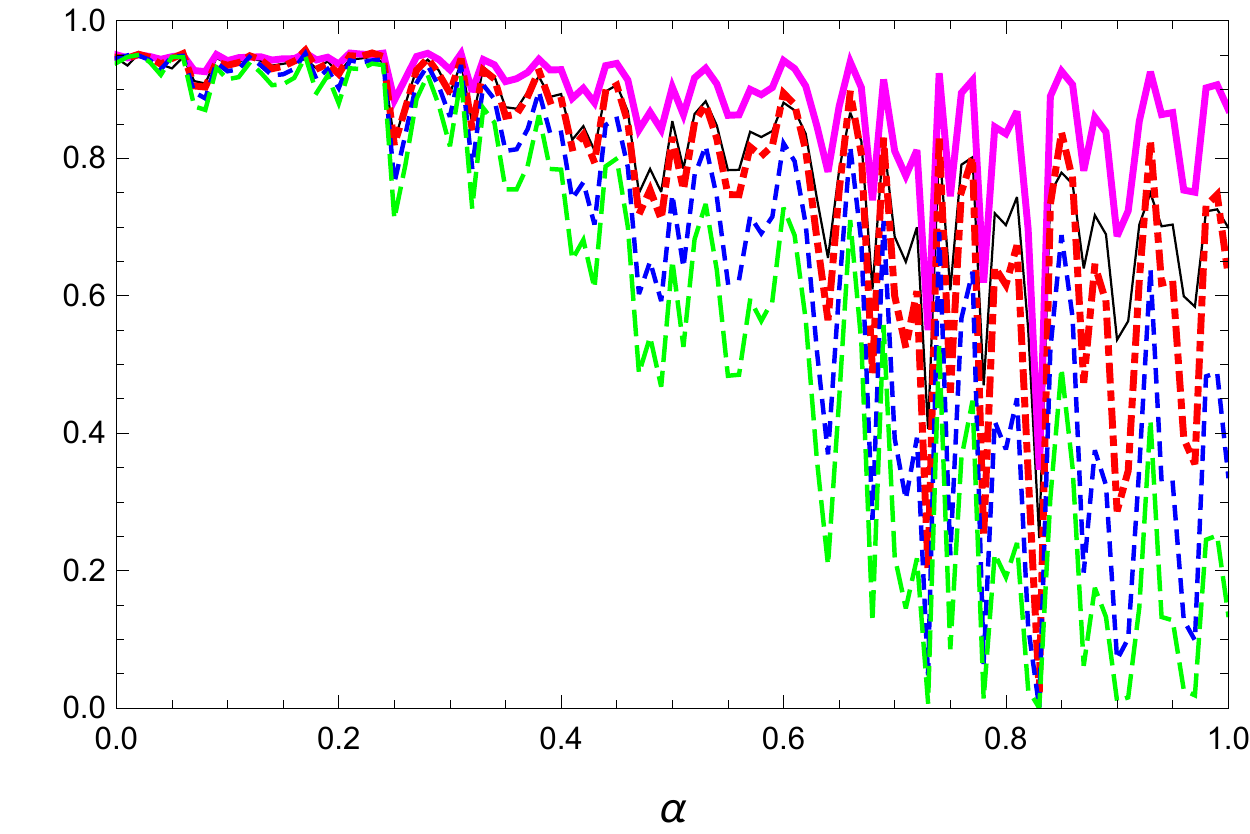}\label{fig:11:2}
}\\
\subfigure[${\rho}_{|R|^{0.25}}(h)$]{
\includegraphics[width=0.45\linewidth]{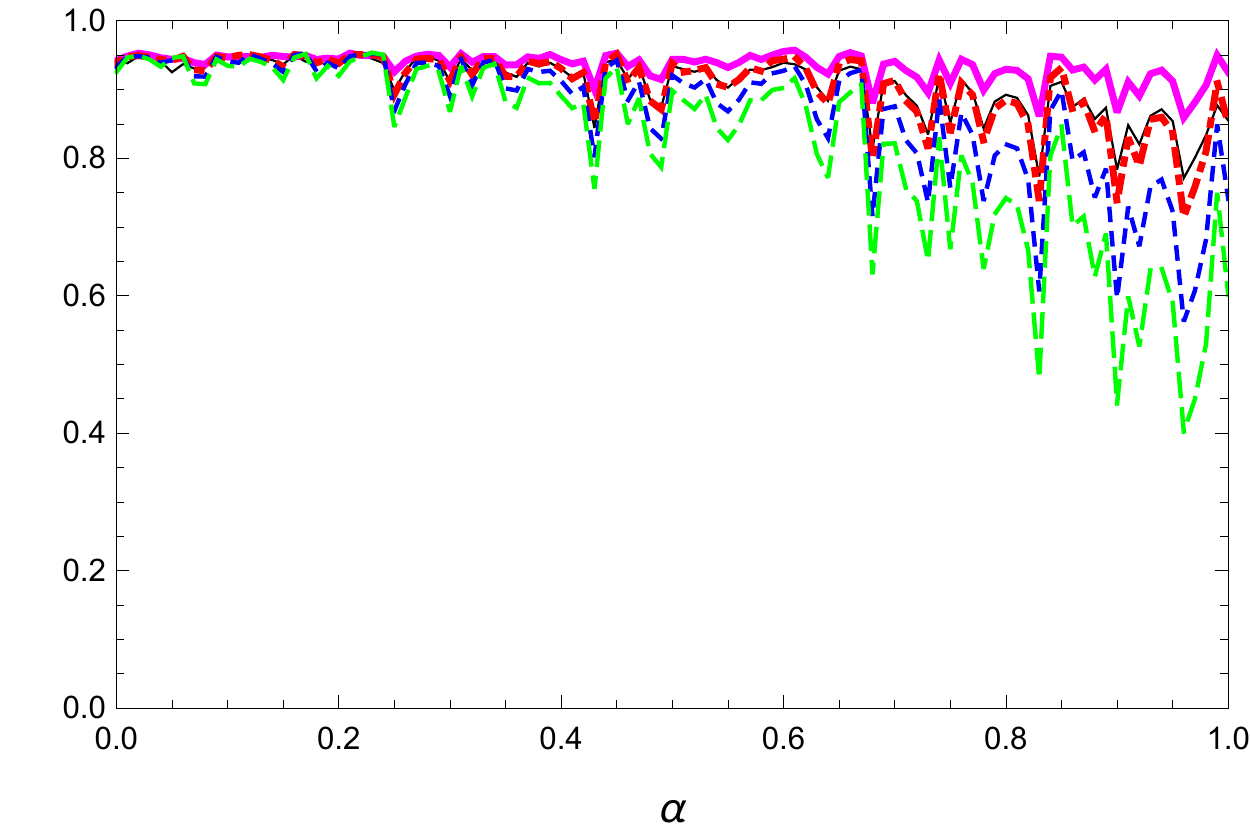}\label{fig:11:3}
}
\subfigure[${\rho}_{|R|^{0.1}}(h)$]{
\includegraphics[width=0.45\linewidth]{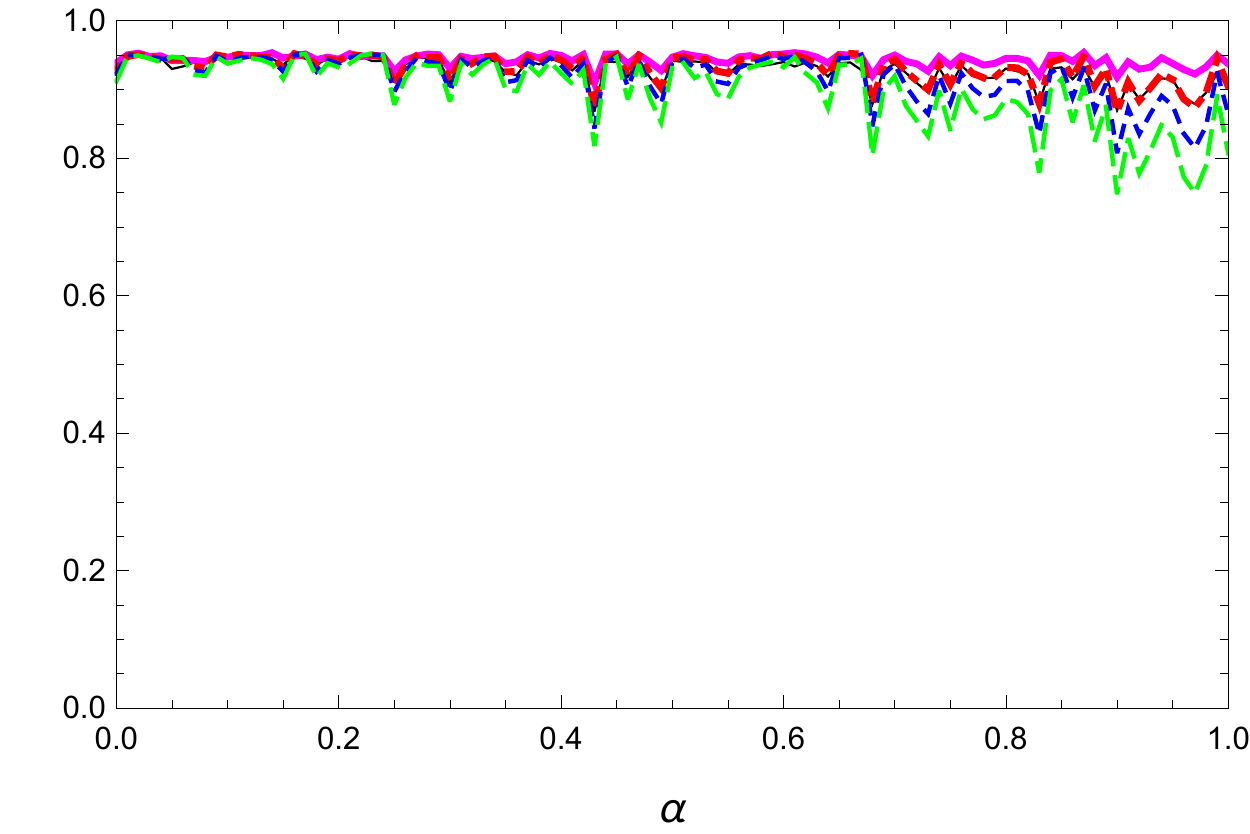}\label{fig:11:4}
}\\
\subfigure[${\rho}_{|R|^{0.1}}(h)$]{
\includegraphics[width=0.45\linewidth]{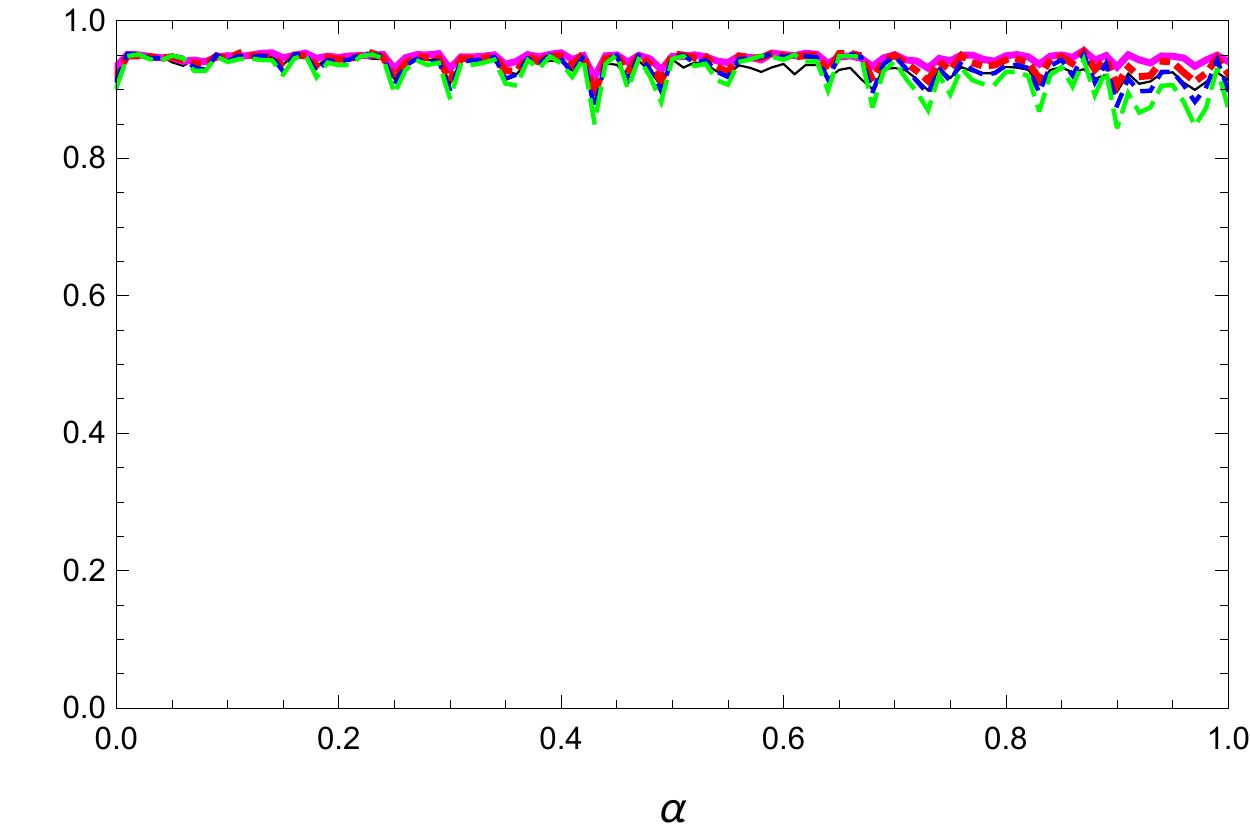}\label{fig:11:5}
}
\end{center}%
\caption{Coverage level for ARCH(1) with $t(50,0.5)$ noise}
\label{fig11}
\centering
HAC:$\textcolor{black}{\rule[0.25em]{2em}{1.6pt}\ }$,
$q=4$:$\textcolor{magenta}{\rule[0.25em]{2em}{1.6pt}\ }$,
$q=8$:$\textcolor{red}{\rule[0.25em]{0.6em}{1.7pt} \ \mathbf{\cdot} \ \rule[0.25em]{0.6em}{1.7pt} \ }$
%$A^*(t_{HLT},s_\alpha):\rule[0.25em]{2em}{0.5pt}\ $
$q=12$:$\textcolor{blue}{\rule[0.25em]{0.4em}{1.6pt} \ \rule[0.25em]{0.4em}{1.6pt}\ }$, $q=16$:$\textcolor{green}{\rule[0.25em]{0.8em}{1.6pt} \ \rule[0.25em]{0.8em}{1.6pt}\ }$
\end{figure}

\newpage

\begin{figure}[h]%
\begin{center}%
\subfigure[${\rho}_{R^2}(h)$]{
\includegraphics[width=0.45\linewidth]{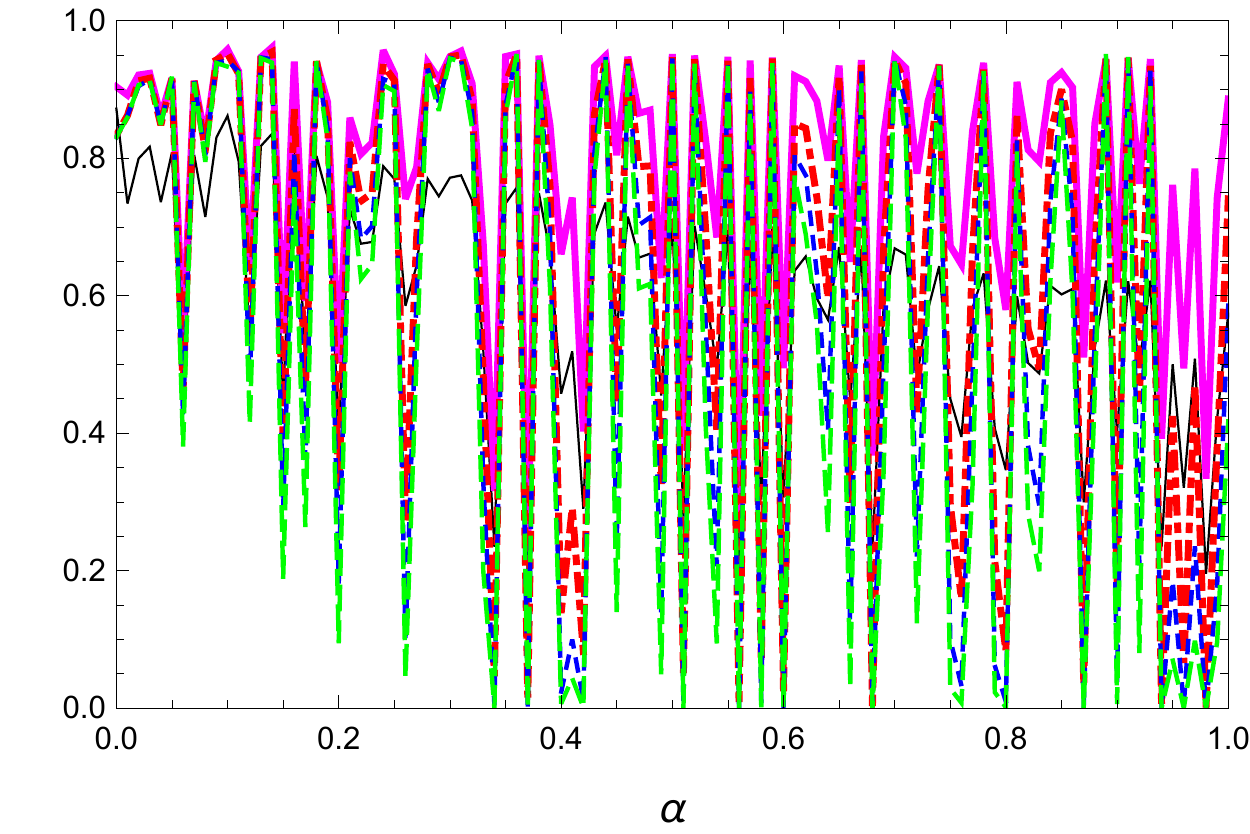}\label{fig:12:1}
}
\subfigure[${\rho}_{|R|}(h)$]{
\includegraphics[width=0.45\linewidth]{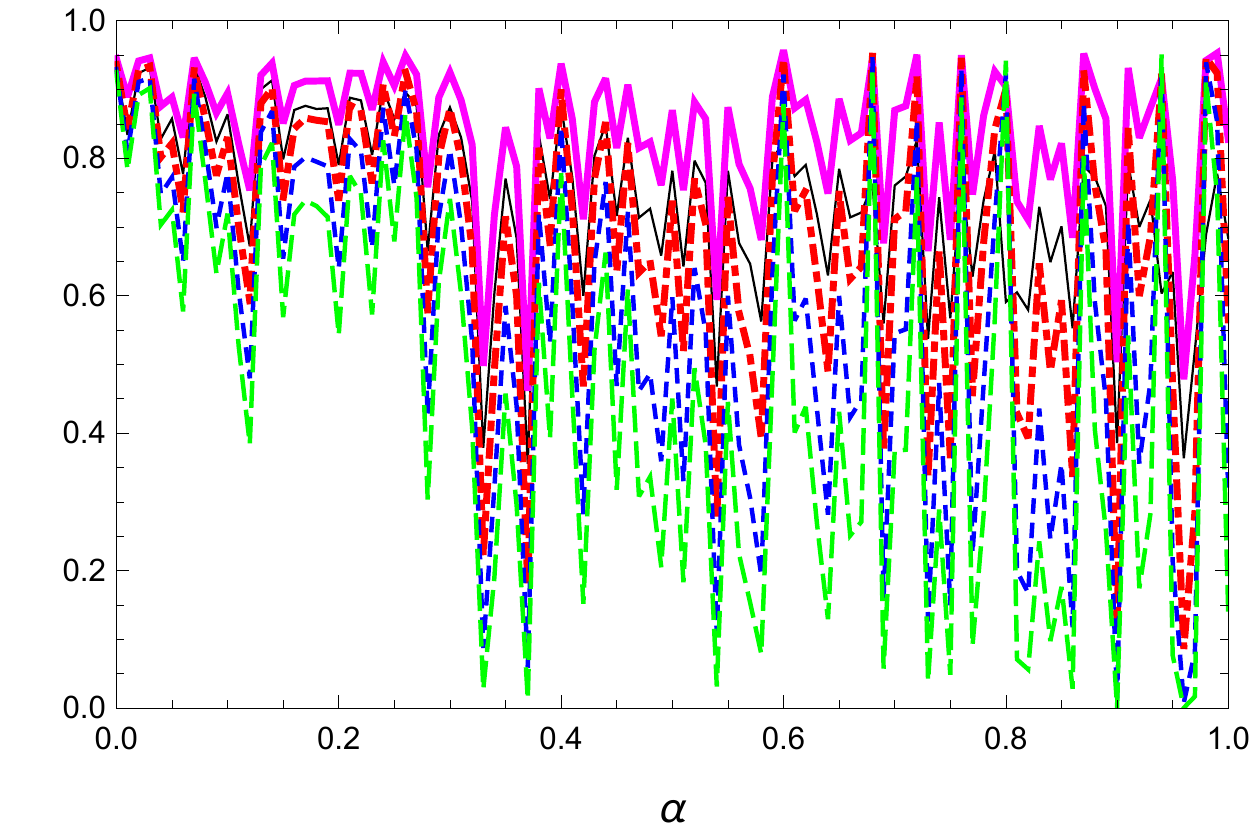}\label{fig:12:2}
}\\
\subfigure[${\rho}_{|R|^{0.5}}(h)$]{
\includegraphics[width=0.45\linewidth]{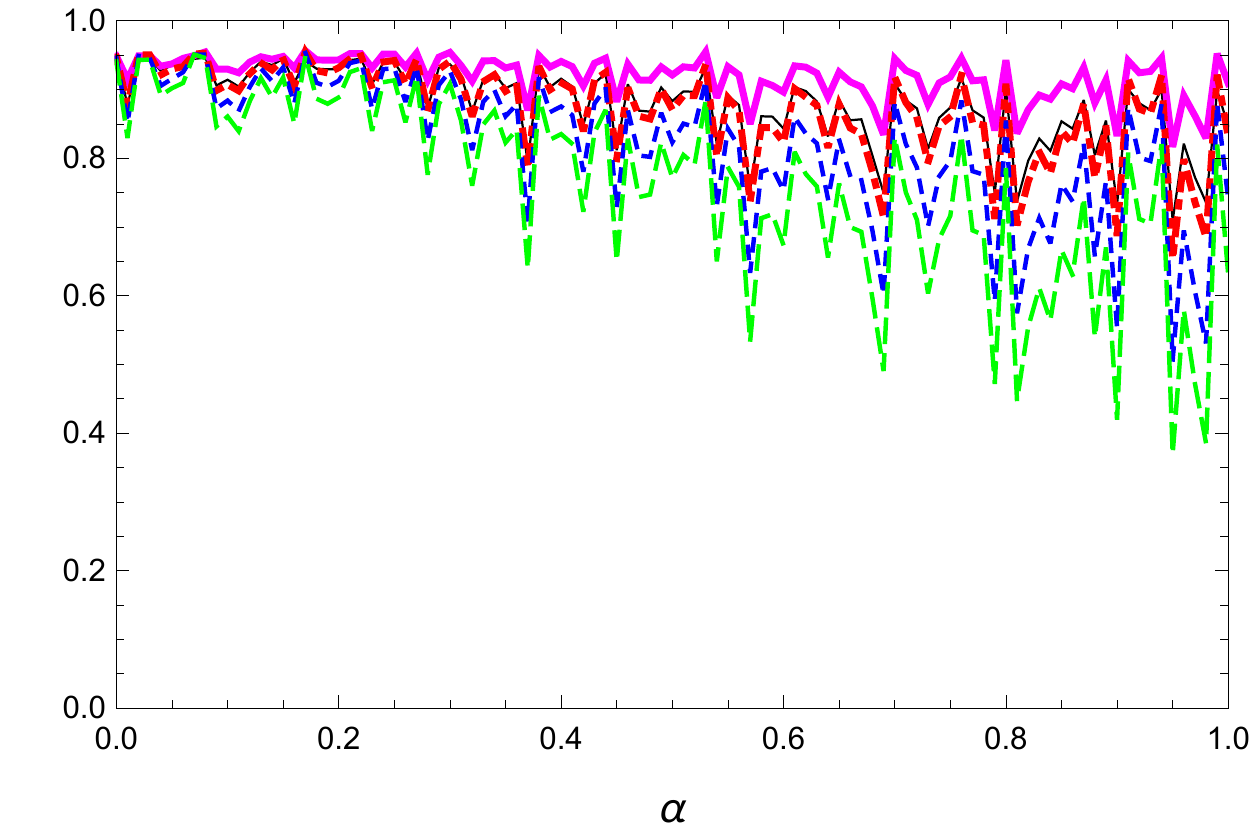}\label{fig:12:3}
}
\subfigure[${\rho}_{|R|^{0.25}}(h)$]{
\includegraphics[width=0.45\linewidth]{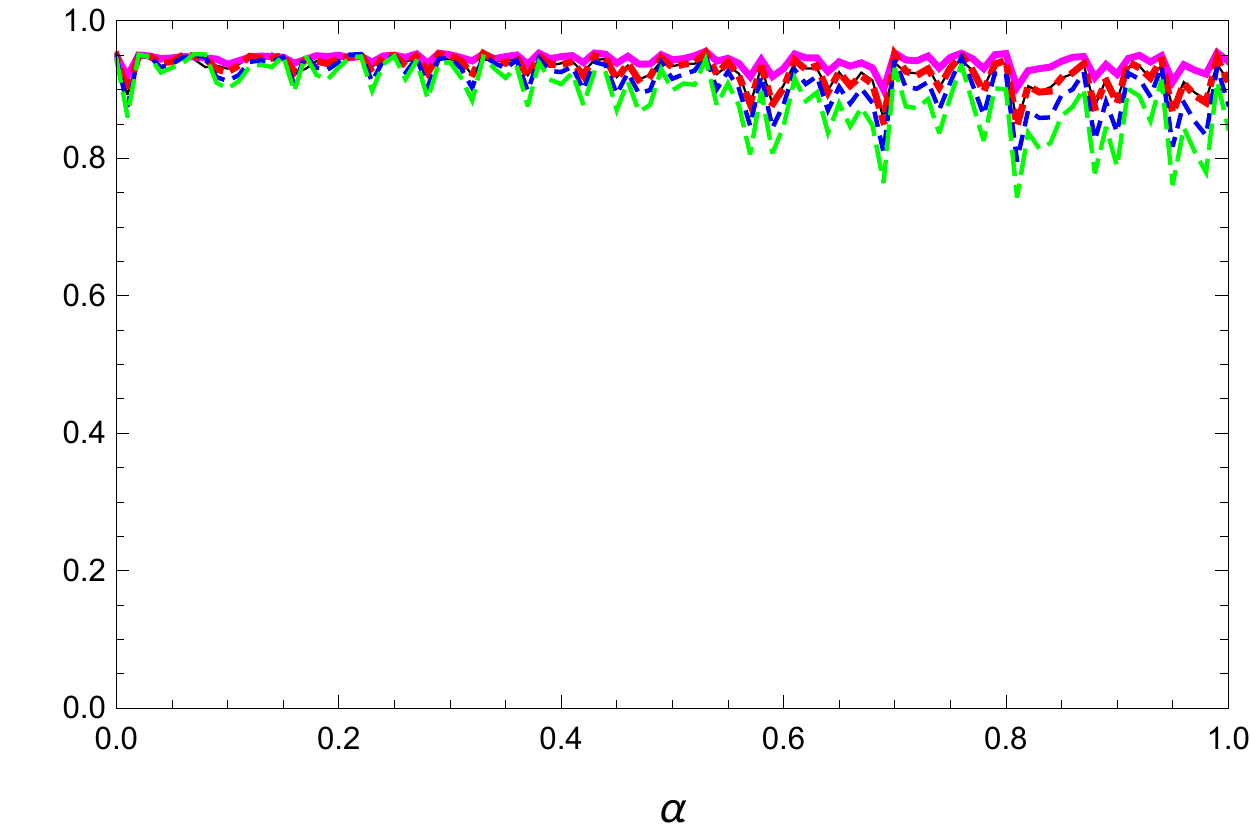}\label{fig:12:4}
}\\
\subfigure[${\rho}_{|R|^{0.1}}(h)$]{
\includegraphics[width=0.45\linewidth]{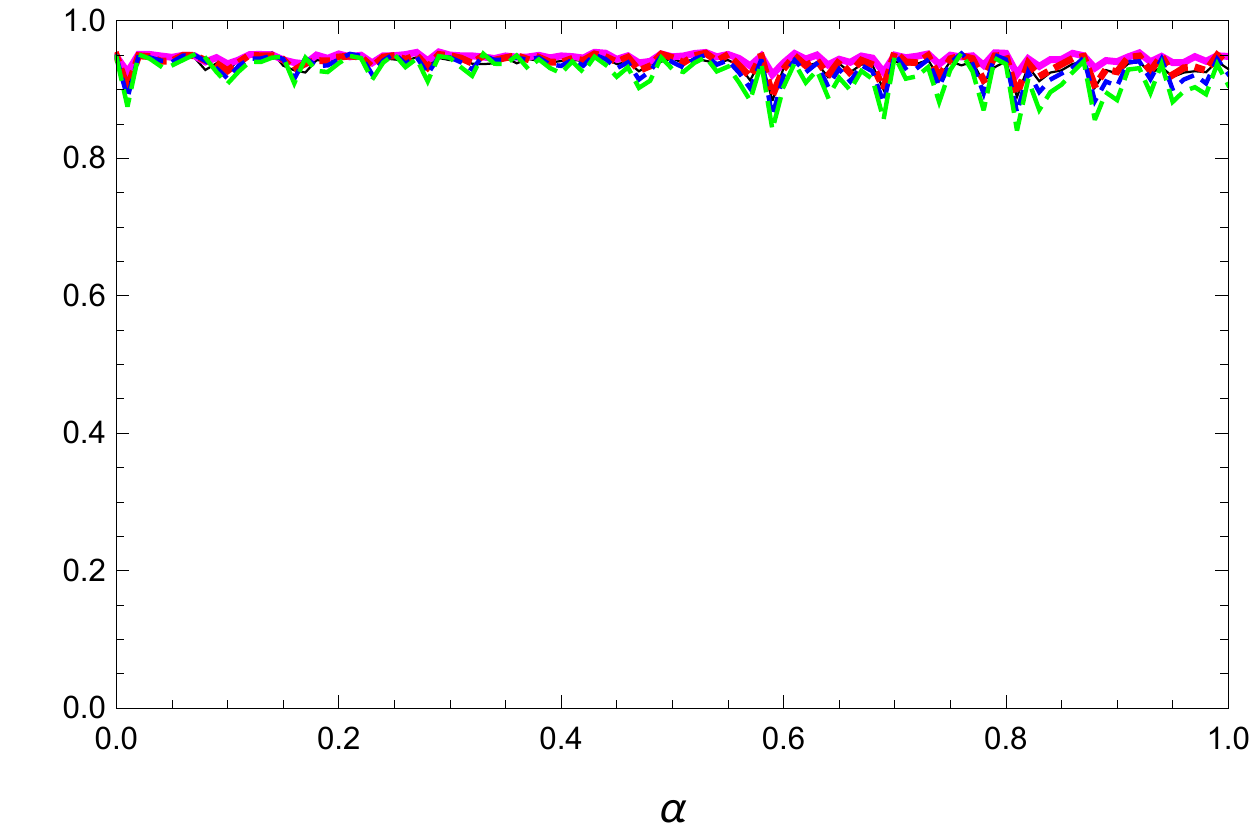}\label{fig:12:5}
}
\end{center}%
\caption{Coverage level for ARCH(1) with $t(3,0.5)$ noise}
\label{fig11}
\centering
HAC:$\textcolor{black}{\rule[0.25em]{2em}{1.6pt}\ }$,
$q=4$:$\textcolor{magenta}{\rule[0.25em]{2em}{1.6pt}\ }$,
$q=8$:$\textcolor{red}{\rule[0.25em]{0.6em}{1.7pt} \ \mathbf{\cdot} \ \rule[0.25em]{0.6em}{1.7pt} \ }$
%$A^*(t_{HLT},s_\alpha):\rule[0.25em]{2em}{0.5pt}\ $
$q=12$:$\textcolor{blue}{\rule[0.25em]{0.4em}{1.6pt} \ \rule[0.25em]{0.4em}{1.6pt}\ }$, $q=16$:$\textcolor{green}{\rule[0.25em]{0.8em}{1.6pt} \ \rule[0.25em]{0.8em}{1.6pt}\ }$
\end{figure}

 \end{document}